\documentclass[a4paper,12pt,fleqn]{article}
\usepackage{latexsym}
\usepackage{amsmath}
\usepackage{amssymb}
\usepackage[dvips]{graphicx}
\author{   }
\title{\textbf{ { \LARGE \makebox{ Disk-averaged Synthetic Spectra of Mars }} }}
\date{ Giovanna Tinetti $^{1, 2, 3}$, Victoria S. Meadows  $^{1,3}$, David Crisp  $^{4}$,  William Fong  $^{3}$,  Thangasamy Velusamy  $^{4}$, Heather Snively  $^{5}$ \newline \newline
 { \footnotesize $^{1}$ NASA Astrobiology Institute;  $^{2}$ National Academies;  $^{3}$ California Institute of Technology;  $^{4}$ NASA-Jet Propulsion Laboratory;  $^{5}$ University of California Santa Cruz     } } 
\begin{document}
\maketitle
\renewcommand{\theequation}{\arabic{equation}}
\renewcommand{\thefigure}{\arabic{figure}}
\newcommand{\ud}{\mathrm{d}}
\newcommand{\de}{\partial}
\noindent Key words: \emph{Radiative-Transfer modeling; Remote-Sensing Spectroscopy; }

\emph{Detectability of Extrasolar-Terrestrial-Planets; Planetary Science; Mars }
\\ \\
\underline{Correspondence should be directed to:} \\
Giovanna Tinetti, \\
California Institute of Technology, IPAC MS 100-22,
\\
1200 E. California, Pasadena, 91125 (CA) \\
Tel. (626)3951815;
Fax.  (626)3977021; \\
E-mail: tinetti@ipac.caltech.edu, Giovanna.Tinetti@jpl.nasa.gov
\section*{Abstract}
\begin{tabular}{|p{14.cm}|}
\hline
{\small
\emph{
The principal goal of the NASA Terrestrial Planet Finder (TPF) and ESA Darwin  mission concepts is to directly detect and characterize extrasolar terrestrial (Earth-sized) planets.   This first generation of instruments is expected to provide disk-averaged spectra with modest spectral resolution and signal-to-noise.  Here we use a spatially and spectrally resolved model of the planet Mars to study the detectability of a planet's surface and atmospheric properties from disk-averaged spectra as a function of spectral resolution and wavelength range, for both the proposed visible coronograph (TPF-C) and mid-infrared interferometer (TPF-I/Darwin) architectures.  At the core of our model is a spectrum-resolving (line-by-line) atmospheric/surface radiative transfer model which uses observational data as input to generate a database of spatially-resolved synthetic spectra for a range of illumination conditions (phase angles) and viewing geometries.  Results presented here include disk averaged synthetic spectra, light-curves and the spectral variability at visible + mid-IR wavelengths for Mars as a function of viewing angle, illumination, season.   We also considered the  appearance of an increasingly frozen Mars and simulated its detection versus real Mars with TPF-C and TPF-I as a function of spectral resolving power, 
signal-to-noise, integration time.
   }} \\
\hline
\end{tabular}
\section{Introduction}

In the past decade, over 122 planets  have been discovered orbiting other stars using indirect detection techniques 
[JPL Planetquest website]. These techniques {\it infer} the presence of a planet based on the time-varying position or luminosity behavior of the parent star. The characteristics of the known sample are constrained by the currently achievable detection techniques, which are more sensitive to larger worlds.    With the exception of  Earth-sized pulsar planets [Wolszczan and Frail, 1992], the planets found to date have lower limits on their masses that are consistent with gas giant, or Jovian planets [Udry and Mayor, 2001], rather than the smaller, rocky, terrestrial worlds like our Earth.     To extend our detection ability down to Earth-sized planets, both the National Aeronautics and Space Administration (NASA) and the European Space Agency (ESA) are developing large and technologically challenging spaceborne observatories.    The NASA Terrestrial Planet Finder (TPF) and the ESA Darwin missions are being designed to directly detect and characterize extrasolar terrestrial planets.   To detect a much fainter planet next to its parent star, the proposed design must  suppress the scattered starlight from the parent star by factors of 10$^9$ for observations in the visible and 10$^6$ for mid-infrared observations.  In addition, the design must have sufficient angular resolution  to spatially separate the signal of the planet from its parent star.

The mission architectures proposed for TPF to meet these requirements fall into two broad categories: coronagraphs operating at visible wavelengths, and interferometers operating at mid-infrared wavelengths.  Both mission concepts are expected to launch within the next 10 to 15 years [(TPF website), (Darwin website)].
The visible-light coronograph (TPF-C) concept uses a single moderate-sized  telescope with a  6 meter overture, operating at room temperature.   Very precise, stable control of the telescope optical quality is necessary to address the required billion to one image contrast between star and planet.  It may be launched around 2014.  
The infrared interferometer concept (TPF-I) uses multiple, smaller (3-4 meters) telescopes that are cooled to extremely low temperatures.   For the designs under consideration, these telescopes will be flown on separate spacecraft separated by distances of a few hundred meters.  While the image contrast requirement is only a million to one at mid-infrared wavelengths, a large baseline is required to provide sufficient angular resolution to separate the planet from its parent star. Very exacting stability and optical quality are also required for this design. TPF-I, will be probably conducted jointly with the European Space Agency-Darwin mission, and is planned to be launched before 2020.

These TPF concepts are expected to search for terrestrial planets around a core set of at least 35 nearby stars, with a sample potentially as large as 165 stars.   The target stars are preferentially selected from those nearby stars whose optical spectra indicate sizes and temperatures similar to our own Sun, namely dwarf stars of spectral type F, G and K.   In addition to the F, G and K core set, cooler stars of type M may be added to increase the sample size. TPF-I will  study individual planets orbiting  parent stars previously observed by TPF-C and also new ones beyond the reach of TPF-C.
 In addition to detecting terrestrial-sized planets, TPF will attempt to  determine whether these planets have a high probability of being habitable.   Habitability here is defined as being able to support liquid water on the planetary surface.  First order estimates of habitability will be determined from the observed separation between the planet and its star.   However, a planet's surface temperature depends not only on the incident stellar  radiation, but also on its surface albedo and the presence of an atmosphere, and the  prevalence of greenhouse gases.   In our  solar system, enhancement of planetary surface temperatures by atmospheric greenhouse effects ranges from 5$^{\circ}$C of warming for Mars to 530$^{\circ}$C for Venus.   Measuring the presence and abundance of atmospheric constituents via remote-sensing spectroscopy, searching for spectral signatures from the planet's surface, and monitoring the planet's appearance as a function of time will provide clues not only to habitability, but may reveal imbalancies in the planetary environment that  indicate the presence of life. 

In Earth and planetary science, spatially-resolved spectroscopy is arguably the most powerful technique available for retrieving the characteristics of planetary surfaces and atmospheres via remote sensing. However, the first generation of instruments for studying extrasolar planets will only be able to spatially-resolve the disk of the target planets, and so are expected to provide disk-averaged spectra with modest spectral resolution R  (potentially R$\sim$75 in the visible, and R$\sim$25 in the IR) and signal-to-noise. 

With the launch of missions like TPF/Darwin only a decade away, we must understand the extent to which the environments of extrasolar terrestrial planets can be characterized for habitability or the presence of life using disk-averaged visible or infrared spectra via computer modeling.  As part of the NASA Astrobiology Institute's Virtual Planetary Laboratory  (P.I. V. S. Meadows) we are developing computer models to understand what planetary surface and atmospheric properties can be retrieved from disk-averaged spectra over a range of spectral resolutions.   One of the first steps in this process is to produce comprehensive spatially and spectrally-resolved models of the planets in our own solar system.  The planet Mars provides a relatively simple example of a planet at the edge of a habitable zone.   Although the atmosphere and surface of Mars have been extensively observed from orbit and other models have been generated to describe this spectrum  [Lellouch et al., 2000], these observations and models are not directly applicable to studies of the environments. In particular, the spatial resolution and viewing geometry is substantially different from that for disk-integrated observations of distant worlds.  Besides, more importantly, existing observations cover only a fraction of the solar and thermal spectra. We have developed  a computer model that uses environmental measurements as inputs, and is validated against available observational data. It therefore provides a realistic, versatile, and comprehensive tool to explore specific questions that cannot be answered using the data alone.   For example, we can  alter planetary properties to better understand the detectability of ice on a successively more frozen Mars and determine how we can best detect and distinguish it from present day Mars (c.f. Section 6.3).

To create these planetary models, we are implementing a suite of realistic atmospheric radiative transfer models and observational instrumentation and system simulation tools. These planetary models will be used to generate high spectral resolution, spatially resolved models of Venus, Earth, Mars, and Titan. Spectra generated by these models can then be averaged over the planet's disk, and processed with the observational system simulators to provide a quantitative assessment of the detectability of planetary characteristics.

In this paper, we explore the first results of this modeling effort, which concentrates on the planet Mars.   Mars has sufficient observational data to provide adequate input and validation for the models, and provides a good example of a likely abiotic planet.   A subsequent paper will discuss our recent results for the more complicated Earth model [Tinetti et al., 2004a], [Tinetti et al., 2004b].   In our approach section, we describe the radiative transfer algorithms, their data requirements and the observational system modeling tools adopted for this investigation.    In the result section we explore the detectability of Martian atmospheric and surface features in disk-averaged spectra as a function of wavelength observed, solar illumination and viewing angle for two days on Mars, one near Northern summer solstice ($L_{s}$ = 105$^{\circ}$) and another corresponding to $L_{s}$ = 211.6$^{\circ}$, closer to Northern autumn equinox.

\section{Modeling Approach}

The objective  of our modeling approach was to create a  model that could provide realistic representations of the spectral appearance of a planet for a range of different observed spatial scales and wavelengths, as a function of solar illuminations and observer viewing angles.    To create this model, we needed to carefully choose our spatial sampling scheme, the input data, and its resolution, radiative transfer  models to create realistic spectra  over large wavelength ranges and for different solar illumination conditions, and the algorithms required to average results over the planet's disk to produce disk-averaged spectra for different viewing geometries.   The techniques that we chose to implement in this model are described in detail in the following sub-sections.  

\subsection{Spatial sampling}

The first step in implementing this model was to adopt an architecture for the spatial sampling of the sphere, to establish the relationship between individual soundings and disk-averaged properties of the atmosphere and  surface.

For this model, the choice of spatial sampling scheme was particularly important because we did not wish to introduce spurious spectral artifacts associated with the number of density of spatial samples on the sphere that could be viewed from different viewing geometries.   Since we may see an extrasolar planet from almost any viewing angle, including pole-on, many classical spatial grids used for Earth science such as latitude-longitude grids were not appropriate for this model, as they had built-in preferential directions for viewing and/or spatial grid boxes of unequal size when viewed from different directions.  For this application, it was also important to have a mathematical structure able to support analytical and geometrical transformations  on the sphere (integrals, rotations, projections etc.),
 that would  facilitate fast and accurate tests of increased and degraded spatial resolving power.

Therefore, the principal criteria driving our selection of a spatial grid on the sphere were: 1) a geometry showing little or no change in mapping as a function of viewing direction -  this criterion could be satisfied either by same size/same shape samples, or segments of different sizes and shapes, but in a uniform pattern across the sphere; 2) a geometry that can be broken up into a reasonable number of bins/points and is robust to multiple scales.  Our criteria were best satisfied by the existing Healpix formalism.  

\paragraph{Healpix}

\begin{figure}
\begin{center}
\includegraphics[width=14 cm]{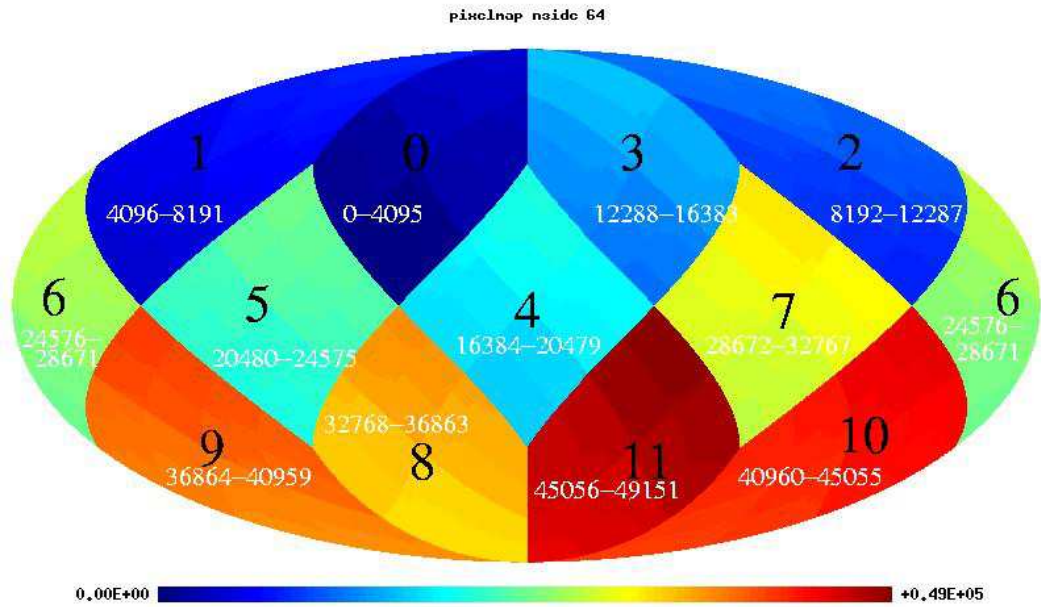} 
\caption{ { \footnotesize \emph{ This figure illustrates the Healpix resolution increase from the base level  
 $N_{pix}^{tot} = 12$, to the level where the sphere is partitioned into $N_{pix}^{tot} = 49152$ tiles.  } }}  \label{fig:1}
\end{center}
\end{figure}

Healpix (Hierarchical Equal Area and iso-Latitude Pixelisation) was originally designed for the ESA Planck and NASA WMAP (Wilkinson Microwave Anisotropy Probe) missions [G\'orski, et al., 1998], to map the cosmic microwave background.   However, Healpix also satisfies our requirements for mapping terrestrial planets via the following three essential properties: 
1) under this scheme, the sphere is hierarchically divided into curvilinear quadrilateral tiles. The lowest resolution partition is comprised of 12 base tiles. Resolution  increases by dividing each tile into four new ones, allowing fast and reversible upgrading and degrading of the resolution  (fig. \ref{fig:1}). 
2) areas of all tiles at a given resolution are identical and the same number of tiles is visible from any viewing geometry, and 3) pixels are distributed on lines of constant latitude. This property allows easy conversion between spherical coordinates and Healpix pixels, and facilitates mathematical transformation on the sphere (rotations, projections, integrals).

\subsection{Input data}
To assess the detectability of seasonal variations, we have run our simulations for
 two particular days on Mars.   These have subsolar longitudes of $L_{s}$ = 104.6$^{\circ}$
and  $L_{s}$ = 211.6$^{\circ}$, corresponding to dates just after Northern summer solstice and just after Northern autumnal equinox.  We have assumed [James et al., 1992] polar caps extending for $\sim 15^{\circ}$ to the South in the northern hemisphere and $\sim 40^{\circ}$ to the North in the southern hemisphere for $L_{s} = 104.6^{\circ}$, and $\sim 35^{\circ}$ to the South in the northern hemisphere and $\sim 30^{\circ}$ to the North in the southern hemisphere in the case of  $L_{s} = 211.6^{\circ}$.

\paragraph{Physical-chemical properties of the atmosphere and surface}

\begin{figure}
\begin{center}
\includegraphics[width=6.5 cm]{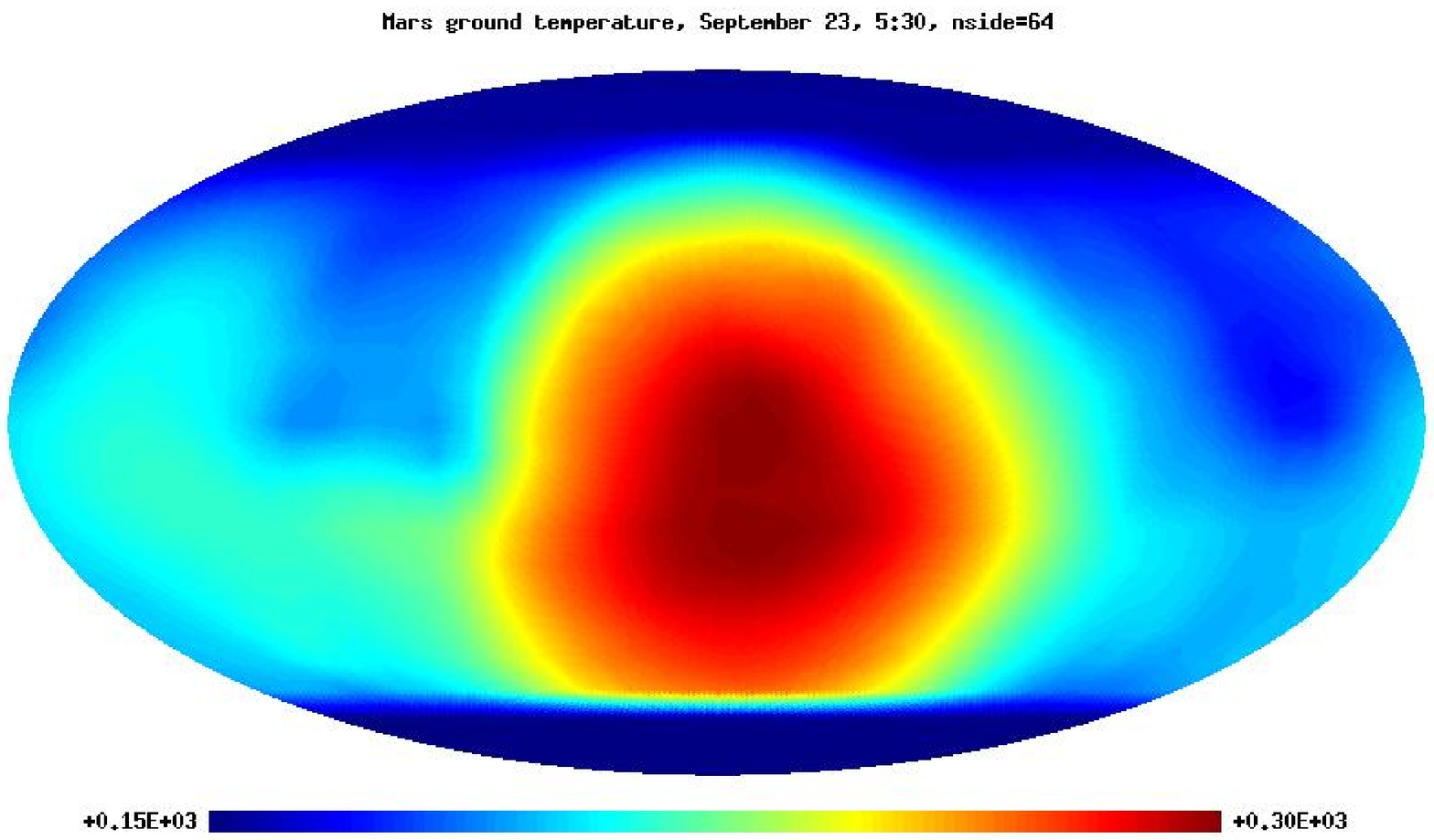} 
\includegraphics[width=6.5 cm]{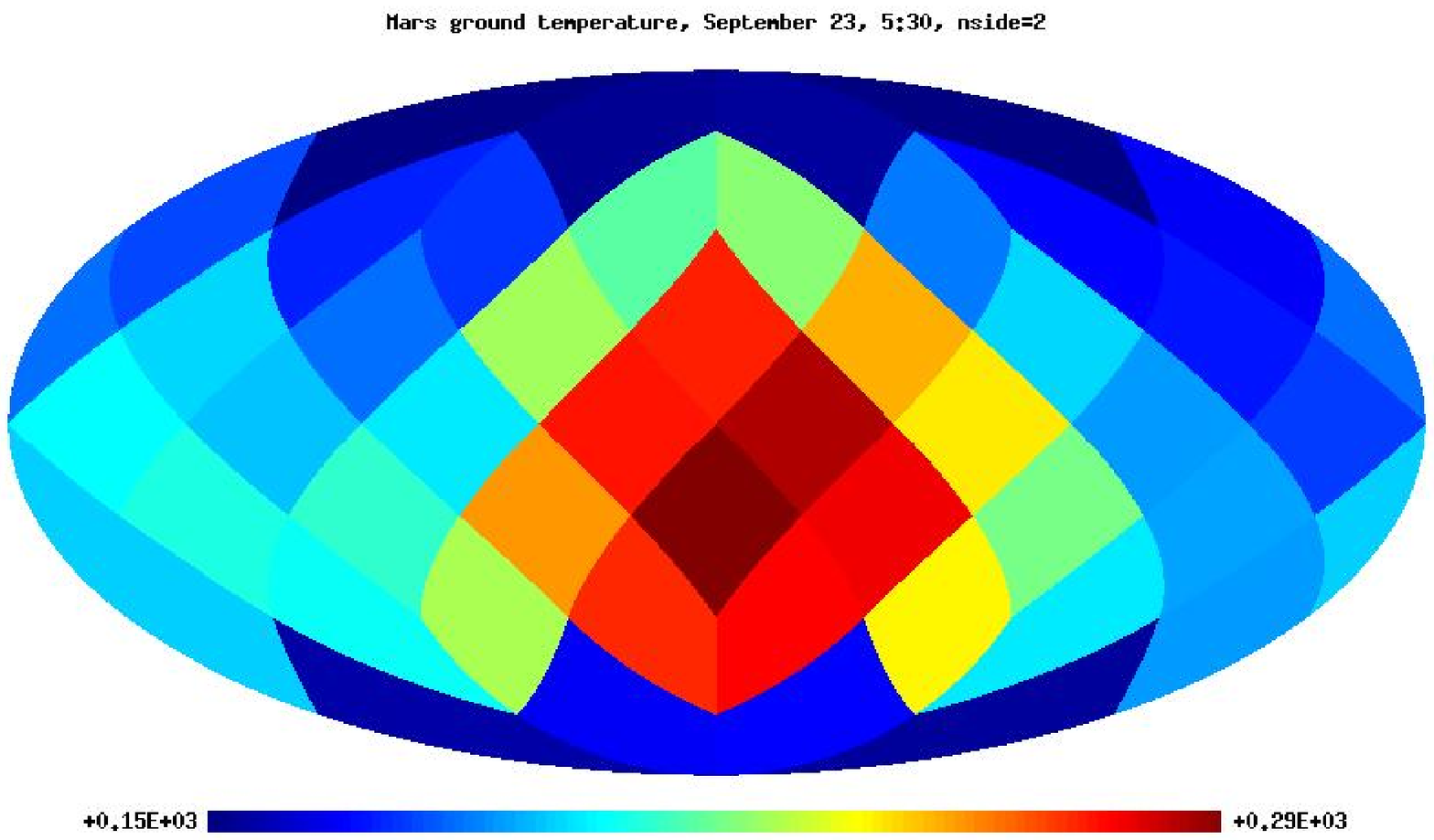} 
\includegraphics[width=9 cm]{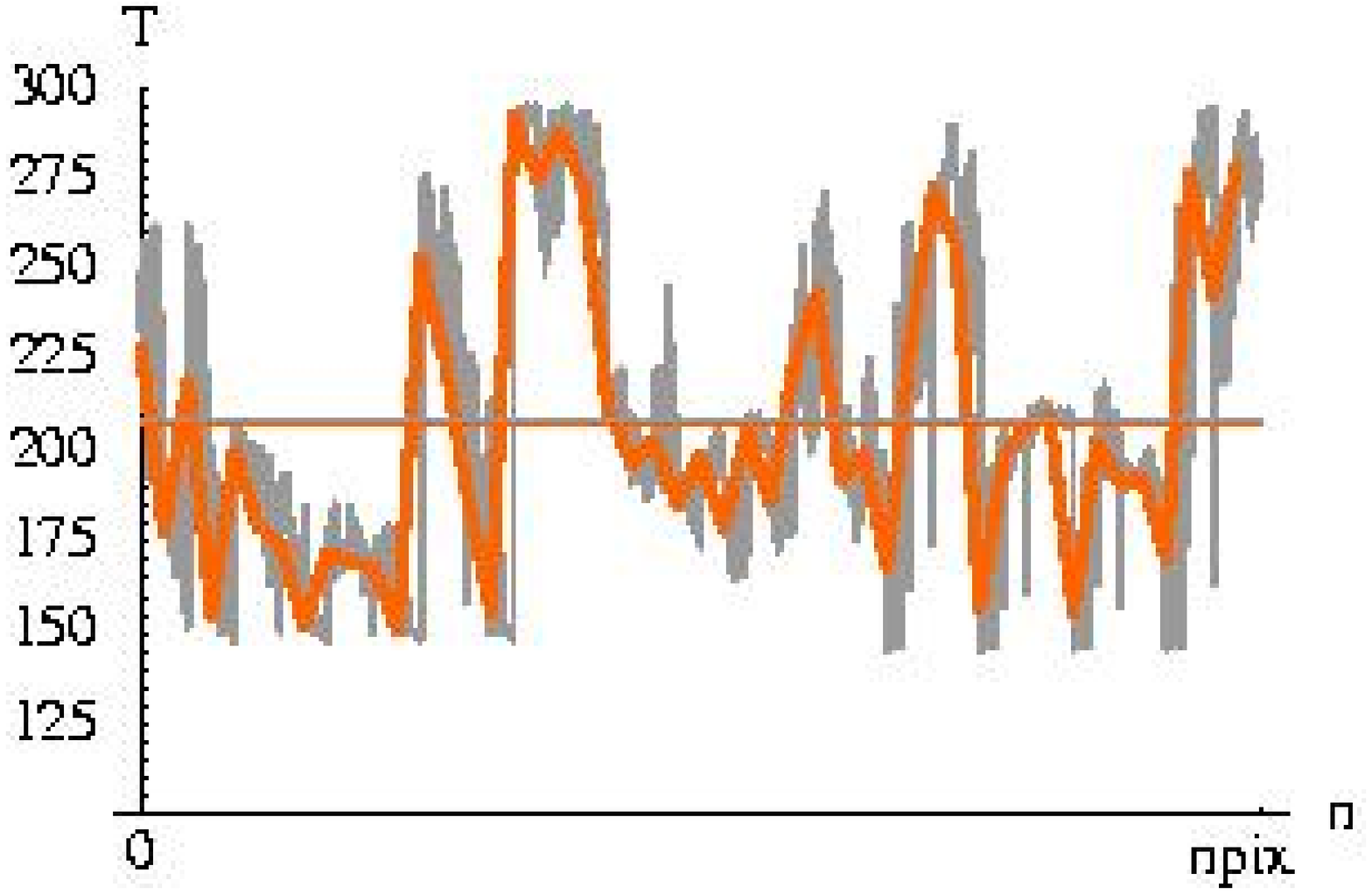}

\caption{ { \footnotesize \emph{ Simulation of Mars surface temperature (in K), with two different spatial resolutions corresponding to $N_{pix}^{tot} = 49152$ and $N_{pix}^{tot} = 48$.
We can see from the lowest diagram $T, n_{pix}$ that the two different resolutions give similar global averages (the lowest resolution is plot by the orange line).  } }}  \label{fig:2}
\end{center}
\end{figure}

\begin{figure}
\begin{center}
\includegraphics[width=12. cm]{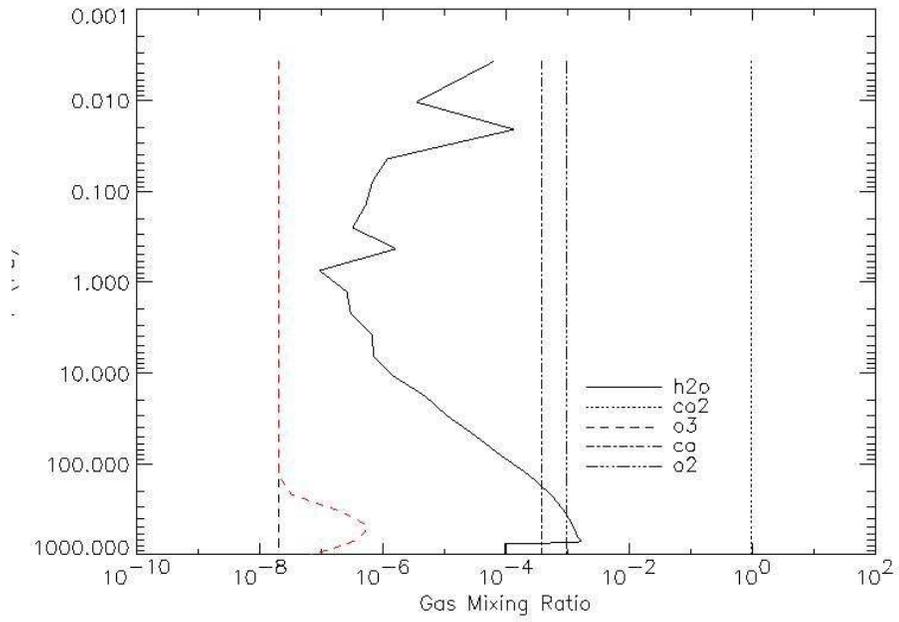}
\caption{  {\footnotesize  \emph{
Example of gas concentration profiles used for simulating Mars atmosphere. The ozone concentration is constant for all tiles away from the polar-cap area, and has a gaussian profile centered at a particular height for polar-cap tiles 
 (red line).  } }}  \label{fig:5z}
\end{center}
\end{figure}

To generate synthetic spectra, we require both information on the physical and chemical state of the atmosphere and surface, such as the horizontal and vertical distributions of surface and atmospheric temperature, trace gas concentration, clouds and aerosols, and surface cover, as well as information on the spectral properties of planetary constituents.   For the physical and chemical state of the atmosphere, we have used GCM results, which we mapped to  48  tiles using Healpix.  The choice of 48 bins (fig. \ref{fig:2}) for mapping the atmospheric properties was made as a compromise between sensitivity to local variations in atmospheric properties, and the computational time required for the radiative transfer model.  Each of the 48 tiles is characterized by an illumination (day, night or terminator), temperature-pressure profile, and vertical profiles for trace gases (CO$_{2}$, CO, O$_{2}$, O$_{3}$, H$_{2}$O) throughout 31 atmospheric levels.   Except for ozone on polar-cap pixels and water vapor, the atmospheric components have a constant mixing ratio with altitude (fig. \ref{fig:5z}), which is a reasonable assumption for Mars.

\begin{figure}
\begin{center}

\includegraphics[width=12cm]{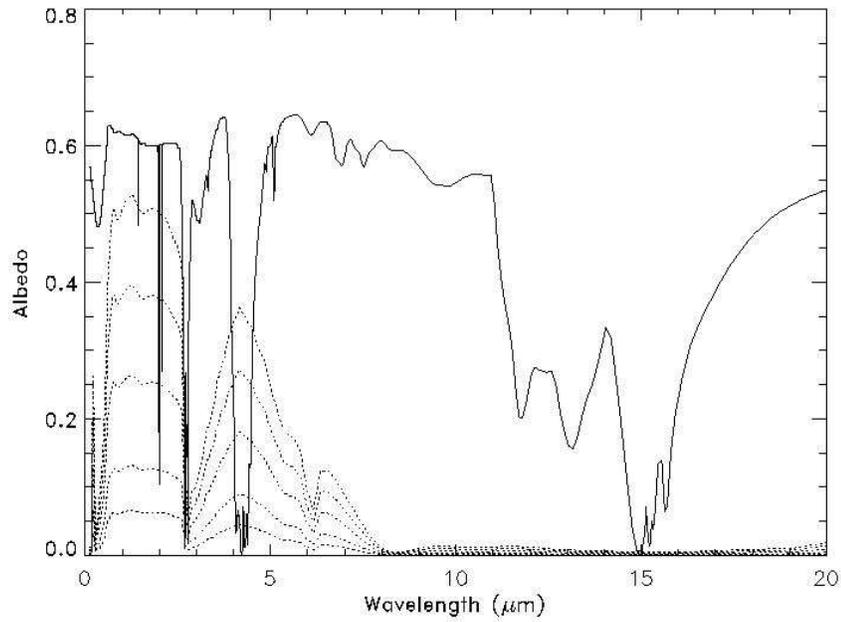}
\caption{  {\footnotesize  \emph{Albedo, $\alpha$, as a function of wavelength, $\lambda$, for dirty-ice (solid line) and
basalt (dotted lines). 
SMART was run for each of the tile for all of these cases, the intermediate cases can be obtained with a linear interpolation. An observational albedo map from Viking was used to rescale at the end the generated spectra .
   }   }}  \label{fig:5zz}
\end{center}
\end{figure}

The surface boundary condition is characterized by a surface albedo which is dependent upon the type of material and wavelength (fig. \ref{fig:5zz}).    
To account for spatial variations in surface albedo, we have used a much finer grid (3072 tiles) than the 48 atmosphere tiles and used an observed albedo map  from the Viking mission to interpolate the synthetic spectra generated for a variety of surface albedos.

\paragraph{Solar spectrum}

\begin{figure}
\begin{center}
\includegraphics[width=12cm]{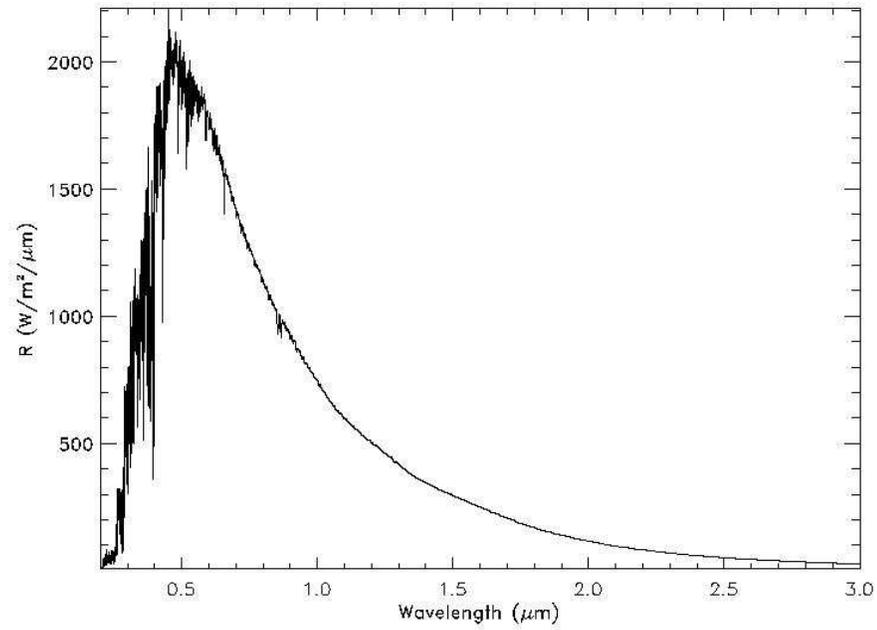}
\caption{  {\footnotesize  \emph{ Solar spectrum used in the simulation. 
The radiance intensity is given for the standard distance of 1 AU, the flux was scaled down by 1/r$^{2}$, where r is the Sun-Mars distance.} }}  \label{fig:5zzk}
\end{center}
\end{figure}

The upper boundary condition of the model is specified by the downward solar flux at the top of the atmosphere.  The solar spectrum used is a high-spectral-resolution spectrum compiled from satellite observations at UV wavelengths (Atlas-2 SUSIM) and stellar atmosphere models at longer wavelengths
[Kurucz, 1979] (fig. \ref{fig:5zzk}). The position of the sun in our simulation was derived from the Jet Propulsion Laboratory Horizons Ephemeris System [Giorgini et al., 1997]. \\

\paragraph{Molecular Line lists}

To model the optical properties of the atmospheric constituents we used molecular line databases to generate the line-by-line description of infrared vibration-rotation bands of gases, and UV and visible cross-sections of gases associated with electronic and pre-dissociation transitions  (e.g. ozone).  

For our model, we have used  the HITRAN2K 
(High resolution Transmission 2K) database [Rothman et al., 2003].  



\paragraph*{LBLABC}

These values were used to generate 
 monochromatic gas absorption coefficients from a line-by-line model  developed by D. Crisp specifically for this application, LBLABC [Meadows and Crisp, 1996].

This model was designed to evaluate absorption coefficients at all atmospheric levels over a very broad range of pressures (10$^{-5}$ to 100 bars),
temperatures (130 to 750 K), and line center distances (10$^{-3}$ to 10$^{3}$ \, cm$^{-1}$). 

The monochromatic vertical gas absorption optical depth, contributed by near-infrared vibration-rotation transitions,
can be expressed in terms of the monochromatic absorption coefficient, $k_{m}$ (the integral extends over the optical path) 
\begin{equation}
\delta \tau_{0} (\nu, \ell) =  \int \, \sum_{m=1}^{M} \, [ 
k_{m} (\nu, z, T) \, N_{m} (z) ] \, \ud z
\end{equation}
In the hydrostatic approximation, we can use pressure $p$ rather than altitude $z$ as the variable for integration.
The coefficient  $N_{m} (p)$ can then be expressed as a function of
the absorbing molecular weight $\mu$,
the mass mixing ratio of the absorbing gas $q_{m}$, Avogadro's number and gravitational acceleration.

The gas absorption coefficient $k_{m}$ includes contributions from all lines that absorb at this wavenumber
\begin{equation}
k_{m} (\nu, p, T) = \sum_{i=1}^{N} \, S_{i} (T) \, f_{i} (\nu \pm \nu_{0i}, p, T)
\end{equation}
where $S_{i}$ is the strength of the $i$th spectral line, centered at number $\nu_{0i}$, $f_{i}$ is the line-shape
function at a distance $\nu - \nu_{0i}$ from the center of line $i$, and $N$ is the total number of lines that produce
absorption at wavenumber $\nu$. The line-shape function is modeled differently in the line-center and far-wing regions.

\subsection{Generating Synthetic Spectra}
\begin{figure}
\begin{center}
\includegraphics[width=10. cm,bbllx=80bp,bblly=24bp,bburx=496bp,bbury=107bp,
clip=]{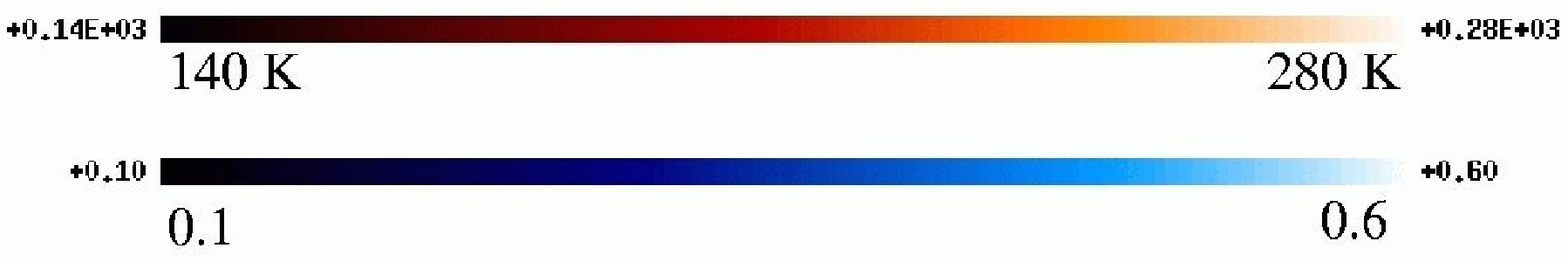} \\
\includegraphics[width=3. cm,bbllx=147bp,bblly=45bp,bburx=443bp,bbury=324bp,
clip=]{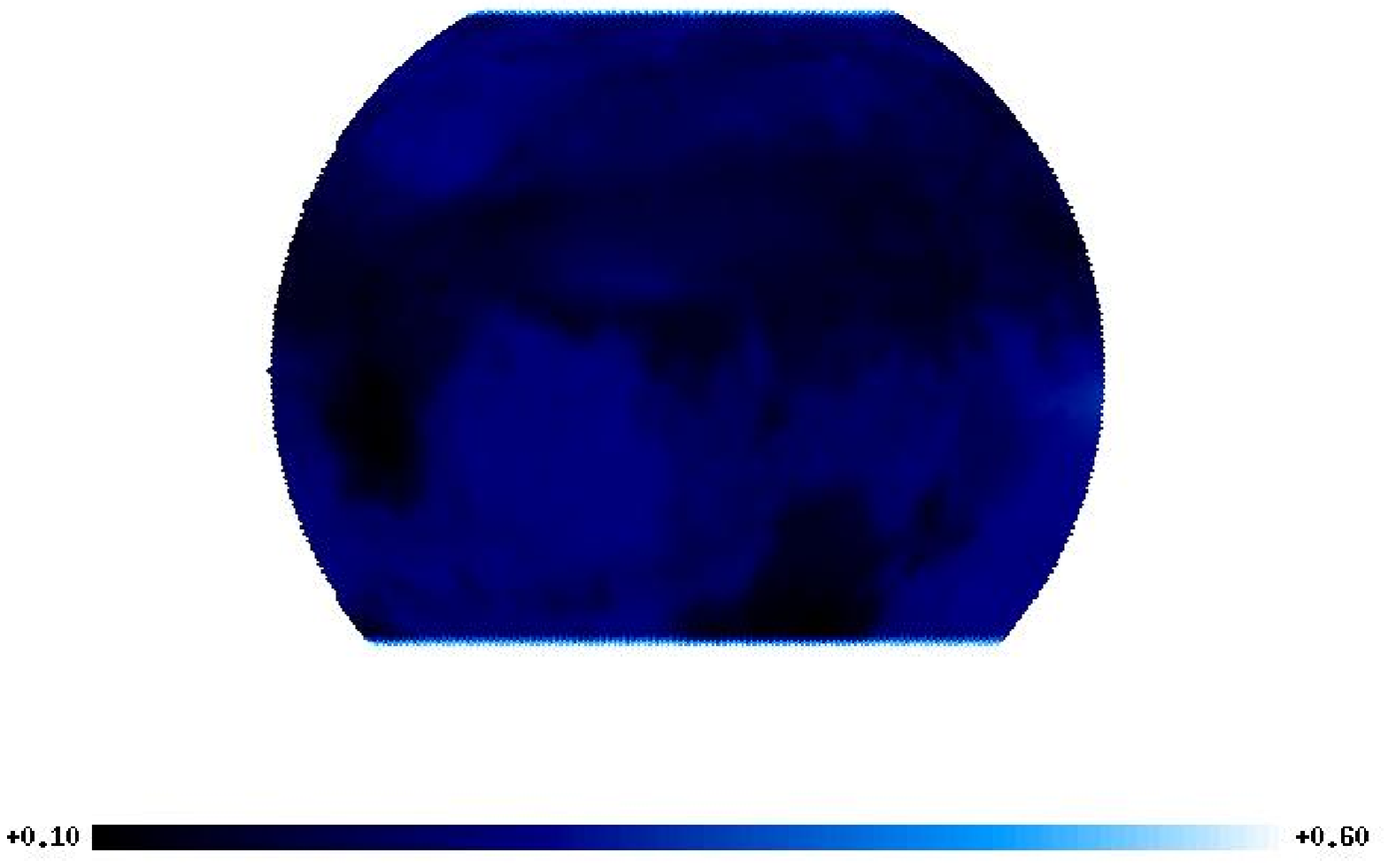}
\includegraphics[width=3. cm,bbllx=147bp,bblly=45bp,bburx=443bp,bbury=324bp,
clip=]{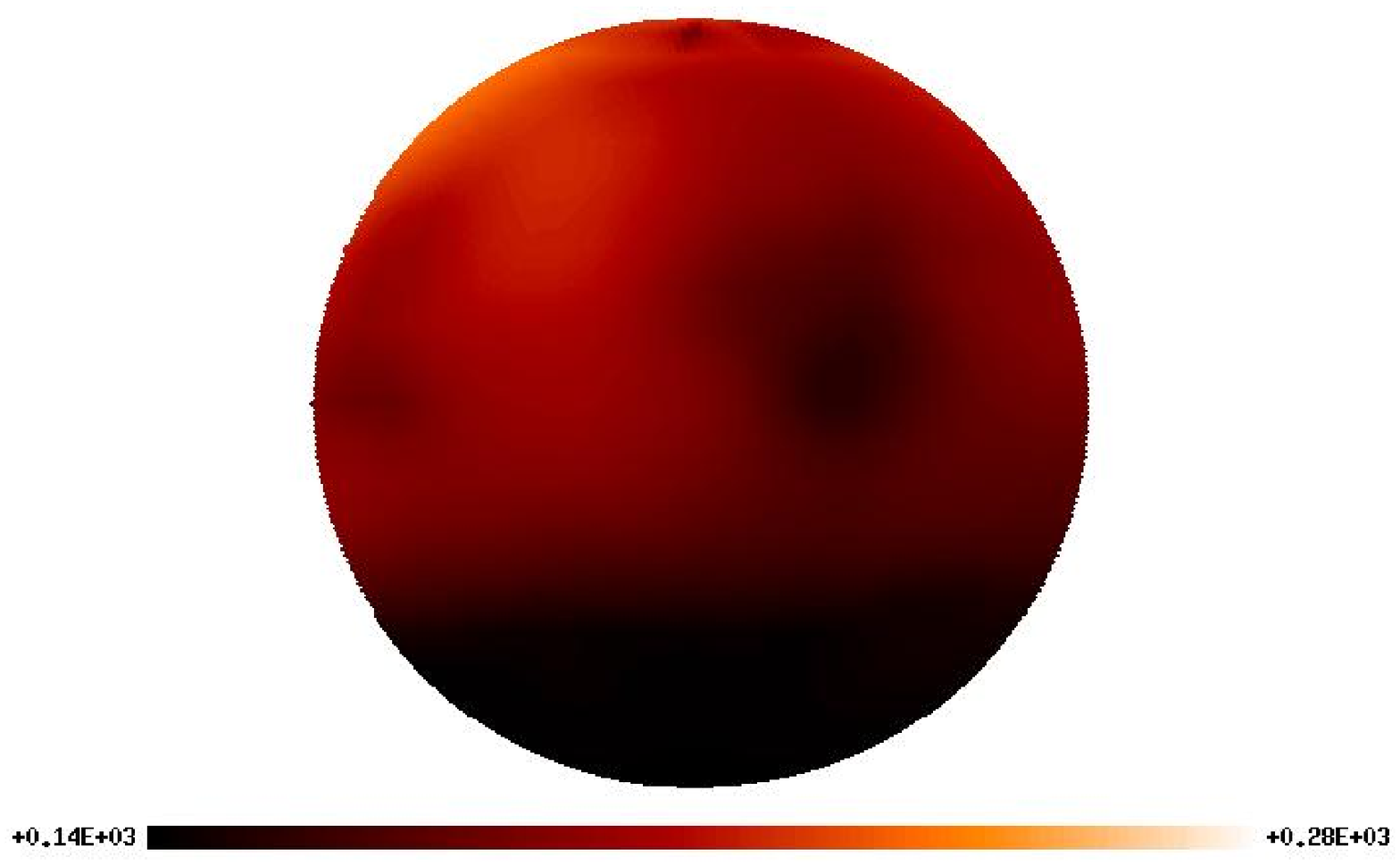} \qquad \qquad 
\includegraphics[width=3. cm,bbllx=147bp,bblly=45bp,bburx=443bp,bbury=324bp,
clip=]{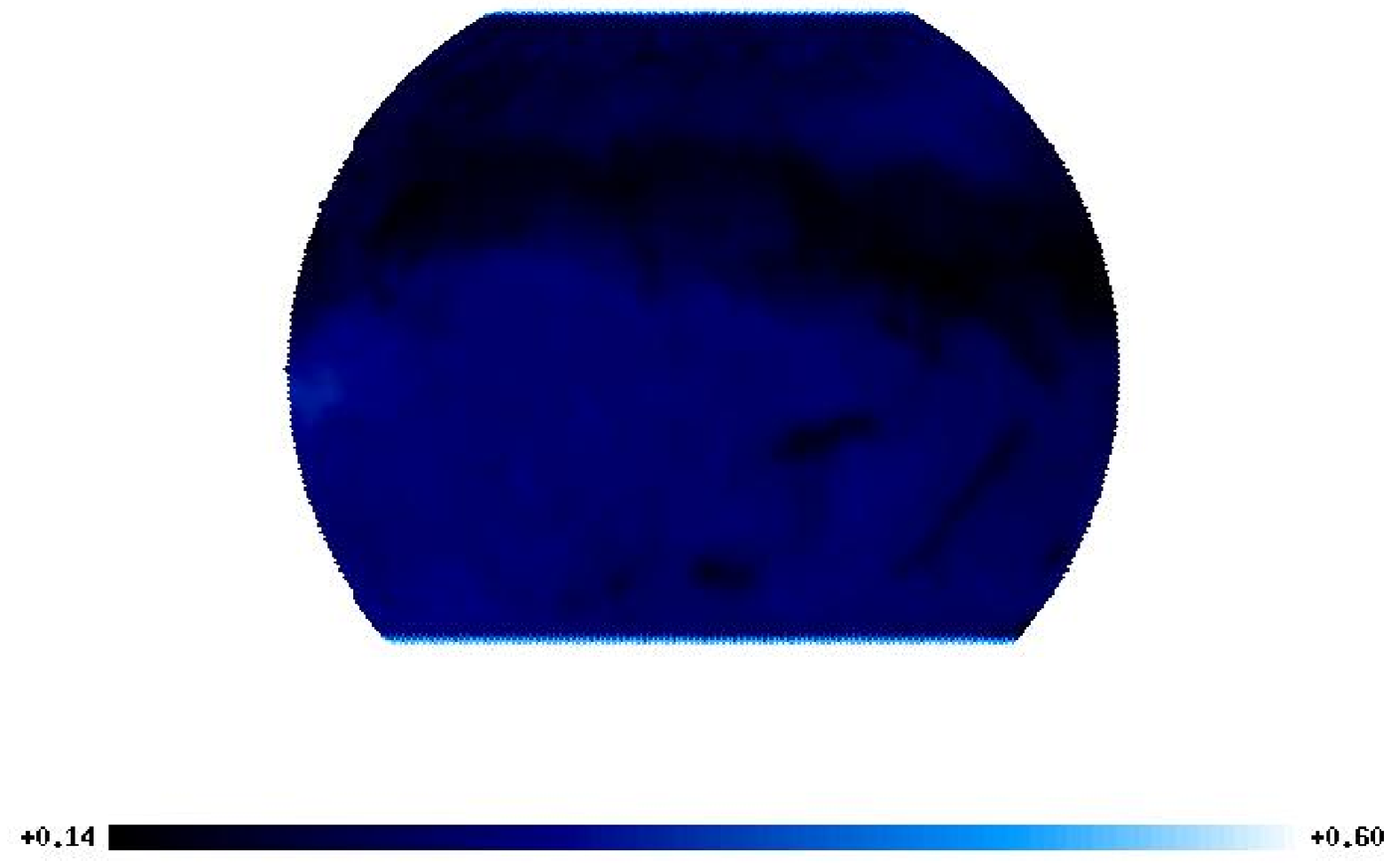}
\includegraphics[width=3. cm,bbllx=147bp,bblly=45bp,bburx=443bp,bbury=324bp,
clip=]{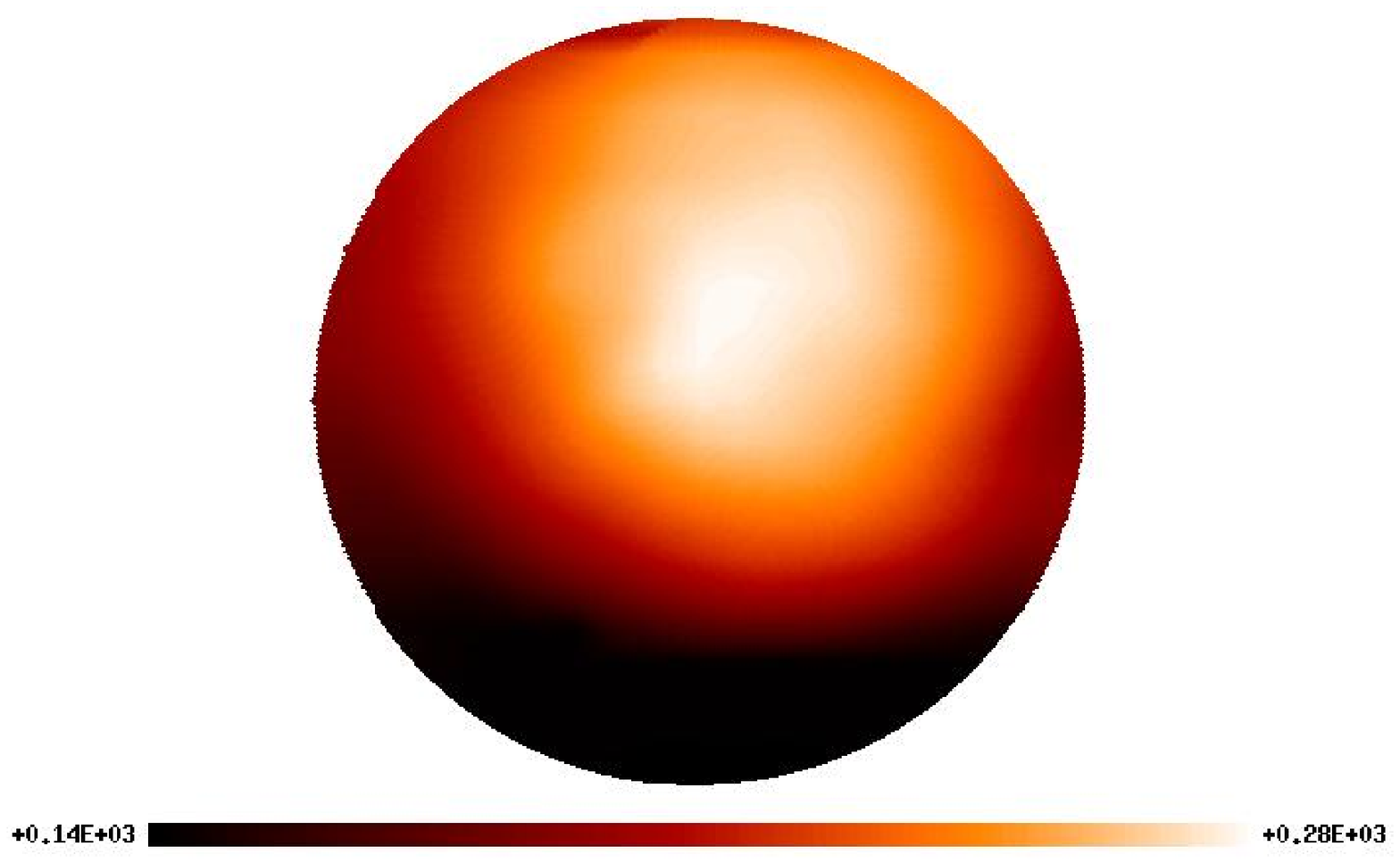} \\
\mbox{  { \footnotesize \emph{ Fig. a) View from the equator,
$\theta = 0^{\circ}, \, \phi = 0^{\circ}$;   
   } } \quad { \footnotesize \emph{Fig. b) View from the equator,
$\theta = 0^{\circ}, \, \phi = 180^{\circ}$  } } } \\ 
\includegraphics[width=3. cm,bbllx=147bp,bblly=45bp,bburx=443bp,bbury=324bp,
clip=]{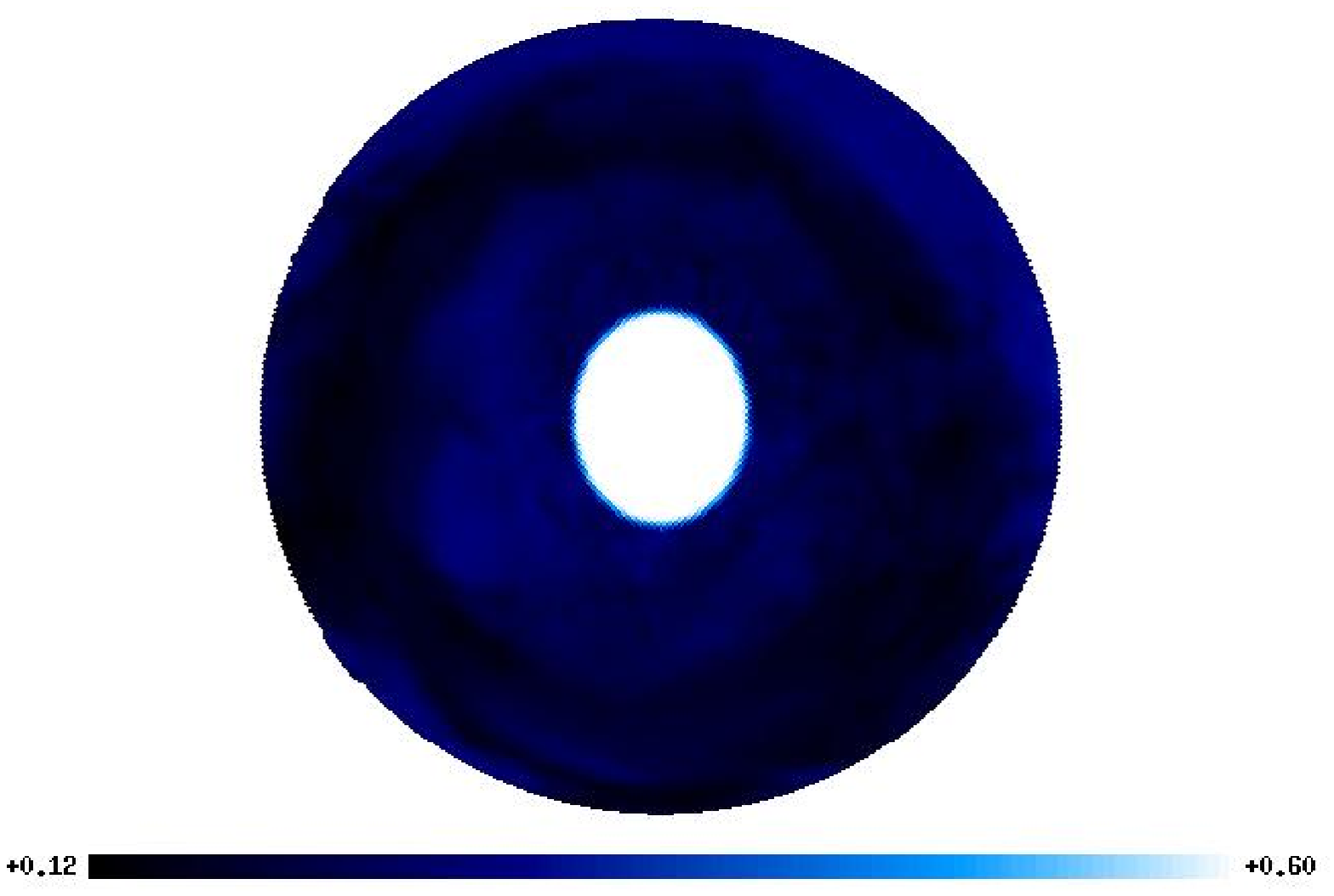}
\includegraphics[width=3. cm,bbllx=147bp,bblly=45bp,bburx=443bp,bbury=324bp,
clip=]{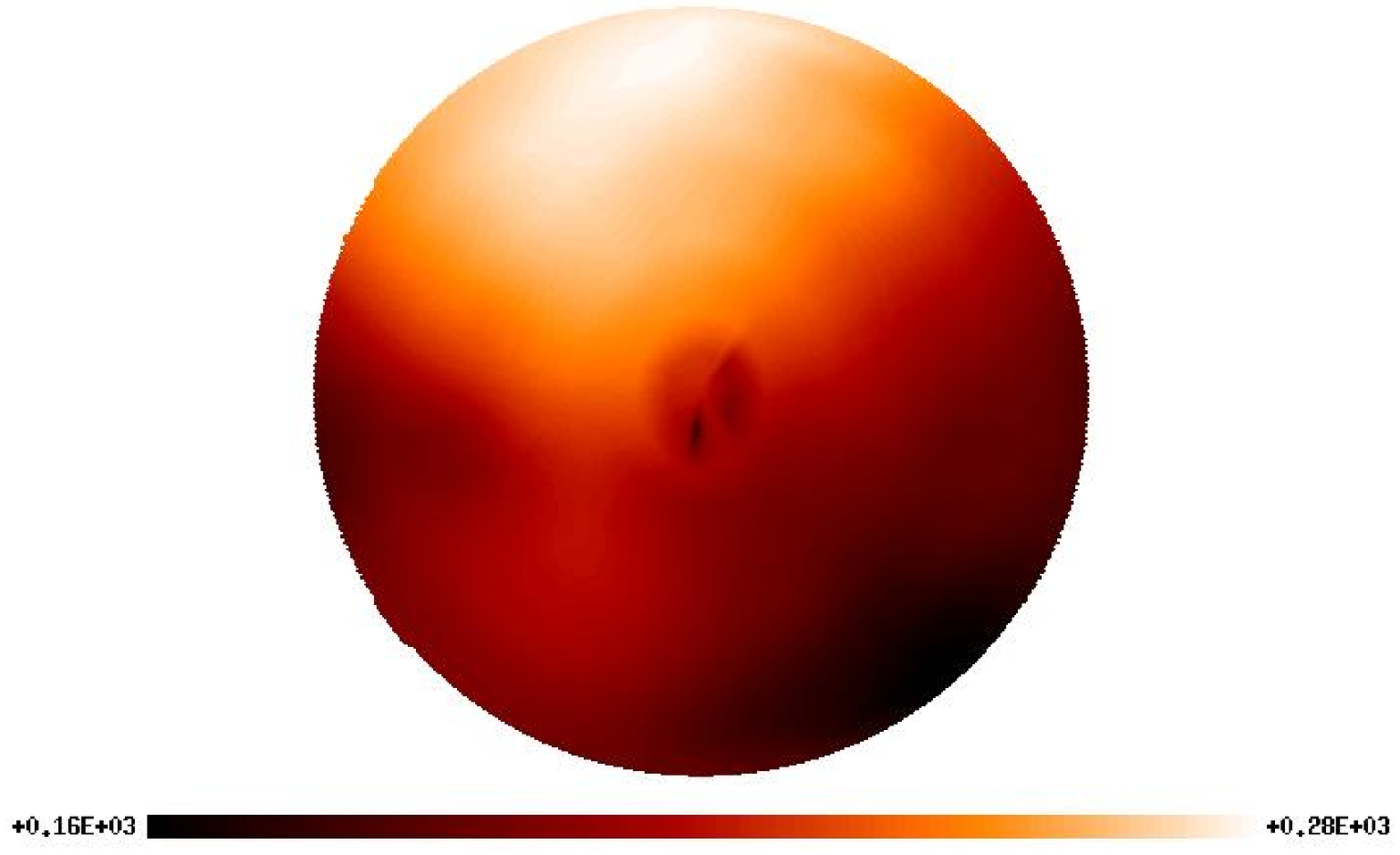}  \qquad \qquad 
\includegraphics[width=3. cm,bbllx=147bp,bblly=45bp,bburx=443bp,bbury=324bp,
clip=]{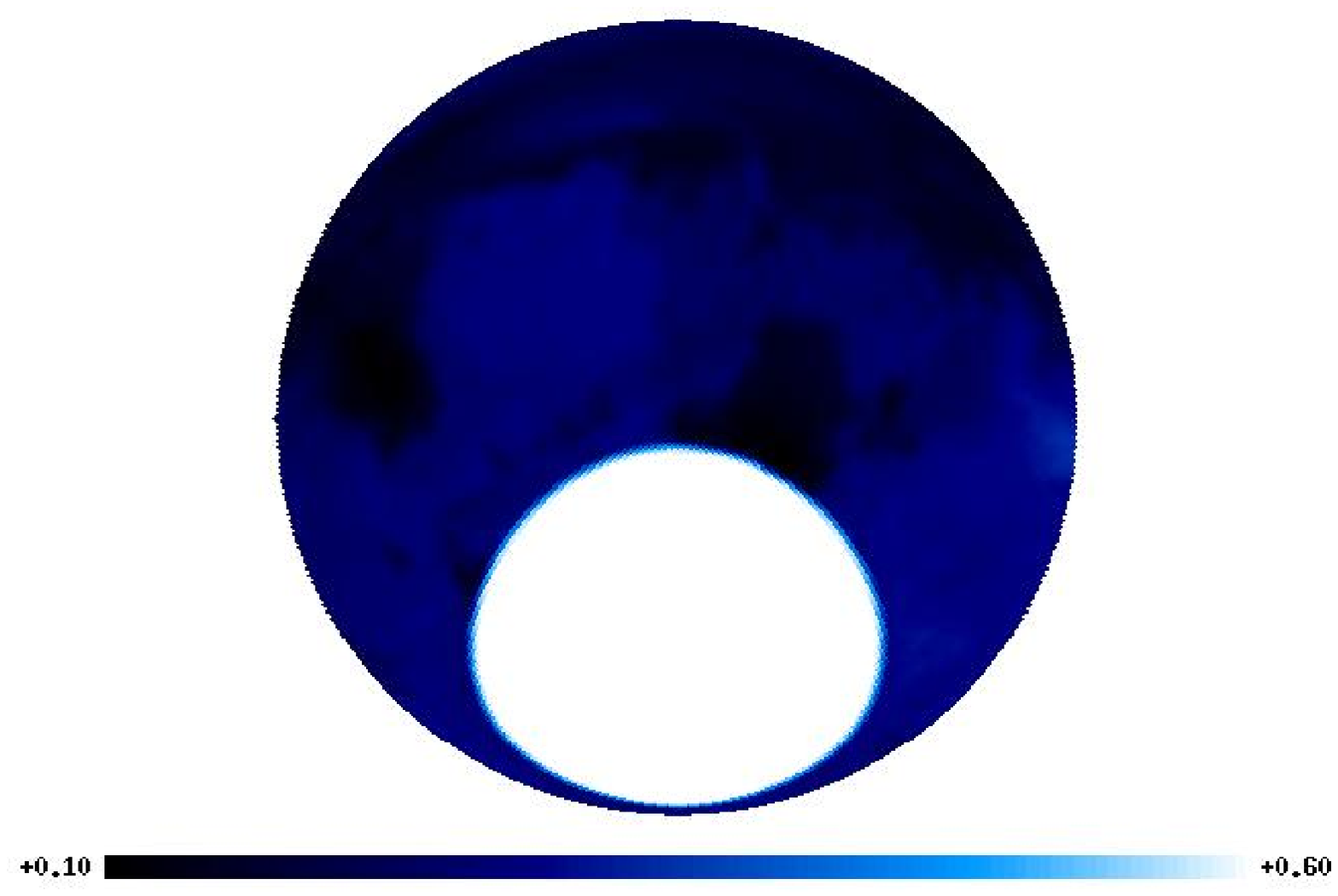}
\includegraphics[width=3. cm,bbllx=147bp,bblly=45bp,bburx=443bp,bbury=324bp,
clip=]{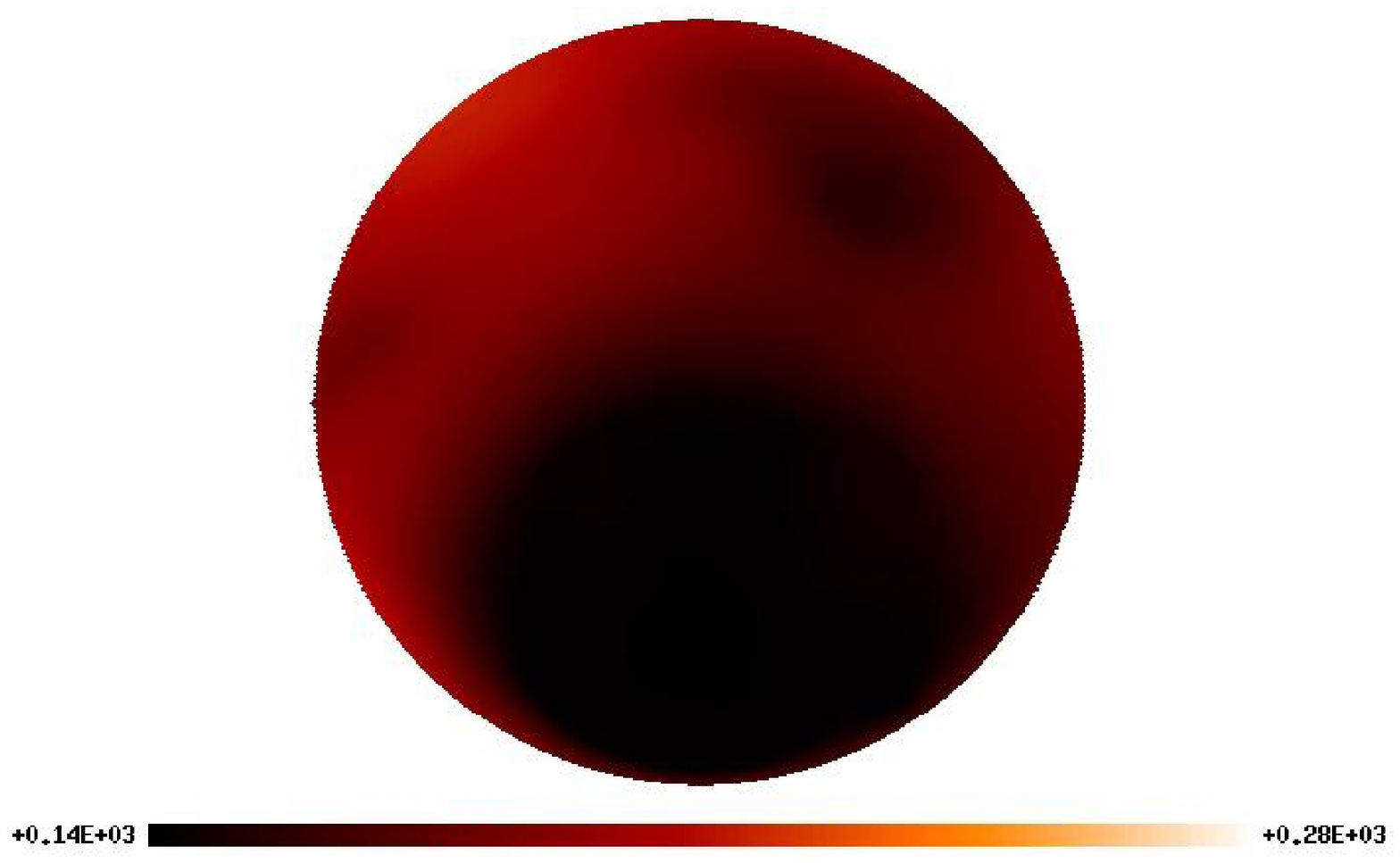}   \\
 \mbox{ { \footnotesize \emph{Fig. c) View from the North Pole,
$\theta = 90^{\circ}, \, \phi = 0^{\circ}$;  } } \quad
 { \footnotesize \emph{Fig. d) View from 
$\theta = -45^{\circ}, \, \phi = 0^{\circ}$  } } } \\
\includegraphics[width=3. cm,bbllx=147bp,bblly=45bp,bburx=443bp,bbury=324bp,
clip=]{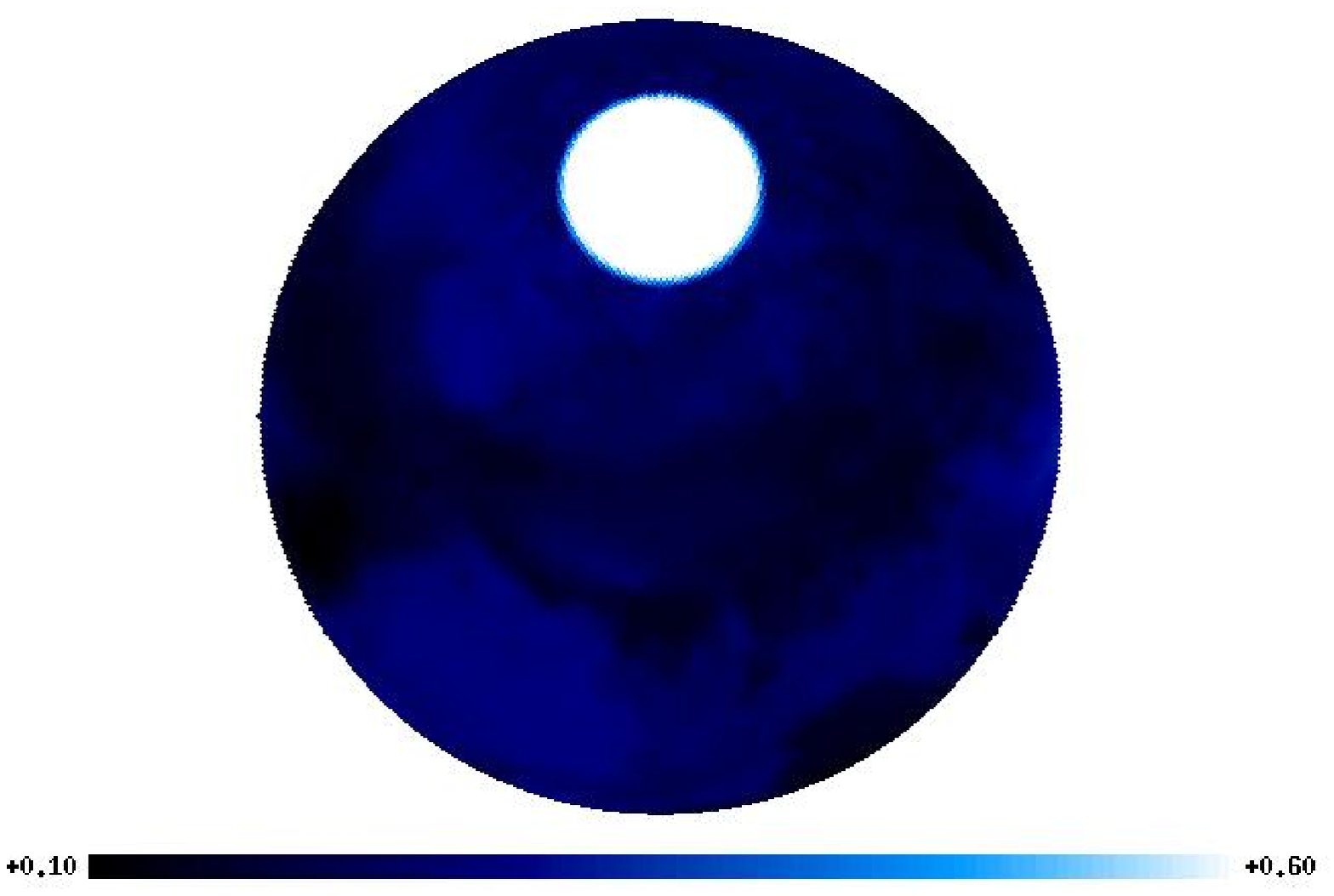}
\includegraphics[width=3.cm,bbllx=147bp,bblly=45bp,bburx=443bp,bbury=324bp,
clip=]{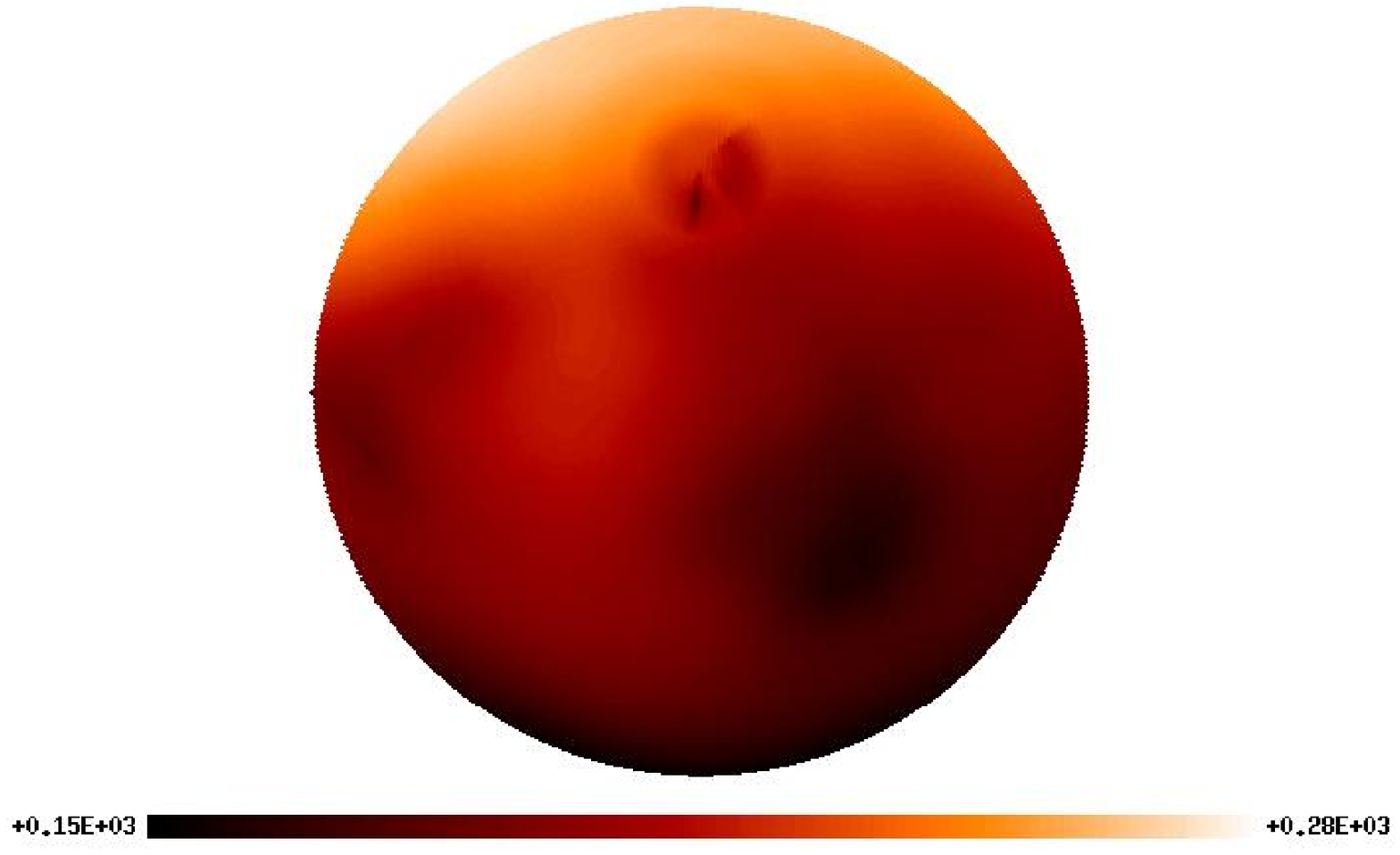}  
\qquad \qquad 
\includegraphics[width=3. cm,bbllx=147bp,bblly=45bp,bburx=443bp,bbury=324bp,
clip=]{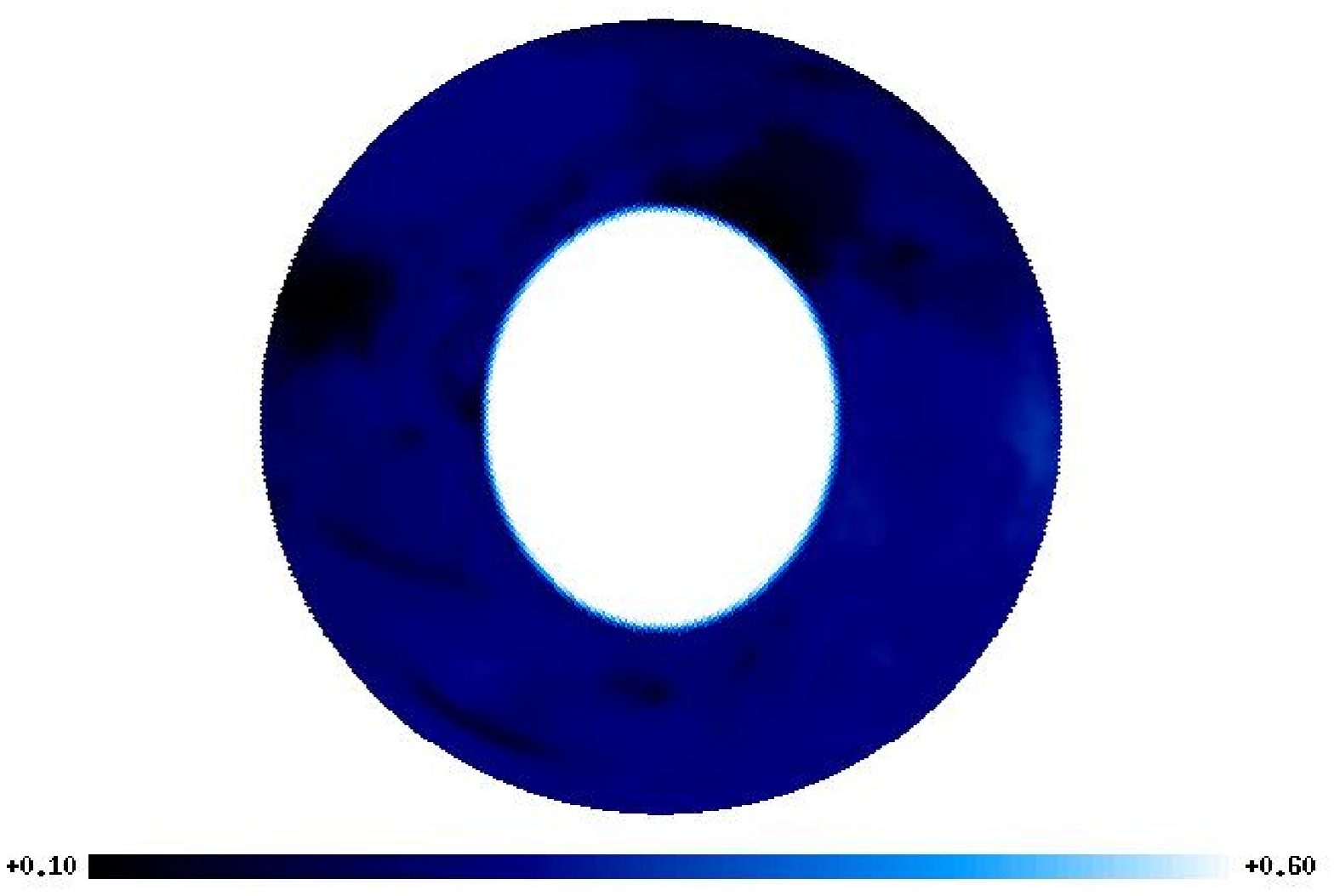}
\includegraphics[width=3. cm,bbllx=147bp,bblly=45bp,bburx=443bp,bbury=324bp,
clip=]{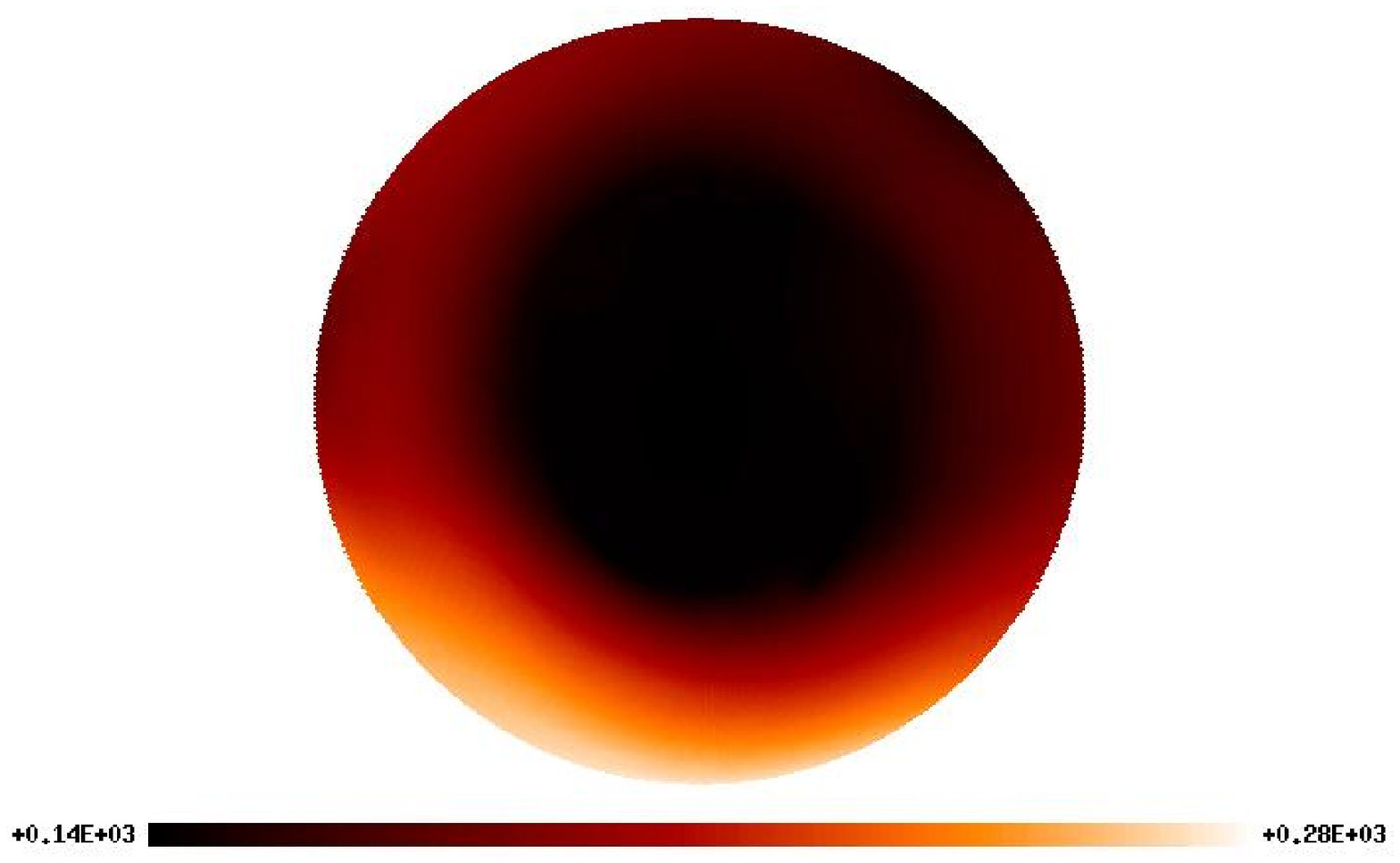}  \\
 \mbox{{ \footnotesize \emph{Fig. e) View from 
$\theta = 45^{\circ}, \, \phi = 0^{\circ}$  } } \quad { \footnotesize \emph{Fig. f) View from the South Pole,
$\theta = -90^{\circ}, \, \phi = 0^{\circ}$ }} }
 \\
\caption{ { \footnotesize \emph{These figures show a simulation of the surface temperature  (figs on the right, 140-280 K) and solar albedo (figs on the left, 0.1-0.6), for a day on Mars 
close to solstice, $L_{s} = 104.6^{\circ}$ and from different viewing angles. The polar caps extend for 
$\sim 15^{\circ}$ in the northern hemisphere, and $\sim 40^{\circ}$ 
in the southern hemisphere.  The sun position derived from the JPL Horizons Ephemeris System is
24.63$^{\circ}$ latitude, and 169.99 $^{\circ}$ longitude.  }} 
 } \label{fig:vp}
\end{center}
\end{figure}

 Using the input data described above, we have then run a radiative transfer model, the Spectral Mapping Atmospheric Radiative Transfer Model (SMART)
[Meadows and Crisp, 1996] to generate  synthetic spectra (for each of the 48 tiles)
 for a range of illumination conditions
(phase angles), viewing geometries (see fig. \ref{fig:vp}), albedos 
and  surface types (see fig. \ref{fig:5zz}). This library of spectra has been used for creating datacubes and disk-averaged spectra from arbitrary viewing positions.

     SMART is a multi-stream, multi-level, spectrum-resolving (line-by-line) multiple-scattering algorithm to generate the high-resolution synthetic spectra of planetary atmospheres. Using a high-resolution spectral grid, it completely resolves the wavelength dependence of all atmospheric constituents, including absorbing gases (infrared absorption bands, UV predissociation bands, and electronic bands) and airborne particles (clouds, aerosols) at all levels of the atmosphere, as well as the wavelength- dependent albedo of the planet's surface, and the spectrum of the incident stellar source. \\
 
This model provides spectrally dependent solutions to an equation of transfer of the form:
\begin{equation}
\mu \, \frac{\de I (\tau, \mu, \phi, \nu)}{\de \tau} = I (\tau, \mu, \phi, \nu) - S (\tau, \mu, \phi, \nu)
\end{equation}
where $I$ is the radiance, $\tau$ is the column-integrated vertical optical depth, measured from the top of the atmosphere downwards, $\mu$ is the cosine of the zenith angle, and $\phi$ is the azimuth angle. The source function $S$ is given by:
\begin{equation} \label{lab}
\begin{split}
 S (\tau, \mu, \phi, \nu) = & \frac{\omega (\tau, \nu)}{4 \, \pi} \, \int_{0}^{2 \, \pi} \, \ud \phi' \,
\int_{-1}^{1} \, \ud \mu \, P (\tau, \mu, \phi, \nu, \mu', \phi') \, I (\tau, \mu', \phi', \nu) + \\
& + (1 - \omega (\tau, \nu)) \, B (\nu, T(\tau)) + \\
& + \frac{\omega (\tau, \nu)}{4 \, \pi} \, F_{\odot} \, P (\tau, \mu, \phi, \nu, - \mu_{\odot}, \phi_{\odot}) \, \text{e}^{- \tau/\mu_{\odot}}
\end{split}
\end{equation}
where $P (\tau, \mu, \phi, \nu, \mu', \phi')$ is the scattering phase function, $ \omega (\tau, \nu)$ is the single scattering albedo, and $B (\nu, T(\tau))$ is the Planck function at wavenumber $\nu$ and temperature $T (\tau)$.
$F_{\odot}$ is the solar irradiance at the top of the atmosphere.

SMART uses the discrete ordinate method for evaluating numerically the integrals in eq. (\ref{lab}),
more specifically the integrals are calculated using Gaussian quadrature to yield radiances at a number of discrete zenith and azimuth angles for each atmospheric layer.
This method can provide accurate solutions to the equation of transfer only when applied to spectral regions  sufficiently narrow that the optical properties and source functions are roughly constant across each layer. 
Spectral mapping methods reduce the number of computations needed by identifying monochromatic spectral regions with similar optical properties, and combining these into bins, such that only one multiple scattering calculation is needed in each bin.
The computed radiances are then mapped back to their original wavelengths to form a high resolution spectrum. Even though this approach typically reduces
the number of monochromatic calculations by a factor of 50 to 100, such calculations still require substantial amounts of computing time.

\subsection{The interpolation model}

This model takes a user specified spatial resolution and observed vantage point
 and creates a simulation of the spatially and spectrally-resolved planet and disk as seen from this direction. It does this by interpolating the angle-dependent radiances generated by SMART contributing to
 the final synthetic view, which can then be averaged on the visible disk. 
For the simulations described in this paper, as we have already mentioned, we have adopted a relatively high spatial resolution (3,072 tiles)
  to include 
surface and albedo inhomogeneities  and only 48  atmospheric profiles.

Given the coordinates of the observer,  the sun position and the 
surface albedo map, we can calculate  very precisely the contribution of the single tile (or of a particular fraction of the tile) 
to the disk. Even though this contribution is position dependent
and area weighted, we can explore whether local 
properties can still be detected after having averaged on the disk. 	
In the following section we describe the details of the calculations we have performed.
 
 \paragraph{Datacubes and disk-averaged spectra}

Before a disk-averaged synthetic spectrum can be produced, a three-dimensional data cube, two dimensions spatially and one dimension for the
  spectra, must be constructed.  In the spatial dimensions, only the tiles
  that are detectable by the remote observer will be considered. Therefore the
  data cube will be comprised of tiles up to 90 degrees in each direction
  from the sub-sensor position on the planet.  The resultant data cube will
  display one half of the planet's surface with the corresponding spectra at
  each of the detectable tiles.

  The first step is to determine which tiles are detectable by the sensor
  from the vantage point selected by the user.  This can be achieved by using a
  Healpix subroutine which lists the tiles that are within a certain radius
  of a central point.  In this case, the central point is the sub-sensor
  point and the radius is 90 degrees. The  same procedure is used for the sub-solar position in order to determine the tiles that are illuminated by the sun.
 
   To calculate the fraction of the tile that is
  detectable by the observer, those
  tiles completely inside the frame of view are given a weight of 1, whereas those on the edge
  are first subdivided into a higher resolved tesselation  (4,096 tiles), so that their weight  can be calculated by the ratio between the number of 
detectable sub-tiles and the total number of sub-tiles in the whole tile.
A similar method is
  used to evaluate what fraction of each tile is illuminated by
  the sun.  The result of these two numerical computations is a list of
  detectable tiles with the fraction detectable by the sensor, the fraction
  illuminated by the sun, and the fraction that is dark. 
 
The  calculation of the radiance spectra requires a center point for each tile.
 The centers of the whole tiles 
can be calculated using a Healpix
  subroutine, for the fractional tiles these can be computed by taking an average of the
  latitudes and longitudes of the centers of the subtiles included in the
  region being considered (the longitude is treated as a linear factor and
  the latitude is treated as linear with respect to cosine).    With these values, the next step is to calculate the solar
  zenith angle, the viewing zenith angle, and the viewing azimuth angle for
  each whole tile and fractional tile. The two zenith angles can be found  
using the distance formula for
  spherical coordinates, for the viewing azimuth angle we have to use
 the law of cosines.

  Now with all of the necessary input values, the spectrum at each tile can
  be calculated using the SMART output radiance file library.  In doing the
  SMART runs, certain values are selected for four factors: albedo, solar
  zenith angle, viewing zenith angle, and viewing azimuth angle.  To
  obtain a radiance value for arbitrary albedo values and angles,
  we must interpolate.  We  used linear interpolation for the albedo factor.  
Since the radiance is very strong near the solar zenith angle
  but much less so in the surrounding angles, a suitable function has yet to
  be found to model the near discontinuity at the solar zenith angle, hence
  linear interpolation was also used here as a best available approximation.
  Lastly, for the viewing angles, a
  bicubic spline interpolation is used since it approximates the surface of the
  sphere almost as accurately as spherical functions (e.g. Legendre polynomials) but
  with fewer computations.  Applying these four interpolations, the radiance
  can be given at any albedo and any set of angles.  The interpolation for the albedo
is the last to be computed, and
has been performed on a higher resolution map (3,072 tiles) to be compared  with an
observational  map. This last step is necessary to calibrate the overall spectral response of the planet.

The radiance at a given
  wavenumber is calculated for each whole tile and fractional tile from the
  SMART output spectra available in the library.  Repeating this four-step
  interpolation for each wavenumber, a radiance spectrum is generated for
  each tile so that a data cube can be produced with a list of
  detectable tiles and their corresponding spectra.

   The disk-averaged synthetic
  spectrum can be constructed  from the data cube by  summing the
  radiance values for all of the detectable tiles at a given wavenumber and
  dividing by the number of tiles,  and then repeating this process for each wavenumber.

\newpage

\section{Results}

\subsection{Disk-averaged synthetic spectra}

In this section we describe the validation of the model, and the experiments that were run to assess the information content of the disk-averaged synthetic spectra of Mars generated with the method described in \S 2, for a range of viewing angles and illumination geometries. 


\subsection{Model Validation}

\begin{figure} 
\begin{center}
\includegraphics[width=12 cm]{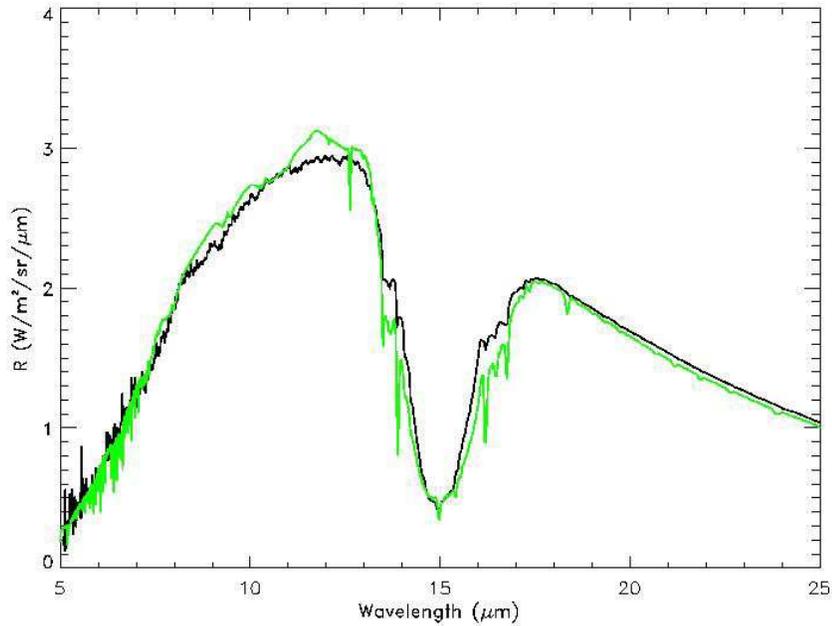} 
	\caption{{\footnotesize  \emph{Disk-averaged IR spectrum of Mars observed by IRIS-Mariner 9 in July 1972 (black line, after [Hanel et al., 1992]).   The sub-observer point is in the mid-latitudes, and the atmosphere was relatively free of dust when the data spectrum was taken. For comparison, we show one of our simulated spectra (green line). \newline
The discrepancies in the 13-17$\mu$m band are certainly due 
to a slightly different atmospheric temperature profile, a better agreement with the data could be achieved using specific algorithms to retrieve the temperature distribution from the observed spectrum. Moreover, in the Mariner 9 spectrum there is still a small quantity of dust, absorbing in the 8-13$\mu$m band;
our synthetic spectrum is dust-free. Finally, since the synthetic spectrum has a higher resolution, some absorption features
are more visible. }}  } \label{fig:mar}
	    \end{center}
\end{figure}

To validate the model, we have compared our synthetic Martian spectra 
with observations of Mars by the IRIS instrument on board the Mariner 9 
spacecraft (Fig. \ref{fig:mar}).   

\begin{figure}
\begin{center}
\includegraphics[width=14 cm]{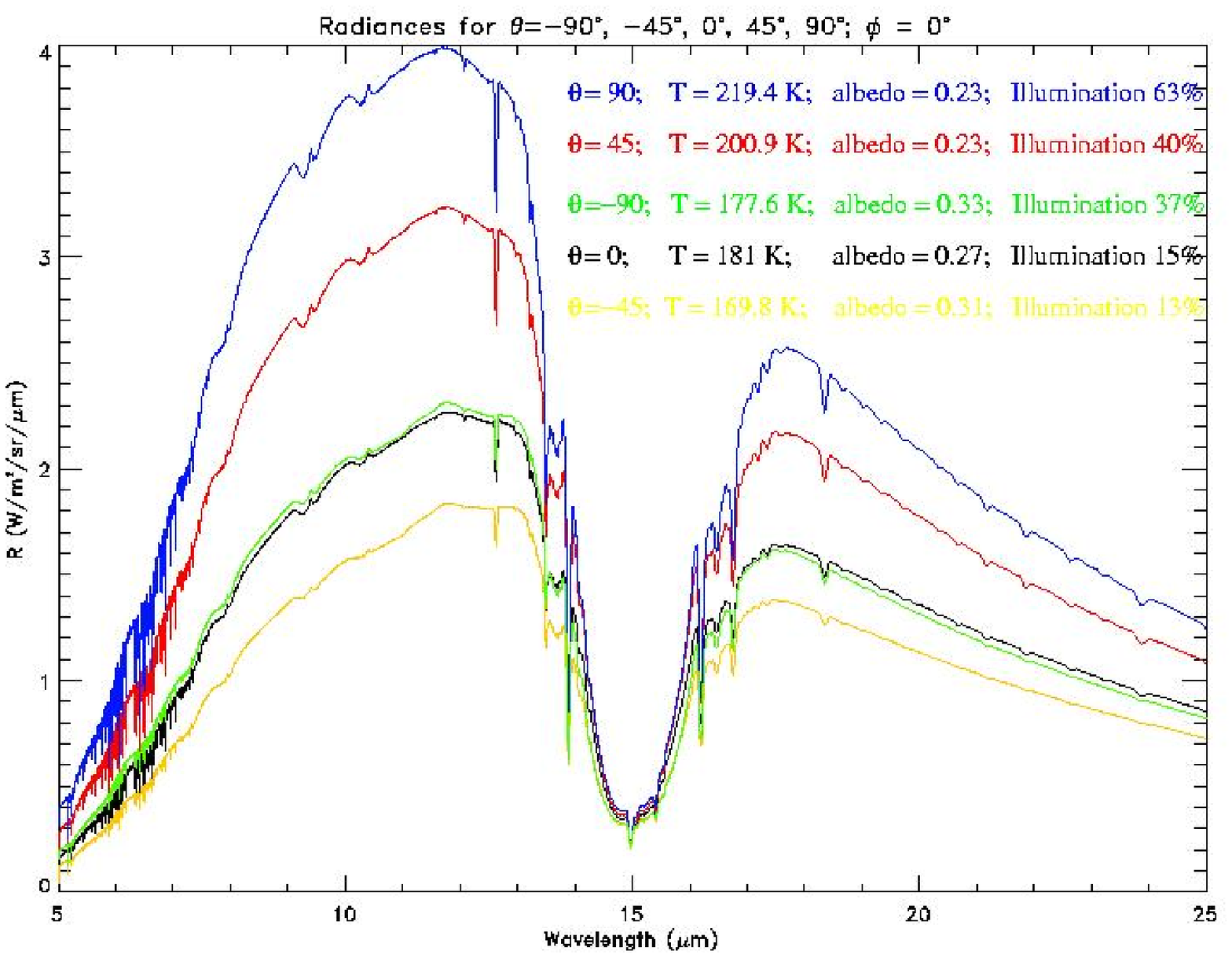}
\includegraphics[width=14 cm]{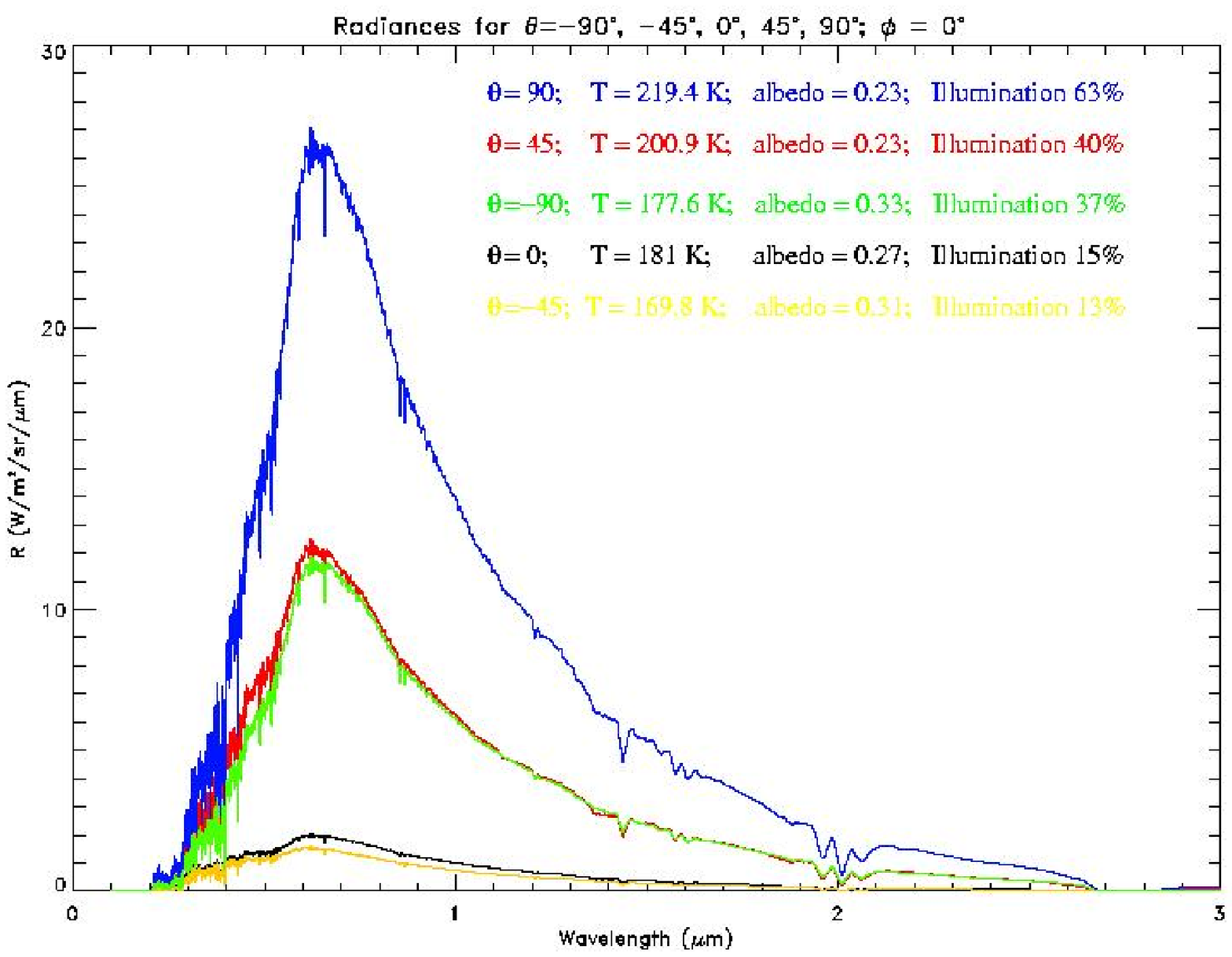}
\caption{ {\footnotesize \emph{  Variability of disk-averaged synthetic spectra.  Synthetic spectra for sub-viewer points with different latitudes (longitude, $\phi = 0^{\circ}$):
$\theta = 0^{\circ}$ (fig. \ref{fig:vp} a), $\theta = 45^{\circ}$ (fig. \ref{fig:vp} e), 
$\theta = 90^{\circ}$ (fig. \ref{fig:vp} c), $\theta = -45^{\circ}$ (fig. \ref{fig:vp} d), $\theta = -90^{\circ}$ (fig. \ref{fig:vp} f) for a day corresponding to $L_{s} = 104.6$. The disk-averaged surface temperature, solar albedo and illumination are indicated as well. \newline The polar caps extend 
$\sim 15^{\circ}$ to the South in the northern hemisphere, and $\sim 40^{\circ}$ 
to the North in the southern hemisphere. The sun position derived from the JPL Horizons Ephemeris System is
24.63$^{\circ}$ latitude, and 169.99 $^{\circ}$ longitude. } } } \label{fig:1a}
\end{center}
\end{figure}

Both the observed and synthetic spectra show a set of features that are characteristic of the CO$_2$ atmosphere of Mars. In the MIR, the spectrum is dominated by the $\nu_2$ fundamental of CO$_2$, centered around 15$\mu$m, and spanning the range from $\sim$ 12 to 19 $\mu$m, at longer wavelengths we have the water-vapor continuum absorption.  Other weaker CO$_2$ absorption features can be seen near 9.4 and 10.4 $\mu$m. At this high spectral resolution, weak absorption from the $\nu_2$ fundamental of water vapor can be seen centered at 6.27$\mu$m and spanning the region 5-7.5$\mu$m. At visible wavelengths (Figure  \ref{fig:1a}, bottom panel) the most prominent features at 1.43, 1.57 and 1.6$\mu$m, and around 2.0$\mu$m (1.98, 2.01 and 2.06$\mu$m) are also CO$_2$ absorption bands [Goody and Yung, 1996].

With the exception of the features described above, the atmosphere of Mars 
is mostly transparent, and emission or reflected solar radiation from 
the solid surface can be observed over a wide spectral range. In our 
model, we have not included contributions from spatially and temporally 
variable phenomena such as  clouds, or airborne dust. 
Airborne dust in particular would have produced a broad silicate 
absorption feature centered around 9.4$\mu$m.

There is only a limited amount of information on the observing geometry of the Mariner 9 IRIS spectrum used in Fig. \ref{fig:mar}, which is an average of four mid-latitude spectra [Hanel et al., 1992]. However, the 
spectrum was taken in July 1972, after the 1971 dust storm, when most of 
the dust had settled to the surface and was no longer prominent in the spectrum. 
We compare this spectrum with an example of mid-latitude spectrum from our model and 
show that the fit is reasonable, given the uncertainties in the viewing 
geometry and atmospheric temperature for the data spectrum.

Different parts of the spectrum provide information about different parts of the surface-atmosphere system.  For example, because the Martian 
atmosphere is relatively transparent at 12 $\mu$m (in dust and cloud-free conditions), most of the thermal radiation at that wavelength originates at the surface.  At wavelengths near 15 $\mu$m, where the atmosphere is relatively opaque, most of the thermal emission observed above the atmosphere originates at high altitudes (25 to 50 km). In other words, the corresponding contribution functions for these two spectral regions sample relatively narrow altitude ranges, with their peaks at the surface and in the upper atmosphere, respectively. Spectral regions with intermediate levels of absorption have contribution functions peaked at intermediate levels. In the absence of multiple scattering, the emission intensities observed at a given MIR wavelength corresponds to the black body emission temperature at the peak of the contribution function.  Hence, if there are large temperature differences between the surface and the atmosphere, the atmospheric absorption features will show up prominently in the spectrum.  As the surface-atmosphere temperature contrast decreases, the amplitude of atmospheric spectral features also vanishes.  Remote sensing retrieval algorithms can therefore exploit the atmospheric opacity differences to provide information about the vertical temperature structure in planetary atmosphere.  

Because CO$_2$ is a long-lived, chemically stable species that is often well-mixed throughout the atmospheres of terrestrial planets,
the broad 15 $\mu$m CO$_{2}$ band is useful for atmospheric temperature
retrieval. Although
details of the atmospheric temperature structure can
be obtained only from spectra with high resolution and a high signal to noise ratio, some information about the temperature structure can be retrieved
from the
relatively low spectral resolution observations of this band that are
likely to be made with TPF/Darwin. For disk-averaged 
Mars, the
overall shape of this band (a "U-shape") indicates a relatively hot 
planetary surface that
is consistently warmer than the overlying atmosphere, which also has a
steady decrease in atmospheric temperatures with altitude.  Note however 
that in
the polar regions, where the vertical temperature gradients are smaller, or of the opposite sign, the shape of the CO$_2$ band, and other spectral 
features, can be
quite different (e.g. Figure \ref{fig:5a}  polar spectrum with water in emission). 
For soundings over the CO$_2$ ice cap gas absorption features appear in emission, indicating that the surface is in fact cooler 
than the overlying
atmosphere at this point on the planet.

High altitude clouds or an opaque dust layer can limit range of altitudes that can be observed using remote sensing techniques.  These sources of opacity can also limit the amount of radiation that reaches the surface, limiting the heating there. These factors will often suppress the spectral contrast seen at both solar and thermal wavelengths.   All the spectra shown here will exhibit a relatively high spectral contrast since the model does not 
include opaque atmospheric dust or water ice clouds.

 \subsubsection{Spectral Variability as a Function of Viewing Geometry}

\begin{figure}
\begin{center}
\includegraphics[width=14 cm]{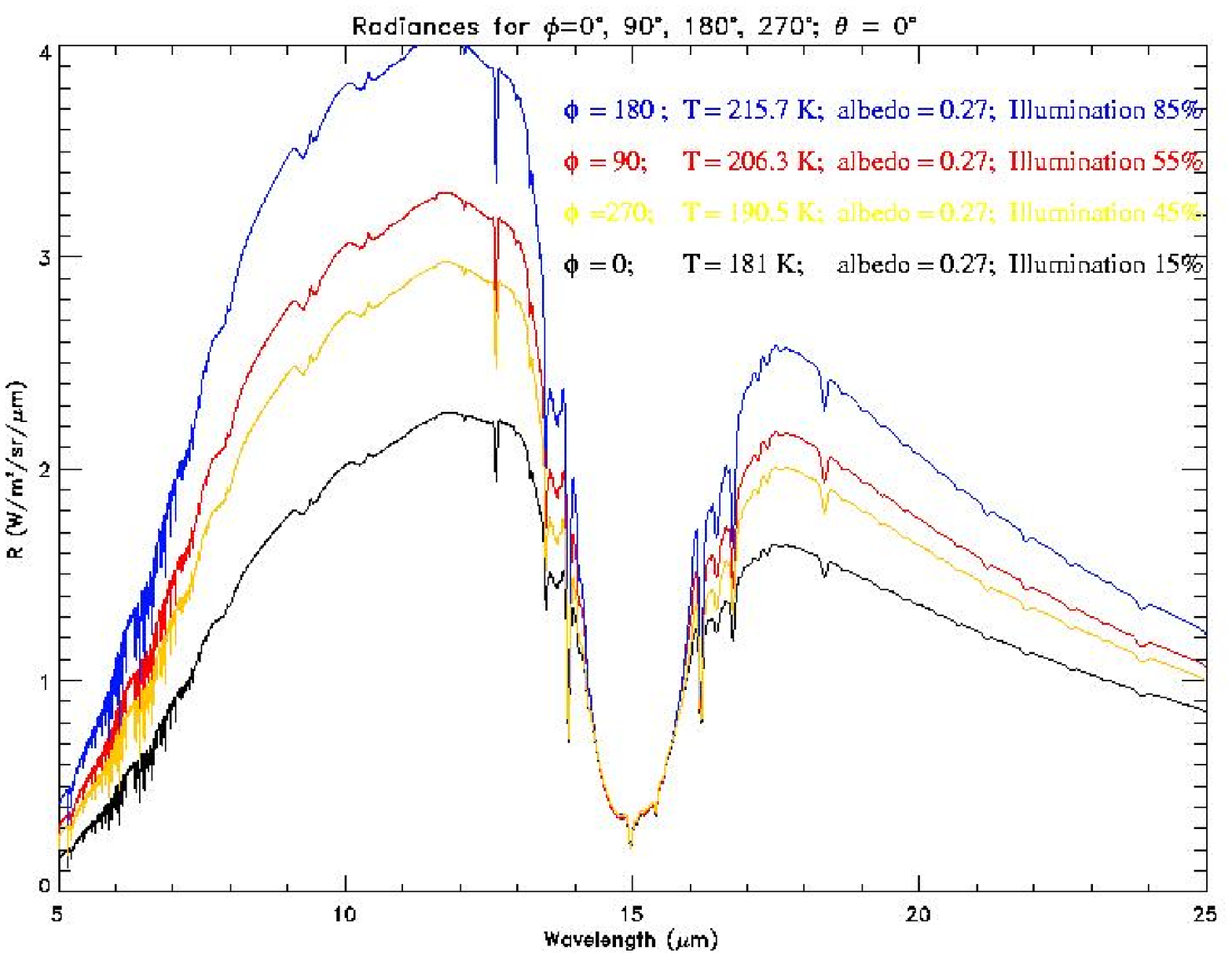}
\includegraphics[width=14 cm]{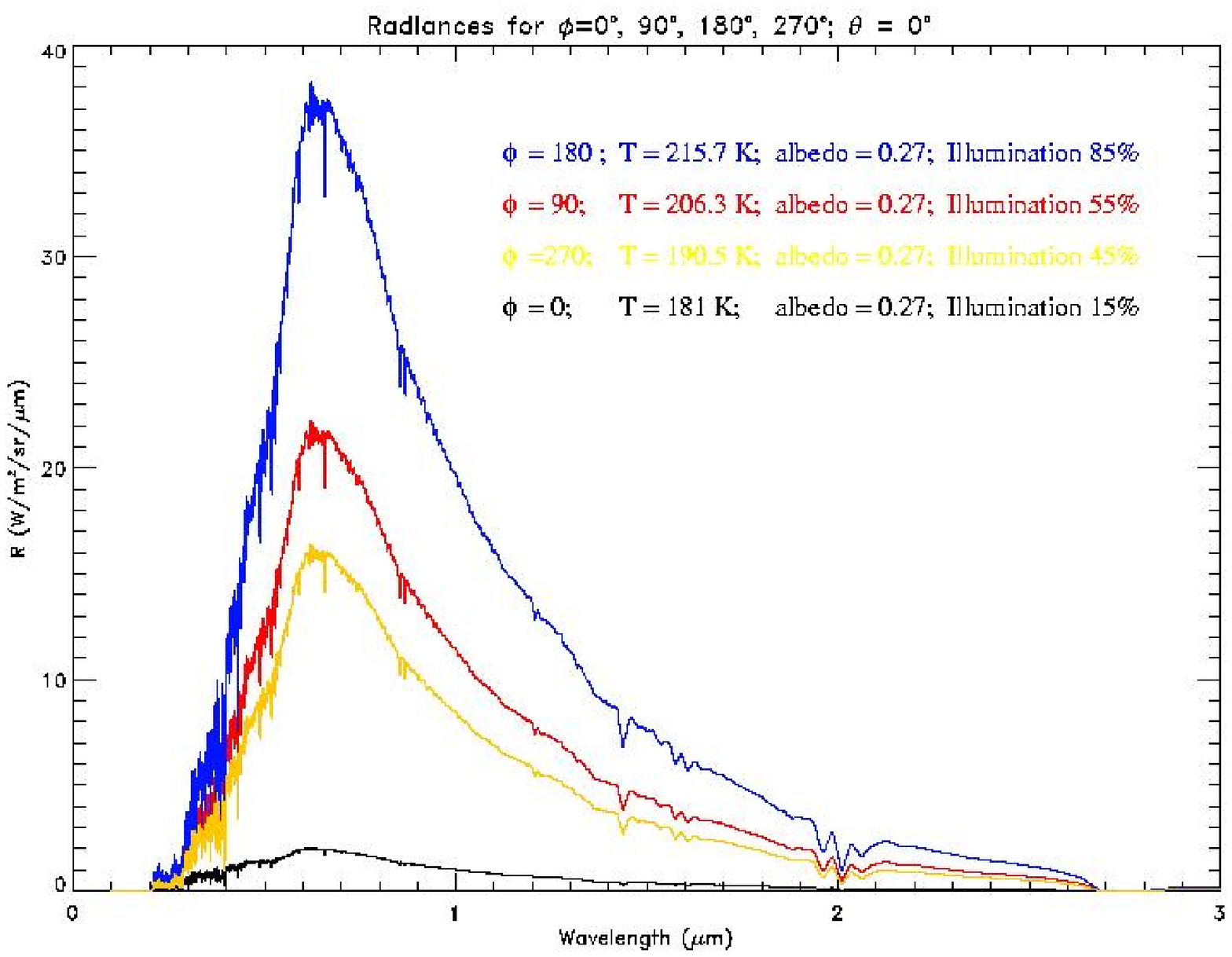}
\caption{  \emph{  {\footnotesize  Variability of disk-averaged synthetic spectra. Synthetic spectra for sub-viewer points
 with different  longitudes (latitude  $\theta = 0^{\circ}$):
$\phi = 0^{\circ}$ (fig. \ref{fig:vp} a),  $\phi = 90^{\circ}$, $\phi =  180^{\circ}$ (fig. \ref{fig:vp} b), $\phi = 270^{\circ}$ for a day corresponding to $L_{s} = 104.6$. The  disk-averaged surface temperature, solar albedo and illumination are indicated as well. \newline The polar caps extend 
$\sim 15^{\circ}$ to the South in the northern hemisphere, and $\sim 40^{\circ}$ 
to the North in the southern hemisphere, but the polar-cap contribution is hardly visible in the disk-average. The sun position derived from the JPL Horizons Ephemeris System is
24.63$^{\circ}$ latitude, and 169.99 $^{\circ}$ longitude.}  } }  \label{fig:}
\end{center}
\end{figure}

\begin{figure}
\begin{center}
\includegraphics[width=14 cm]{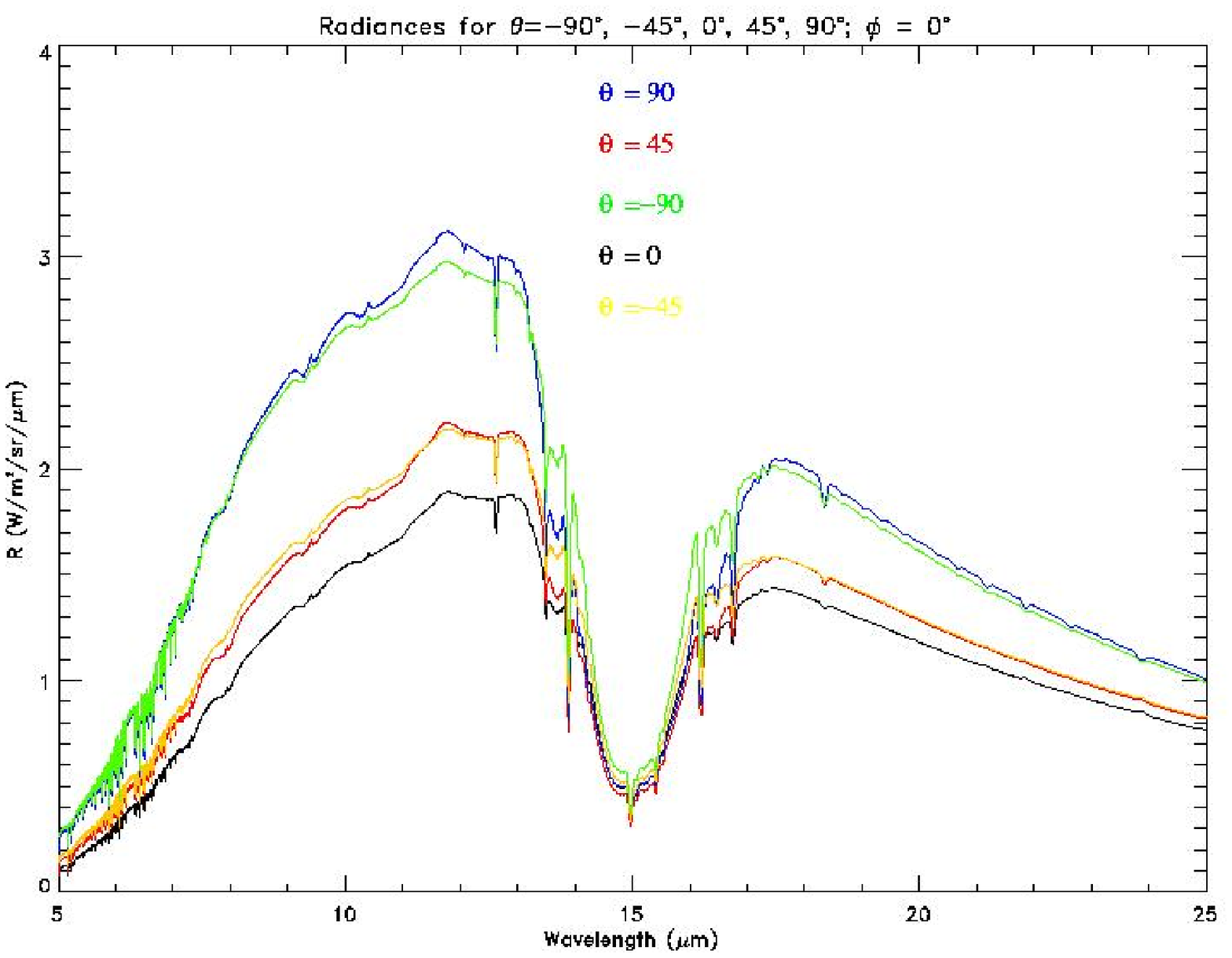}
\includegraphics[width=14 cm]{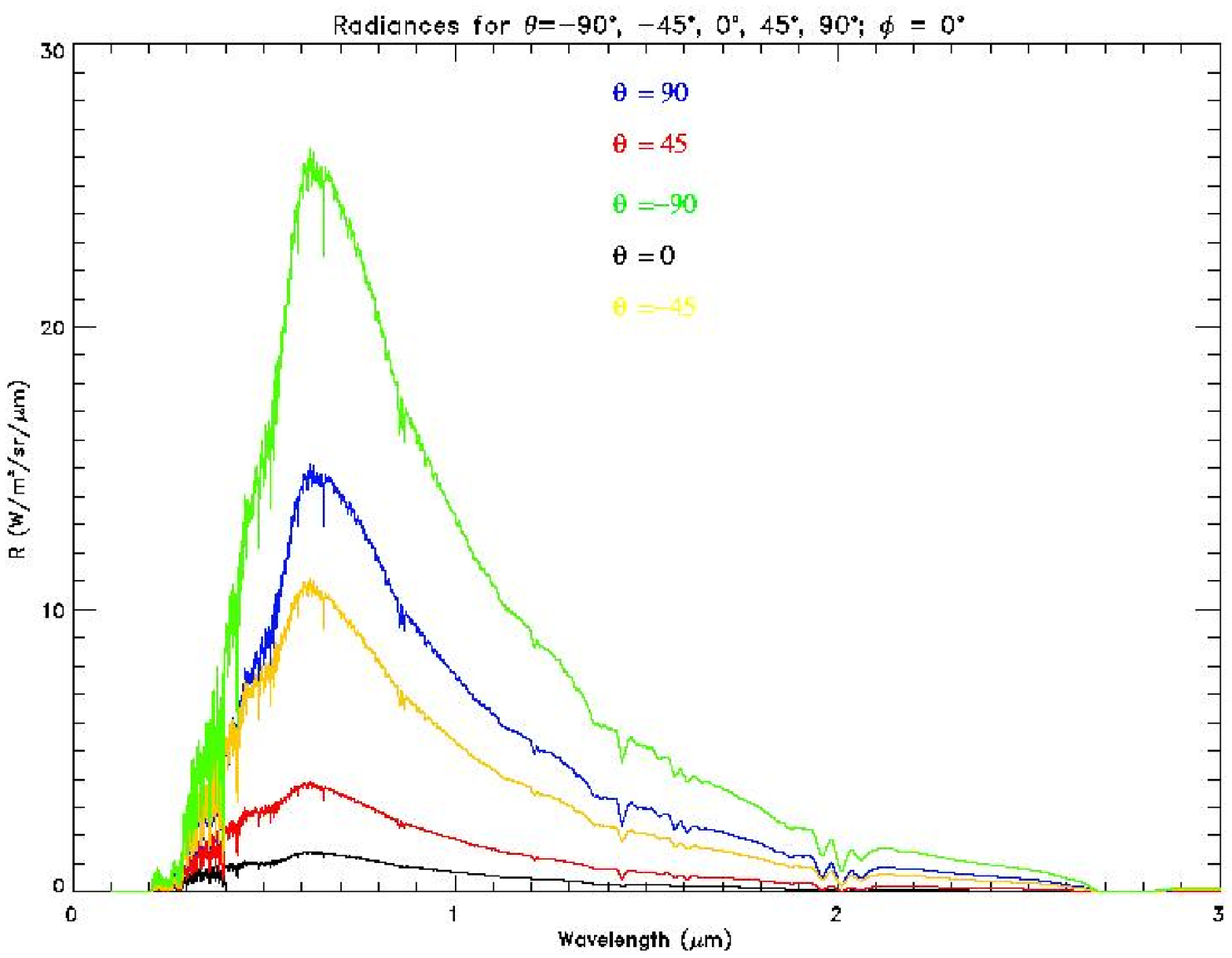}
\caption{ {\footnotesize \emph{  Variability of disk-averaged synthetic spectra.  Synthetic spectra for sub-viewer points with different latitudes (longitude  $\phi = 0^{\circ}$):
$\theta = 0^{\circ}, 45^{\circ}, 90^{\circ}, -45^{\circ}, -90^{\circ}$, for a day corresponding to $L_{s} = 211.6$. \newline The polar caps extend 
$\sim 35^{\circ}$ to the South in the northern hemisphere, and $\sim 30^{\circ}$ 
to the North in the southern hemisphere. The sun position derived from the JPL Horizons Ephemeris System is
-12.97$^{\circ}$ latitude, and 198.13 $^{\circ}$ longitude. } } } 
\label{fig:1aa}

\end{center}
\end{figure}

\begin{figure}
 \begin{center}
\includegraphics[width=14 cm]{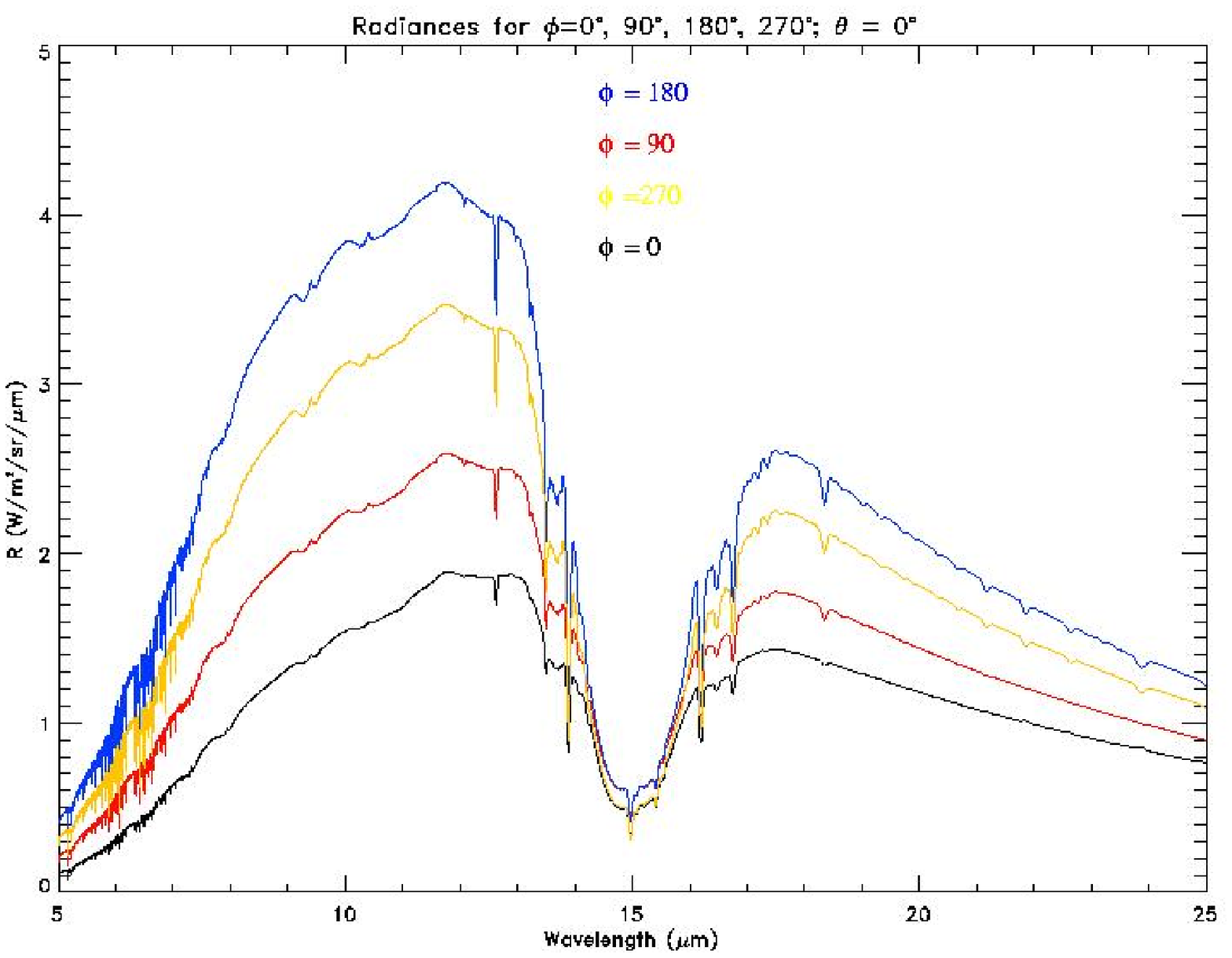}
\includegraphics[width=14 cm]{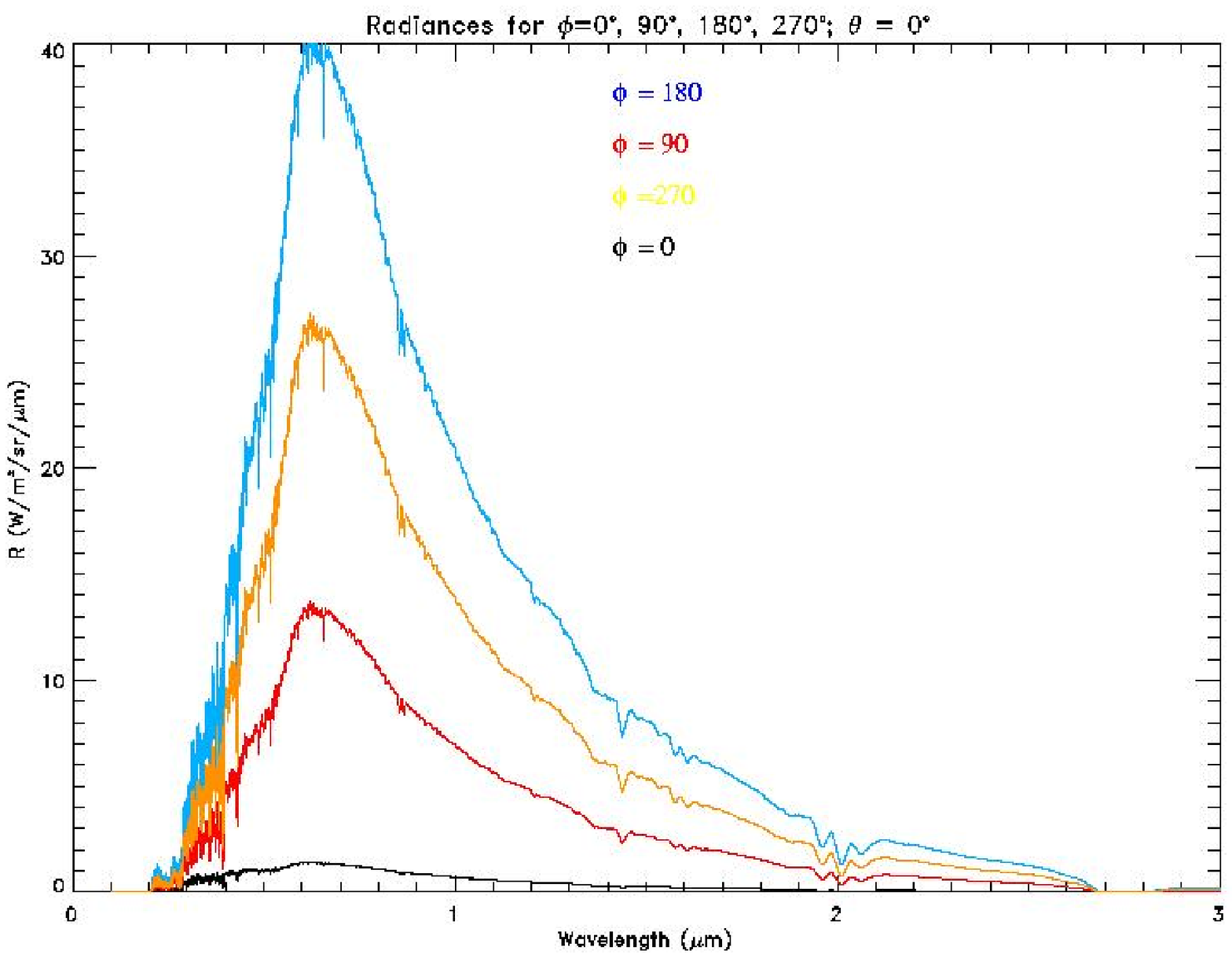}
\caption{ {\footnotesize \emph{  Variability of disk-averaged synthetic spectra.  Synthetic spectra for sub-viewer points with different longitudes (latitude  $\theta = 0^{\circ}$):
$\phi = 0^{\circ}, 90^{\circ}, 180^{\circ}, 270^{\circ}$, for a day corresponding to $L_{s} = 211.6$. \newline The polar caps extend 
$\sim 35^{\circ}$ to the South in the northern hemisphere, and $\sim 30^{\circ}$ 
to the North in the southern hemisphere. The sun position derived from the JPL Horizons Ephemeris System is
-12.97$^{\circ}$ latitude, and 198.13 $^{\circ}$ longitude. } } } \label{fig:1ab}
\end{center}
\end{figure}

   Illumination, temperature, albedo and surface type all contribute to the 
   intensity of the observed planetary radiation.  These properties, in turn, are functions of 
   planetary characteristics and viewing geometry (figs \ref{fig:vp}).  To 
   explore the dependence of disk-averaged spectra on the viewing geometry and illumination, 
   we first considered the example of Mars illuminated with a fixed solar 
   geometry, but viewed from several different vantage points.  In Figs. 
   \ref{fig:1a} and \ref{fig:1aa} we show changes in the spectral appearance
of disk-averaged Mars for days in the two different seasons ($L_s$=104.6 and 211.6) as a function of subviewer latitude, i.e. we have kept the longitude constant and 
   varied the latitude.  This simulates the appearance of the planet for an observer viewing at different angles to the plane of the planet's orbit, from vantage points varying from pole-on to equator-on.    In all cases the top panel shows the resultant MIR 
   disk-averaged spectrum, and the lower panel shows the disk-averaged 
   UV-NearIR spectrum for the planet.   In Figs. \ref{fig:} and \ref{fig:1ab} 
   we have kept the latitude of the sub-viewer point constant (at 
   0$^{\circ}$, the equator), and varied the longitude.    This 
   simulates observations of the planet as it goes through 
   different illumination phases from partly to fully illuminated, for each of the two seasons.    For both seasons considered here, not only does the visible portion of the planet's surface change with viewing 
   angle, the fraction of the planet that is illuminated also changes.

   As anticipated, the visible spectra show large variations in intensity 
   as a function of planetary phase.   Similarly, large variations in 
   overall intensity are seen for the MIR spectra, although for the same 
   range of viewing geometry, the total variation is a factor of 10 less 
   than that seen in the visible spectra.   The MIR variability is driven 
   by the difference in day-night surface temperatures on Mars, and the 
   resulting disk-averaged temperature seen by the observer.    This 
   variability in Martian surface temperature is relatively high for a 
   terrestrial planet with an atmosphere, as Mars lacks an ocean, and its 
   atmosphere is thin, so it has a very limited heat capacity to buffer its 
   climate and even out day/night variations.

\begin{figure}
\includegraphics[width=15 cm]{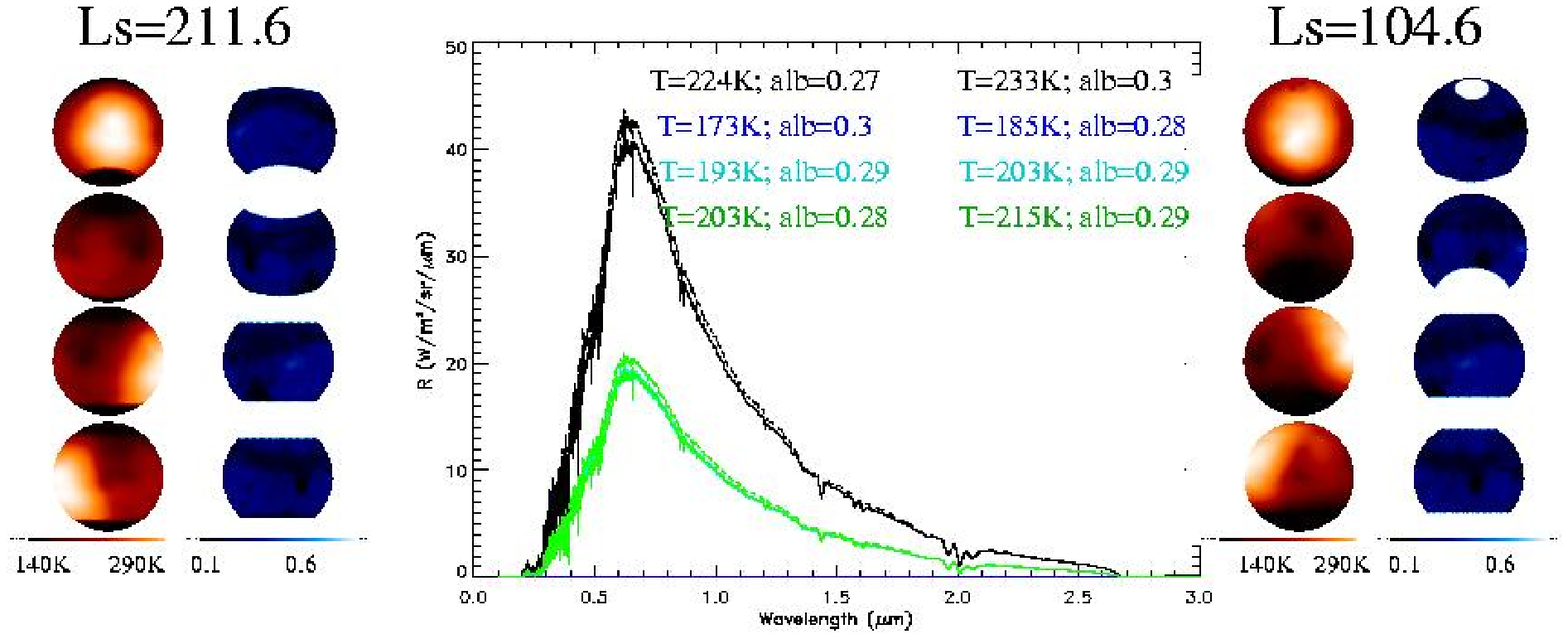} \\
\includegraphics[width=7 cm]{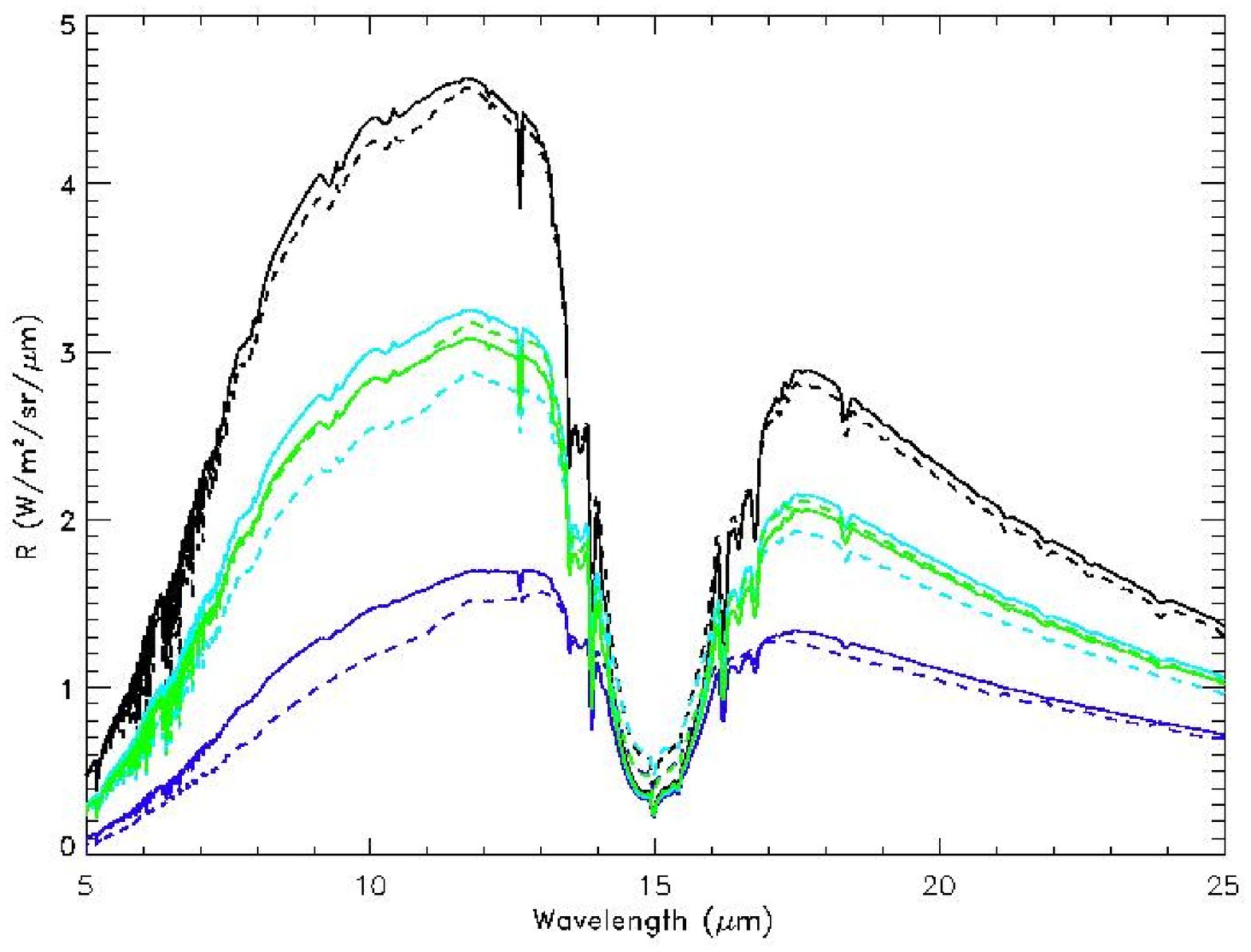}
\includegraphics[width=7 cm]{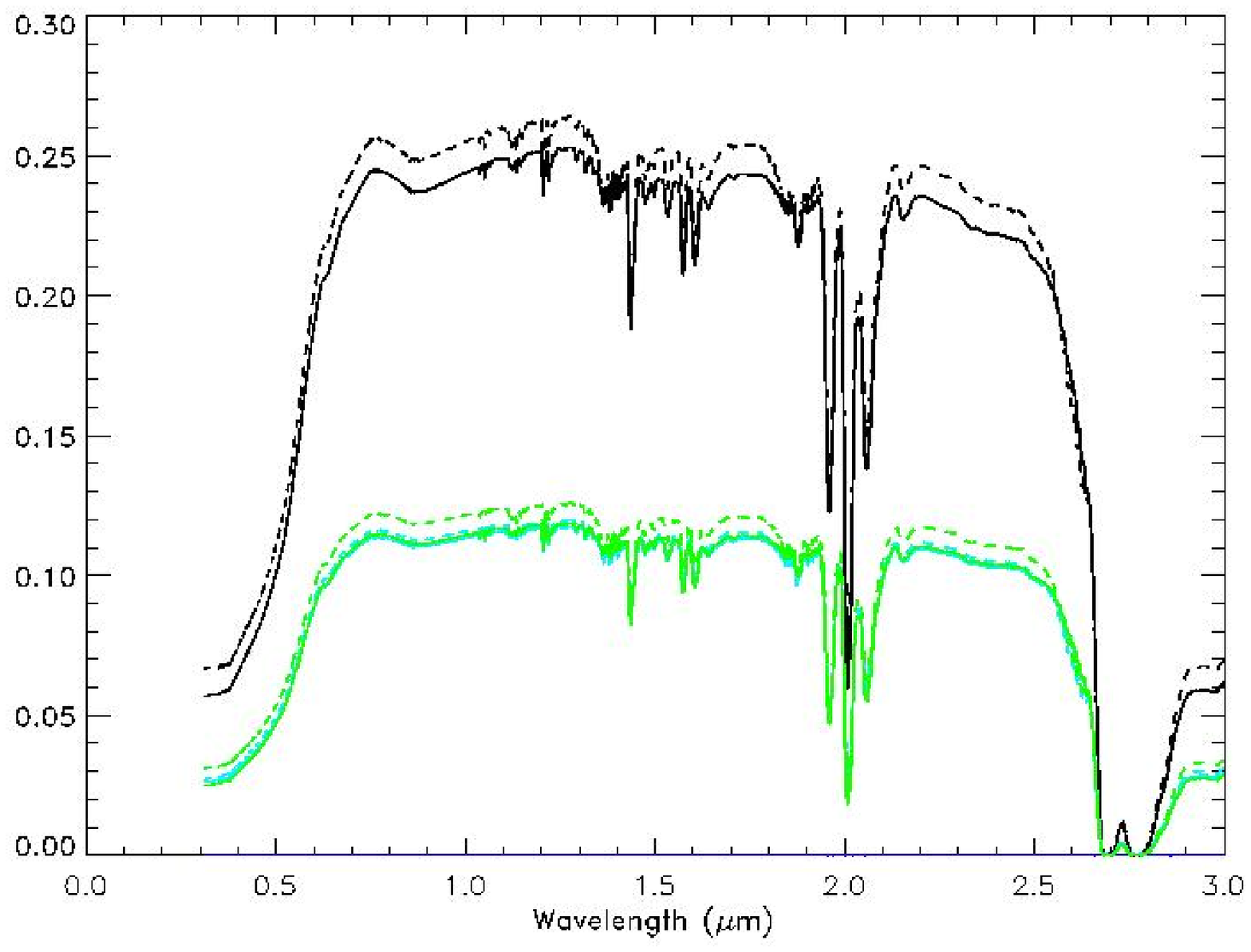}
\caption{ {\footnotesize \emph{Disk-averaged spectra for the two chosen days  equally illuminated (totally illuminated -black curve-, totally dark - blue curve-, dichotomies - green and light blue curves-). The dotted lines correspond to $L_{s} = 211.6$ (sun position:
-12.97$^{\circ}$ latitude, and 198.13 $^{\circ}$ longitude) the solid to  $L_{s} = 104.6$ (sun position:
24.63$^{\circ}$ latitude, and 169.99 $^{\circ}$ longitude). The central figure on the top, shows the UV-nearIR part of the spectrum: radiation intensity is given in W/m$^{2}$/sr/$\mu$m. On the sides are shown the surface temperature and albedo distributions for the two days corresponding to different illuminations (100\%, 0\% and 50\%). Disk-averaged values for surface temperature and albedo are indicated as well on the figure. \newline
The figures on the bottom show the IR part of the spectrum (fig. on the left) and the UV-nearIR part divided by the total solar flux at the top of the atmosphere. 
     } } } \label{fig:8acd}
\end{figure}

   In  Fig \ref{fig:8acd} we directly compare disk-averaged spectra of 
   the two seasons ($L_{s} = 104.6^{\circ}$ and $L_{s} = 211.6^{\circ}$) 
   under the same solar illuminations.   This allows us to remove 
   the effects of the illumination condition, and discern differences in 
   the spectra that are due to the planetary characteristics themselves.
   These results can then be used to 
   quantify the total range of expected brightness for a Mars-like planet 
   as a function of phase.    The UV to visible wavelength range is shown both a raw radiance spectrum, and with the solar spectrum divided out, to yield an albedo spectrum.   In the radiance spectrum, the brightness varies from 0, 
   in the totally dark case, to a maximum of 44 W/m$^2$/sr/$\mu$m, 
   disk-averaged, for the totally illuminated case.      For the mid-infrared 
   case the observed flux varies by a factor of only $\sim 3$, 
   being confined between the range 1.5-4.6 W/m$^2$/sr/$\mu$m at its peak 
   near 12$\mu$m. 

   Interestingly, when the effect of solar illumination is accounted for, 
   we can clearly see differences in spectra as a function of 
   season. This manifests itself as changes both in observed intensity and 
   spectral shape, the effect is most pronounced for the MIR spectra. 
   These changes are due primarily to the increased surface coverage by 
   the polar ice caps for the L$_s$ = 211.6 case.  With a larger surface ice 
   fraction, the visible spectra show an increased average 
   albedo and more reflected flux.  At MIR wavelengths, the ice covered region is colder, than the surrounding surface, and emits less thermal radiation.  In addition, some spectral features at MIR wavelengths are due to wavelength dependent variations in the CO$_{2}$ ice emissivity, which become more obvious as the polar ice coverage increases.   This effect can be seen in 
   the range 8-13.5$\mu$m, where the contrast between the two seasons 
   reaches a maximum between 8-11$\mu$m (a $\sim$10\% difference in flux), 
   but is less pronounced over the 11.5-13.5$\mu$m wavelength range.  This 
   effect is due to a spectral feature from CO$_2$ ice, which will be
   discussed in the next section.  For comparison, in the solar albedo spectrum we see what amounts to a disk-averaged, wavelength-dependent albedo, or reflectivity, modulated by the overlying atmospheric transmission.   These spectra show that, unlike the phenomenon observable in the MIR spectra, the compositional changes due to the seasonal polar caps are hardly observable as spectral features in the quasi-albedo spectrum for the visible, although an overall increase in reflectivity is clearly seen. This behavior is consistent with the results presented in the next section (fig. \ref{fig:6ab}).

\subsection{Light-curves}

\begin{figure}
\begin{center}
\includegraphics[width=12 cm]{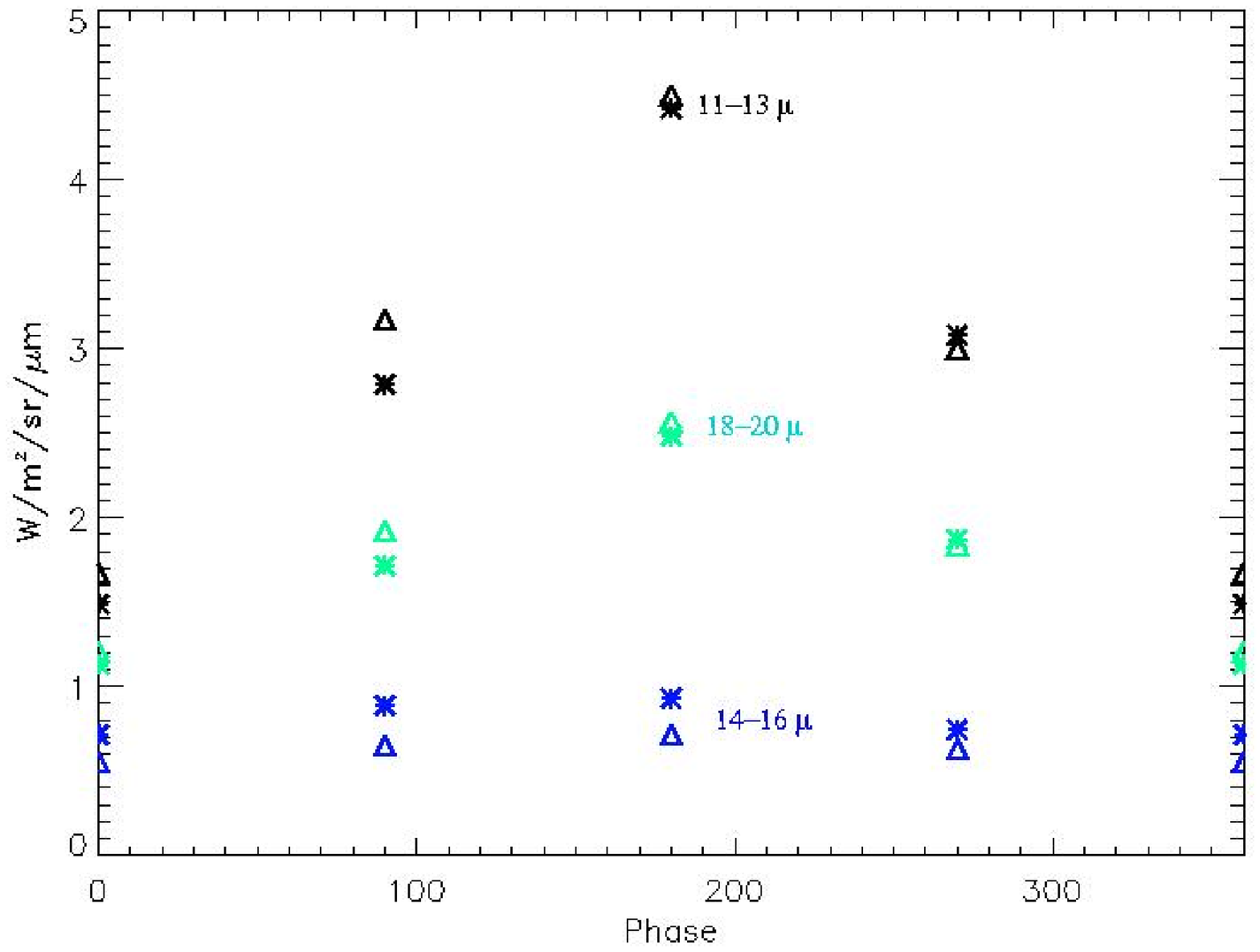}
\includegraphics[width=12 cm]{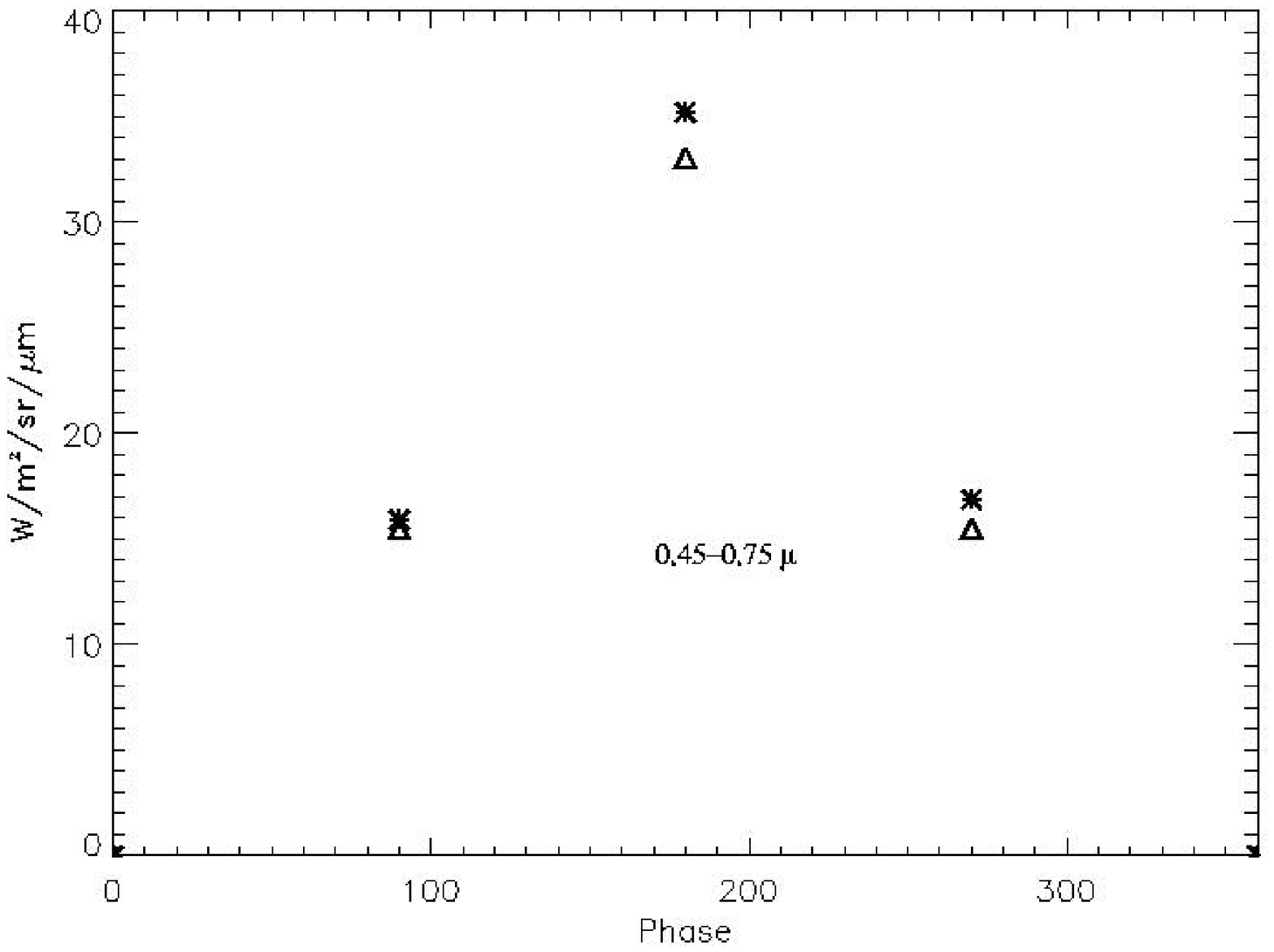}
\caption{ {\footnotesize \emph{Light-curves for the intervals 11-13 $\mu$m,  14-16 $\mu$m, 18-20 $\mu$m and 0.45-0.75 $\mu$m for the two
 chosen days  as a function of diurnal cycle (from antisolar to solar position). The quantities plotted are $ \frac{ \int_{\lambda_{1}}^{\lambda_{2}} \, \mathcal{I} \, \ud \lambda }{\int_{\lambda_{1}}^{\lambda_{2}} \,  \ud \lambda} $, where $\mathcal{I} (\lambda)$ are the disk-averaged radiation intensities shown in fig. \ref{fig:8acd},  $\lambda_{1}$ and $\lambda_{2}$ are the extremes of the chosen interval. \newline The triangle indicates $L_{s} = 104.6^{\circ}$, the star
 $L_{s} = 211.6^{\circ}$. 
Totally illuminated: phase = 180; totally dark: phases= 0, 360; dicotomies: phases = 90, 270.  } } }  \label{fig:8ac}
\end{center}
\end{figure}

   To understand the sensitivity of disk-averaged, broad-band photometric observations to seasonal variations in temperature or surface compositions differences, we have plotted light-curves for the two Martian 
   days/seasons as a function of the diurnal phase (totally illuminated, 
   totally dark and dichotomies) in fig. \ref{fig:8ac}. The quantities 
   plotted are 
\begin{equation*}
 \frac{ \int_{\lambda_{1}}^{\lambda_{2}} \, 
   \mathcal{I}(\lambda,t) \, \ud \lambda }{\int_{\lambda_{1}}^{\lambda_{2}} 
   \, \ud \lambda} 
\end{equation*}
   where $\mathcal{I} (\lambda, t)$ are the disk-averaged radiation 
   intensities shown in fig. \ref{fig:8acd}.

   We selected 11-13 $\mu$m, 14-16 $\mu$m, 18-20 $\mu$m and 0.45-0.75 
   $\mu$m as values for $\lambda_{1}$ and $\lambda_{2}$.

   As was the case for the spectra described above, clear discrepancies are 
   seen between the two different days/seasons. In particular in  the    11-13$\mu$m region, we can register a maximum variation of approximately 15\%.  As discussed in the previous section, this wavelength region is  in fact the most sensitive to changes in 
   surface ice, which produces the largest observable surface-related seasonal variations 
   in this case (increase in visible polar cap region  
   in the L$_s$=211.6 case).

   A consistent difference is also seen in the relative 
   depths of the CO$_2$ absorption band between 14-16$\mu$m for the two seasons.  This difference, of $\sim 20$\% is due  to the change in  atmospheric temperature-pressure, and CO$_2$ column depth between the two seasons, with the L$_{s}$ = 104.6 (near-summer season) showing the largest column depth of CO$_2$.

\subsection{Frozen Mars}
\begin{figure}
\begin{center}
\includegraphics[width=6.5cm]{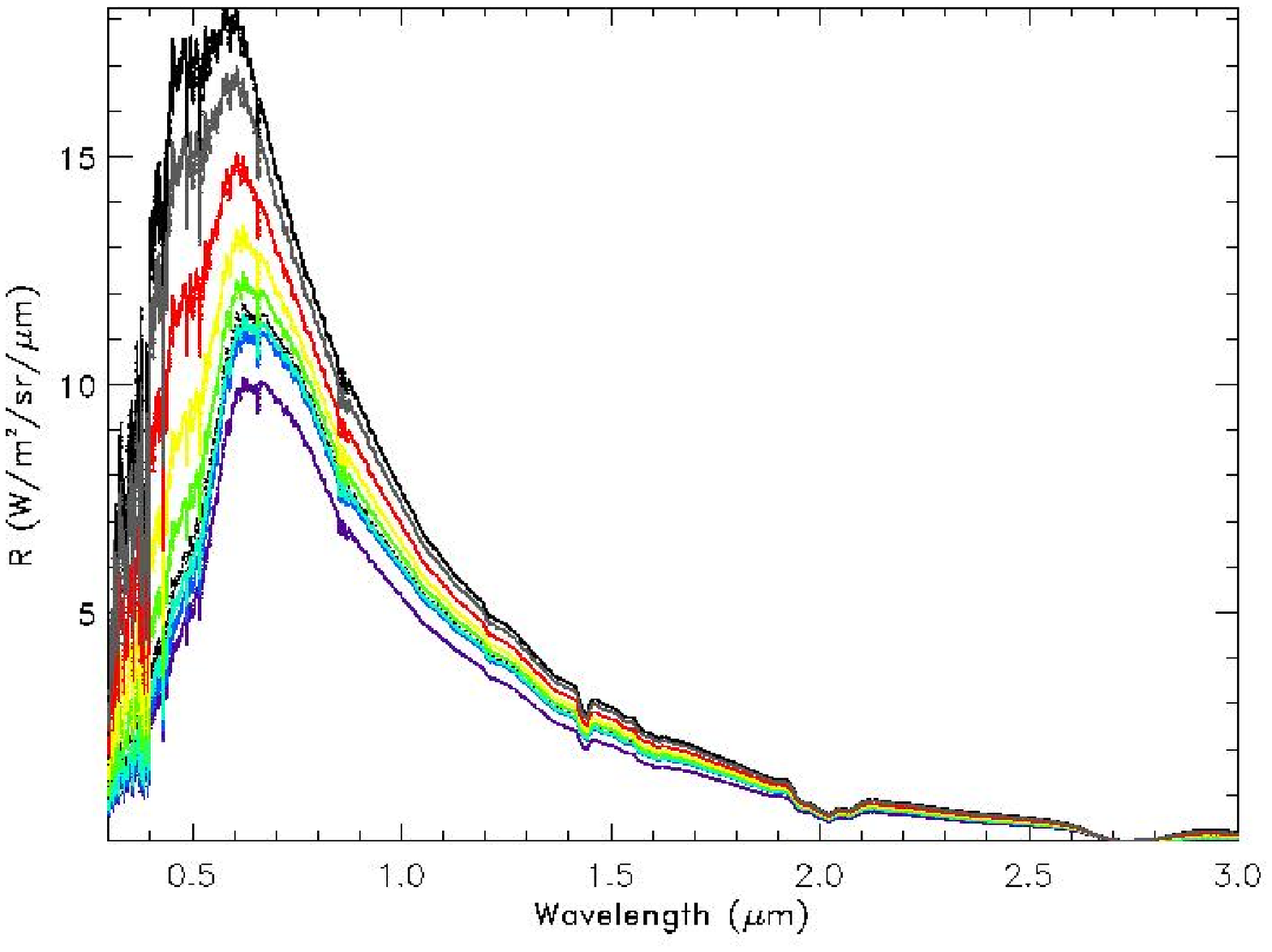}
\includegraphics[width=6.5cm]{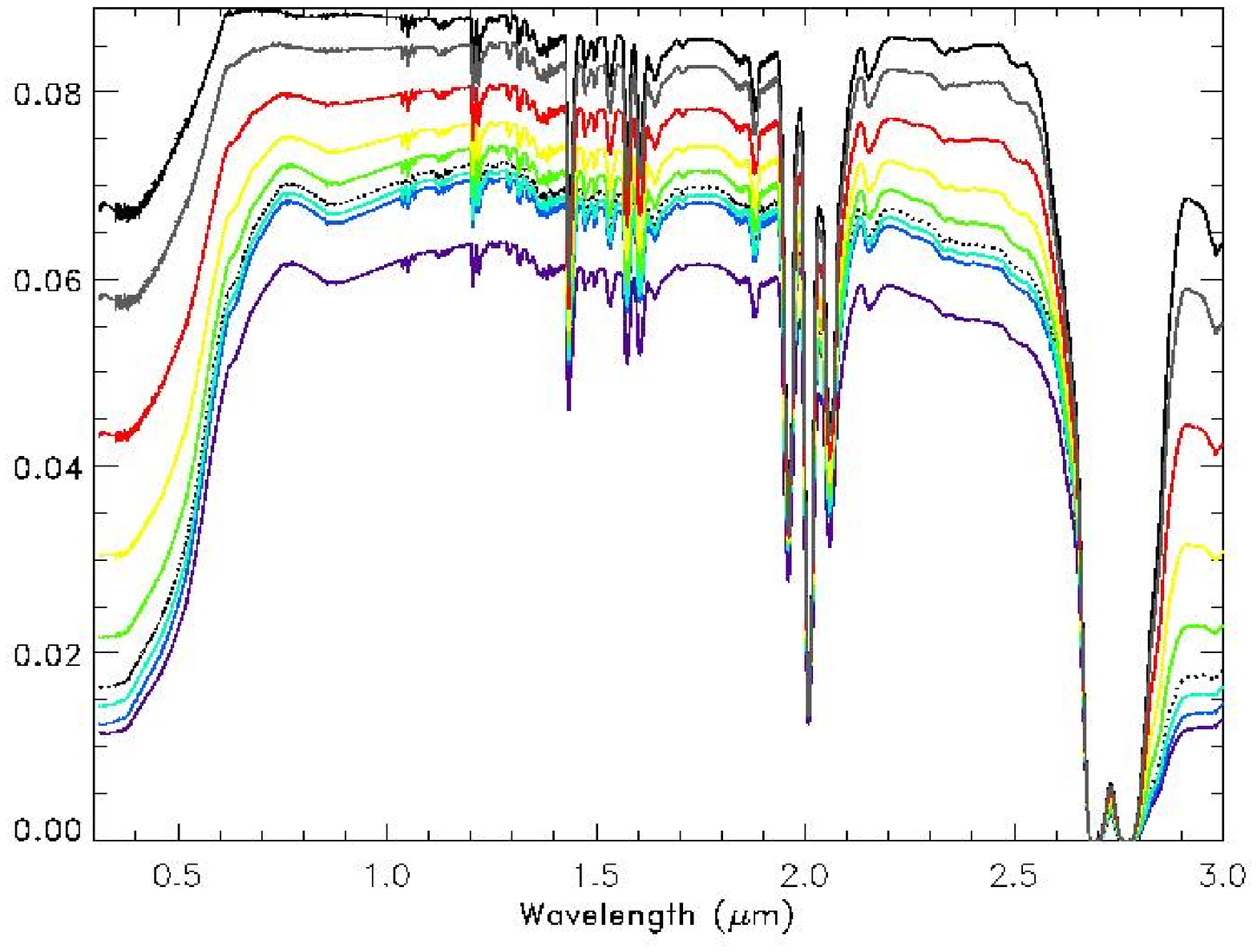} \\
\includegraphics[width=14cm]{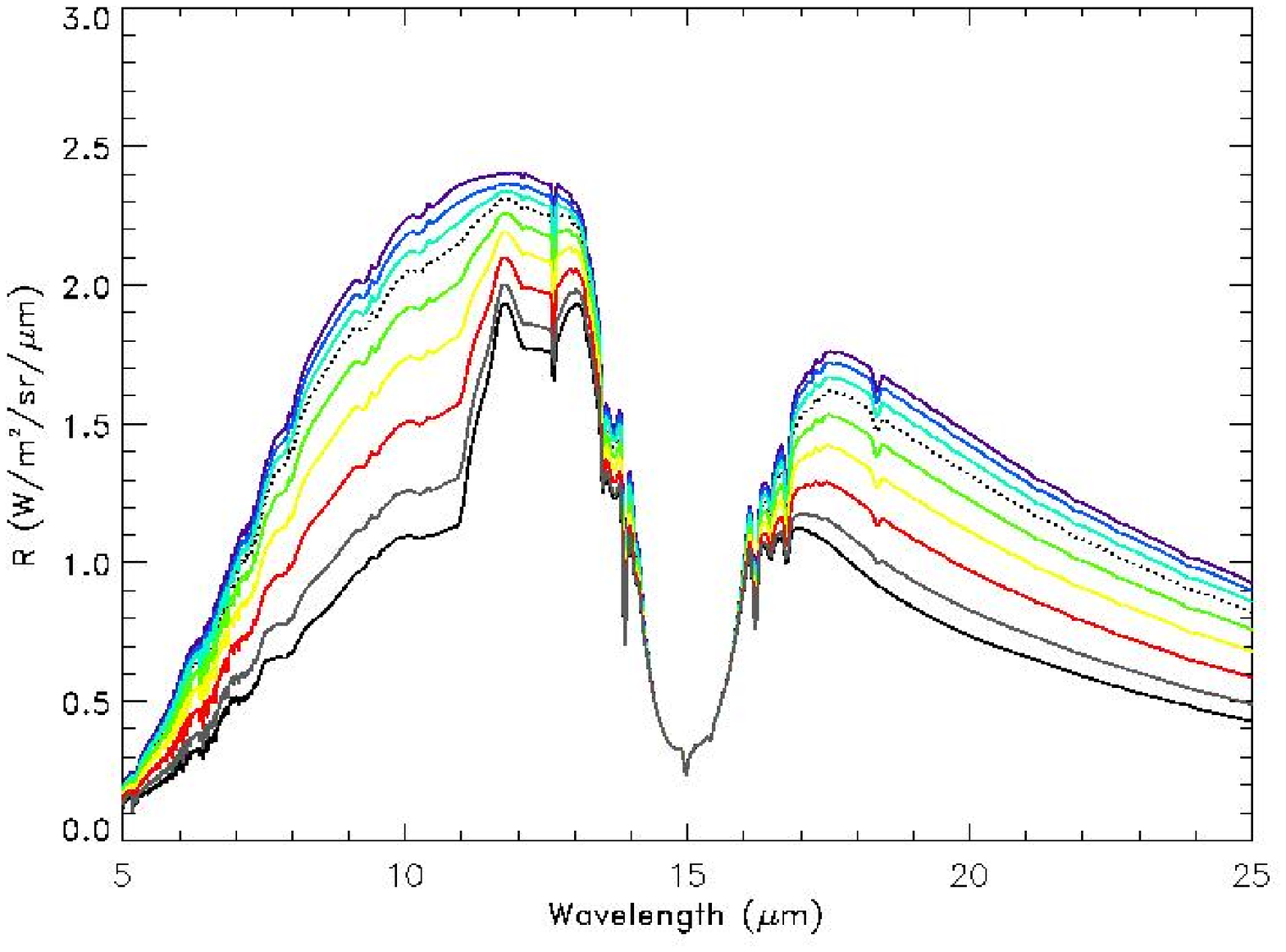}
\caption{  {\footnotesize   \emph{ Disk-averaged spectra of the southern polar-cap area (supposed extending  for $0^{\circ}$ (violet curve),  $20^{\circ}$ (dark-blue curve), $30^{\circ}$ (light-blue curve), $40^{\circ}$  (dotted curve), $50^{\circ}$ (green curve), 
$60^{\circ}$ (yellow curve),  $70^{\circ}$ (red curve),  $80^{\circ}$ (grey curve), $90^{\circ}$ (black curve)
 For increasing extension of the polar-cap area the CO$_{2}$ ice features are more and more detectable. The figures on the top, show the UV-nearIR part of the spectrum: radiation intensity is given in W/m$^{2}$/sr/$\mu$m in the fig. on the left, and divided by the total solar flux at the top of the atmosphere in the one on the right. 
   Abscissas: wavelength in $\mu$m.  }  }}  \label{fig:6ab}
\end{center}
\end{figure}

Although the effect of the realistic Martian polar ice-caps (extending 40$^{\circ}$ in latitude), produce $\sim$10\% variations in the seasonal disk-averaged spectra, for possible future TPF/Darwin measurements, we also wished to explore the detectability of planetary polar-caps in disk-averaged spectra, as a function of increasing surface coverage.   We used our Mars model to simulate a Mars-like planet, viewed pole-on, with increasingly surface coverage of CO$_2$ ice (fig. \ref{fig:6ab}).   This viewing geometry will also result in the maximum detectability of this surface type because the ice is seen near a viewing angle of 0$^{\circ}$, and not near the planetary limb, where areal projection and atmospheric path-length may serve to reduce the overall contribution to the disk-average.

\begin{figure}
\begin{center}
 \includegraphics[width=12cm]{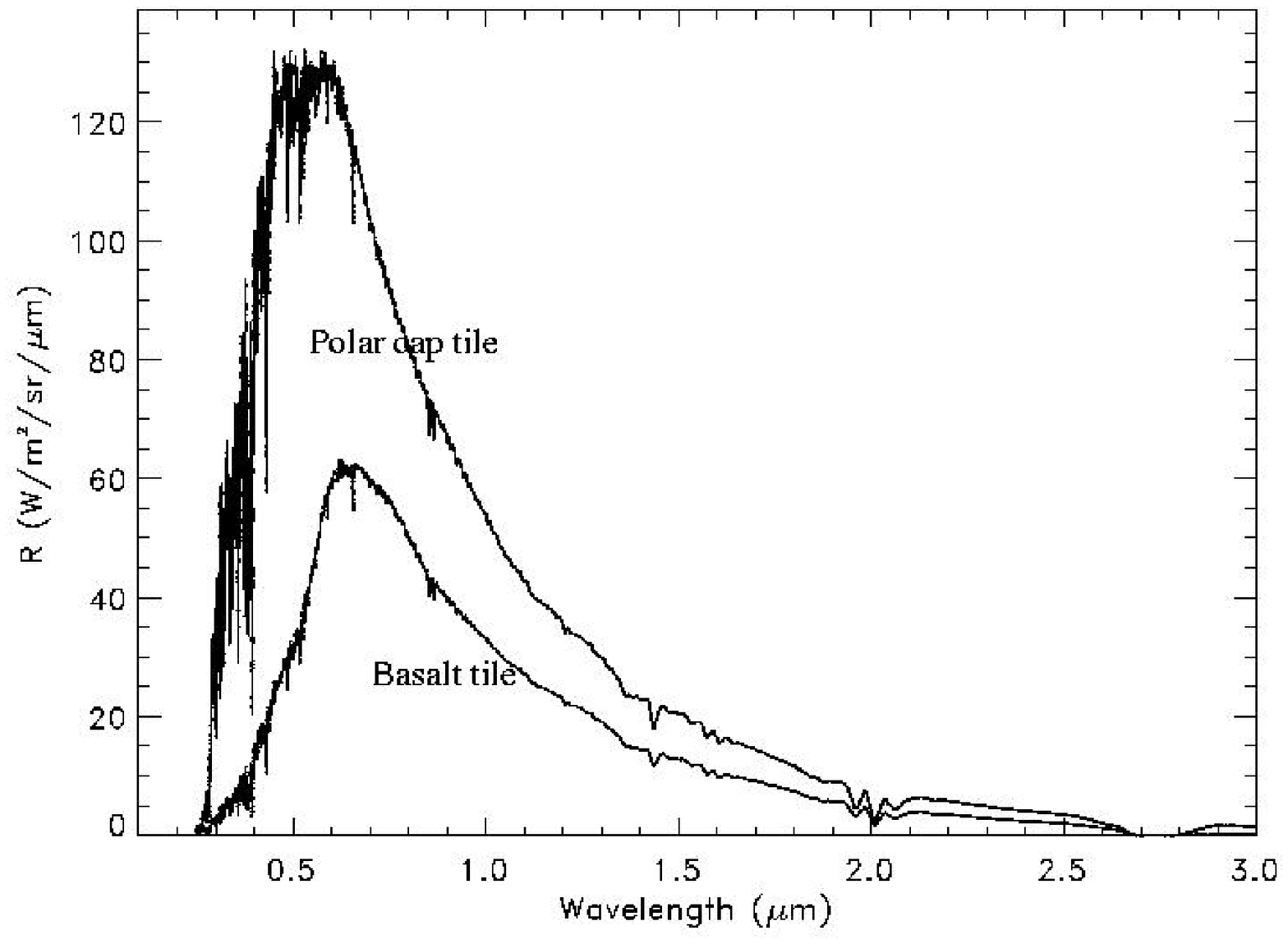} \\
 \includegraphics[width=7.cm]{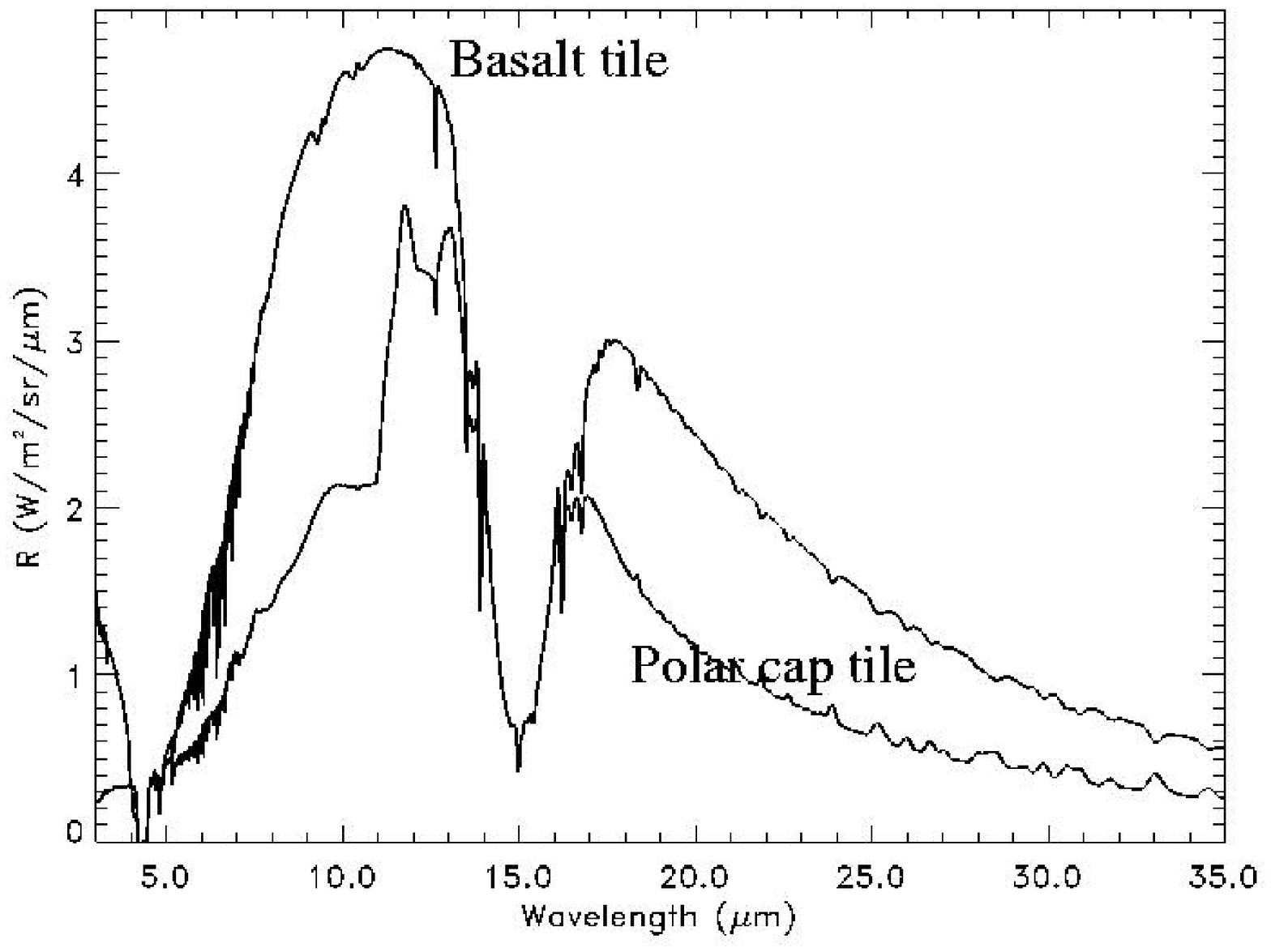} \includegraphics[width=6.cm]{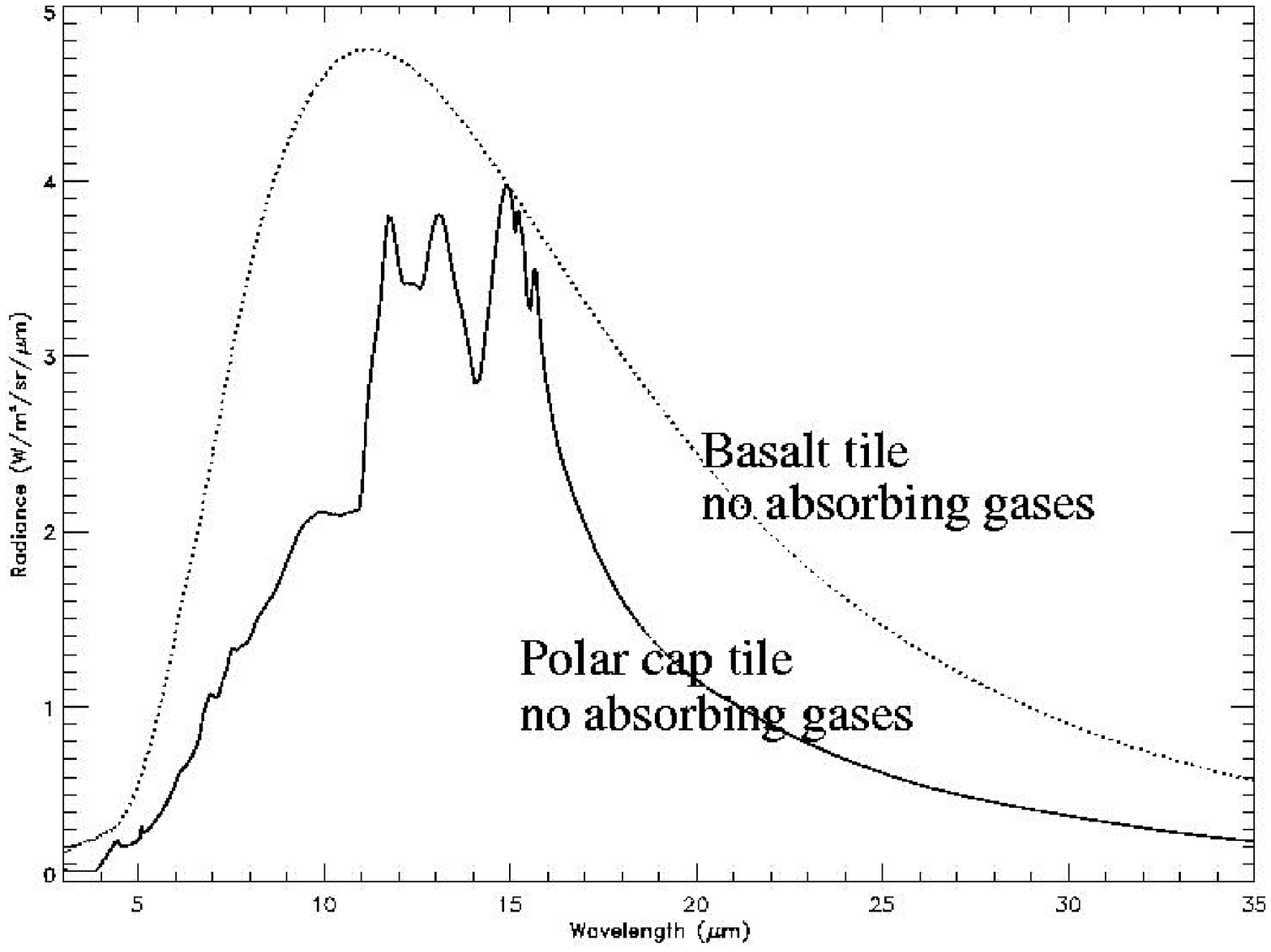}   \\
\caption{  {\footnotesize  \emph{ Day-time spectrum 
of a tile  close to the North Pole. 
 In these figures we have two graphs:
 the tile with and without the polar-cap features included. We have rock and basalt and low albedo in one case, and  CO$_{2}$ ice, high albedo and different ozone profile in the other. 
  \newline  The figure on top shows the spectrum at UV-nearIR wavelengths, at the bottom on the left,
 at mid-IR wavelengths. At the bottom on the right: a synthetic spectrum of the same tile in the same thermal state and illumination but with no absorbing gases in the atmosphere. From the comparison with the real case,  we can see what features are due to the albedo of the surface and what are due to the absorption and emission processes of atmospheric gases.   }   }}
  \label{fig:5a} \end{center} \end{figure}

Unlike the majority of the planet, with its predominantly basalt surface, the poles have a water/CO$_{2}$ ice dominated surface coverage with the resultant high albedo (figs \ref{fig:5zz}, \ref{fig:5a}).  Polar spectra also exhibit increased ozone absorption due to the increased ozone abundance at high latitudes (figs \ref{fig:5z}, \ref{fig:5a}).

In the MIR spectrum, figure \ref{fig:6ab} shows a dramatic increase in the relative prominence of the CO$_2$ ice feature between 11-13\,$\mu$m.   From this plot, the polar ice cap is first visible at 30$^{\circ}$ extent when the polar ice cap is centered on the sub-observer point.   However, this detectability threshold will depend strongly on the spectral resolution and sensitivity of the instrument being used to obtain the disk-averaged spectrum.   Note also that the poles were also visible in the equator-on view, when the North pole extended $\sim$35$^{\circ}$ southward and the South pole extended $\sim$30$^{\circ}$ northward (figs \ref{fig:1aa}, \ref{fig:1ab}).   In the visible light case, differences are seen predominantly in the amount of reflected radiation received by the observer.   Correction for the incoming solar spectrum reveals however that differences in the disk-averaged reflectivity CAN be discerned in this case.   The most notable changes in the visible reflectivity occur between 0.5 and 1.0\,$\mu$m, and especially between 0.6 and 0.8\,$\mu$m, where the mostly flat spectrum of the CO$_2$ ice starts to smooth out the basalt absorption features near 0.65 and 0.85$\mu$m.   It is also interesting to note that, although these wavelength regions are unlikely to be included in a baseline TPF mission, the largest changes in reflectivity with the changing surface composition occur between 0.3 and 0.5\,$\mu$m, and around 2.9\,$\mu$m (where the surface is relatively dark and the frost is bright).

\subsection{Detectability}

The detectability of a spectral feature or a time-dependent variation in a spectral feature is clearly a function of the instrumentation used for the observation.  The synthetic spectra discussed in the paper so far have been noiseless, and at extremely high spectral resolution.   This allows us to understand the intrinsic variability or detectability of the features assuming a perfect instrument.   Here, however, we show the results of processing our synthetic disk-averaged spectra with a realistic observational system simulator (developed by T. Velusamy) for representative designs for both the TPF coronograph and interferometer [Beichman and Velusamy, 1999], [Beichman et al., 1999]. 

\begin{figure}
\begin{center}
\includegraphics[width=14 cm,bbllx=55bp,bblly=219bp,bburx=560bp,bbury=735bp,
clip=]{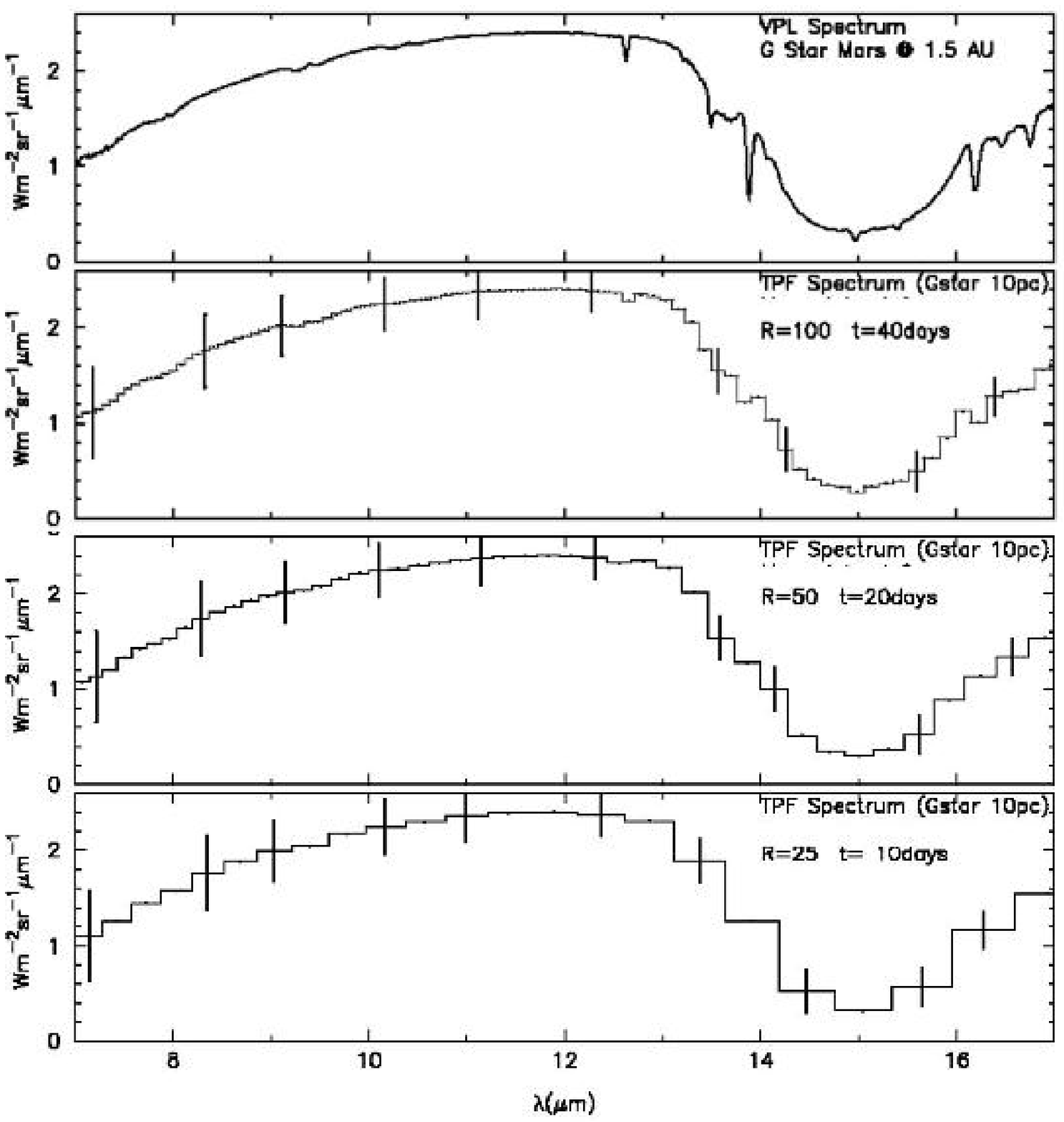}
\caption{ {\footnotesize  \emph{Simulation of TPF detection of a disk averaged spectrum of a basalt-Mars (distance: 10 pc) at MIR wavelengths.
The three panels  show the interferometer instrument simulator results for increasing degradation in the spectral resolving power, R ($\lambda/\Delta\lambda$) = 100, 50 and 25 
  } } } \label{fig:10a}
\end{center}
\end{figure}

Figure \ref{fig:10a} shows the results of this study for a disk-averaged basalt-Mars (distance: 10 pc) at MIR wavelengths.  The interferometer design assumed here has four 4m telescopes on a 75m baseline.  This is not the final design for TPF, but is representative of designs under consideration at the time this paper was drafted.  Note also that at the time of drafting, the full-TPF wavelength range under consideration was 6.5-17\,$\mu$m, which is what we have considered here.   The top panel in Fig \ref{fig:10a} shows the original synthetic spectrum. The three panels underneath show the interferometer instrument simulator results for increasing degradation in the spectral resolving power, R ($\lambda/\Delta\lambda$) = 100, 50 and 25.   For reference, the current nominal resolving power for the TPF interferometer design, is R = 25.   The error bars show 1-$\sigma$ noise levels for the observing configuration given in each panel.  

\begin{figure}
\begin{center}
\includegraphics[width=14 cm,bbllx=55bp,bblly=219bp,bburx=560bp,bbury=735bp,
clip=]{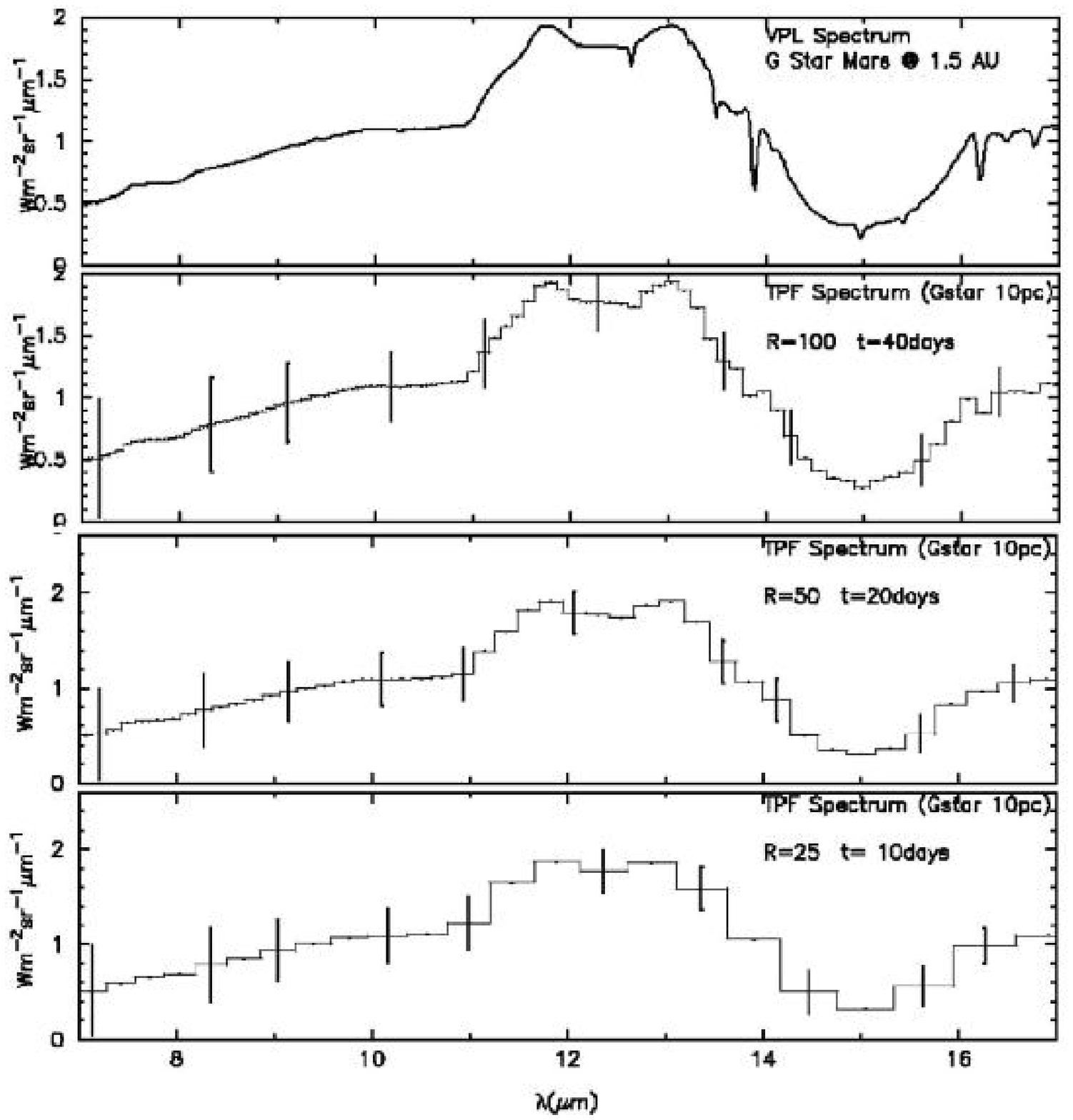}
\caption{ {\footnotesize \emph{Simulation of TPF detection of a disk averaged spectrum of a ice-Mars (distance: 10 pc) at MIR wavelengths.
The three panels  show the interferometer instrument simulator results for increasing degradation in the spectral resolving power, R ($\lambda/\Delta\lambda$) = 100, 50, and 25.
 }
  } }  \label{fig:11a}
\end{center}
\end{figure}

Figure \ref{fig:11a} shows similar results for a pure polar-cap spectrum (simulating the CO$_2$ ice covered Mars).     Note that even with the nominal resolution of R=25, it is still possible to discriminate Mars from a CO$_{2}$ ice covered planet based on the 11-13$\mu$m CO$_2$ spectral feature. 

\begin{figure}
\begin{center}
\includegraphics[width=14 cm,bbllx=40bp,bblly=170bp,bburx=550bp,bbury=590bp,
clip=]{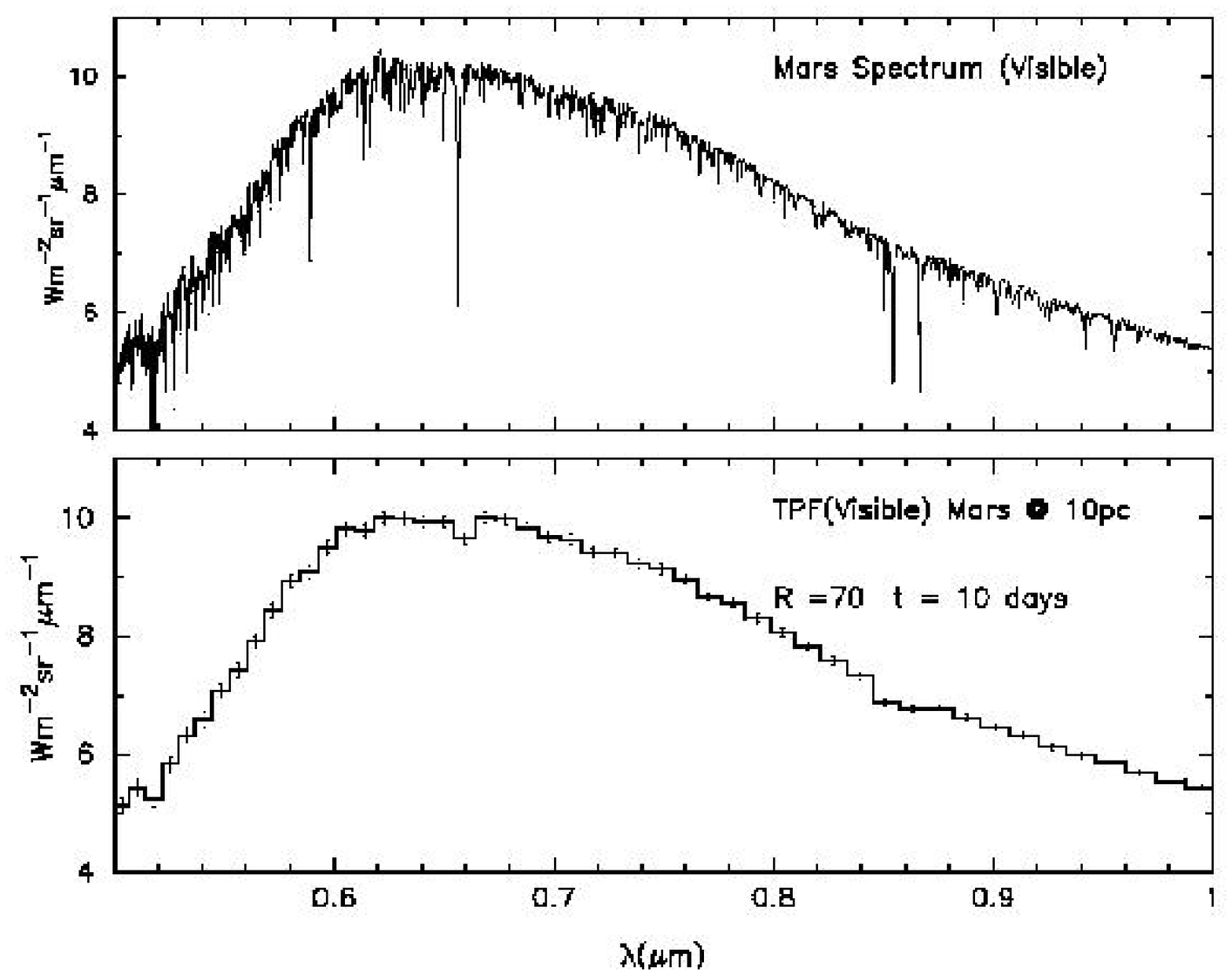}
\caption{ {\footnotesize \emph{Simulation of TPF detection of a disk averaged spectrum of a basalt-Mars (distance: 10 pc) at VIS wavelengths.
The lower panel  shows the coronograph instrument simulator result at a spectral resolving power of R= 70.  Most of the absorption lines are due to the star-spectrum, as it can be inferred from fig. \ref{fig:6ab}, where the solar spectrum has been accounted for.  }
  } }  \label{fig:14a}
\end{center}
\end{figure}

\begin{figure}
\begin{center}
\includegraphics[width=14 cm,bbllx=40bp,bblly=170bp,bburx=550bp,bbury=590bp,
clip=]{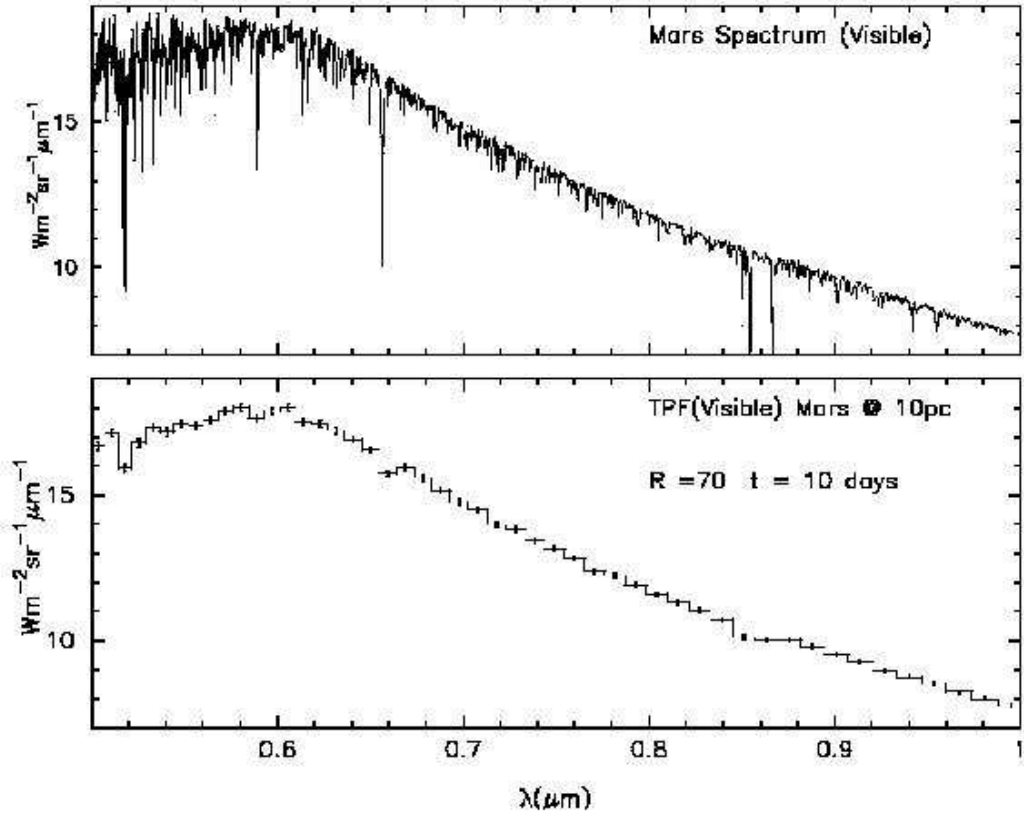}
\caption{ {\footnotesize \emph{Simulation of TPF detection of a disk averaged spectrum of a ice-Mars (distance: 10 pc) at VIS wavelengths.
The lower panel  shows the coronograph instrument simulator result at a spectral resolving power of R= 70.  
Most of the absorption lines are due to the star-spectrum, as it can be inferred from fig. \ref{fig:6ab}, where the solar spectrum has been accounted for. } } }  \label{fig:15a}
\end{center}
\end{figure}

Figures \ref{fig:14a} and \ref{fig:15a} estimate the sensitivity of the TPF chronograph for basalt-icy Mars. These plots have been produced by processing the synthetic disk-averaged spectra with an observational system simulator of the TPF coronograph, at a spectral resolving power of R= 70, and an integration time of 10 days. In the visible, the shape and intensity of the two spectra are considerably different, even at a resolution of R = 70. The more difficult question is whether such differences would yield a unique interpretation.

   \section{Discussion}
   \subsection{Recommendations for Extrasolar Planet Characterization}
   Driven largely by a desire to understand the appearance of extrasolar 
   terrestrial planets, in this
   paper we have attempted to quantify 1) the range of planetary 
   brightnesses and intrinsic variability
   for a Mars-like planet when viewed at arbitrary geometry and 2) whether 
   disk-averaged photometric or
   spectral observations were sensitive to changes in local properties, 
   namely the advance or retreat of
   polar ice cap coverage as a function of season, or planetary type.  Mars 
   was chosen for the development
   of this first model because it is less complex than Earth, and yet 
   provides a good example of a relatively well-measured (probably) abiotic environment.  It is also a 
   particularly challenging target
   for planet characterization and detection missions, due to its small 
   size and low temperature, and is
   likely to represent the extreme lower bound for terrestrial planet 
   detection.  

   To explore these issues, the model developed here has allowed us to 
   present synthetic, disk-averaged
   spectra for present-day Mars, as a function of wavelength, viewing 
   geometry, phase and season, far
   exceeding the wavelength range and viewing geometries available in any 
   homogeneous set of observations
   of this planet.  
      Although variations as large as a factor of 3 were still seen for the 
   MIR spectra as a function of
   viewing angle and phase, these were a factor of 10 less than those seen 
   for visible spectra with non-zero fluxes with the same range of viewing geometry.   In the MIR, the 
   variability with phase and
   viewing geometry seen for Mars is much larger than that for Venus or Earth, as Mars lacks an ocean, and its atmosphere is thin, so it has a very 
   limited ability to buffer its   climate and even out day/night variations.

   Addressing the second issue, on whether disk-averaged photometric or 
   spectral observations were
   sensitive to changes in local properties, we found that although 
   disk-averaging clearly reduced the
   influence of localized features, when the effect of solar illumination 
   was accounted for, we were in
   fact able to detect differences in the observed spectra as a function of 
   season at all wavelengths. 

   This effect either manifested itself as changes in observed intensity, 
   or spectral shape.  
   Seasonal variations were related either to changes in atmospheric 
   density, surface temperature, or
   areal coverage of surface ice.   While changes in overall brightness 
   would be interesting to observe as
   a function of season on an extrasolar terrestrial planet, it is 
   detectable changes in spectral shape
   that will ultimately allow us to understand the process or planetary 
   feature that is responsible for
   the seasonal change.  

   The seasonal effects were most pronounced for the MIR spectra, where a 
   larger surface ice fraction
   reduced the brightness of 
   the planet at wavelengths
   outside the 15\,$\mu$m CO$_2$ absorption band by as much as 15\% between 
   days separated by 1/4 of a
   Martian year.    These modeling results also indicate that it is 
   possible to discern differences due to
   increased polar ice coverage in the spectra, particularly in the 
   wavelength range 8-13.5$\mu$m, where
   the wavelength dependent emissivity of CO$_2$ ice produces a strong 
   spectral feature.  In comparison,
   the disk-averaged visible reflectivity of Mars did not show almost any 
   observable changes in spectral
   features, although clear increases in the disk-averaged brightness were 
   seen at most visible
   wavelengths. This result suggests that MIR observations will 
   be able to distinguish a
   planetary surface with a CO$_2$ ice component if these observations cover 
   the 8-13
   $\mu$m region, which is currently under consideration as the "minimum 
   mission" wavelength range for
   TPF-I.   Broadband photometry could also be used to check for this 
   feature, and it could also be
   distinguished using filters that would also search for ozone.  For 
   example, filters from 8-9, 9-10.5
   and 10.5-13\,$\mu$m could be used both to search for ozone or CO$_2$ 
   ice.   Note however that all
   three bands would be required, to ensure that the CO$_2$ ice emission is 
   not mistaken for continuum
   adjacent to O$_3$ absorption.   

   Between the two Martian seasons, a consistent difference of $\sim20$\% 
   was also seen in the relative
   depths of the CO$_2$ absorption band between 14-16$\mu$m due primarily 
   to the change in overall
   atmospheric temperature profile.

      Although our frozen world has not been allowed to build up 
   O$_3$, we nonetheless examined the
   spectra to see whether even Martian levels of O$_{3}$ were detectable.   
   Although ozone is more plentiful
   over the Martian poles due to reduced photochemical destruction rates [Yung and DeMore, 1999]  it was still not
   visible in the MIR spectrum of the totally CO$_2$ ice-covered planet because the ice covered surface was accompanied by a more isothermal atmosphere.   This 
   however, is not a rigorous test
   of this hypothesis, as the O$_3$ profile used was for a planet that still 
   had active oxygen sinks, and O$_3$
   was not allowed to collect in the atmosphere over time.    Further work 
   on this issue might involve
   using a couple climate-chemistry model to model the rate and extent of 
   O$_3$ buildup on a totally frozen
   world.

\section{Conclusions} 
We have developed a new spatially- and spectrally-resolved model of the 
planet Mars, which is the first
in a set of similar models to be used to explore the detectability of 
planetary characteristics for
terrestrial planets around other stars.   Using this model we have 
presented synthetic, disk-averaged
spectra for present-day Mars, as a function of wavelength, viewing 
geometry, phase and season.  

Although variations as large as a factor of 3 were still seen for the 
MIR spectra as a function of
viewing angle and phase, these were a factor of 10 less than those seen 
for visible spectra with non-zero fluxes with the same range of viewing geometry.   Interestingly, 
although disk-averaging clearly
reduced the influence of localized features, when the effect of solar 
illumination was accounted for
we clearly saw differences in the observed spectra as a function of 
season at all wavelengths,
manifested as either changes in observed intensity or spectral shape.  
These seasonal variations were
related either to changes in atmospheric density, or areal coverage of 
surface ice.
  The seasonal effects were most pronounced for the MIR spectra where a 
larger surface ice fraction
increased the albedo and cooled the planet, reducing the brightness of 
the planet at wavelengths
outside the 15\,$\mu$m CO$_2$ absorption band by as much as 15\% between 
days separated by 1/4 of a
Martian year.    These modeling results also indicate that it is 
possible to discern differences due to
increased polar ice coverage in the spectra, particularly in the 
wavelength range 8-13.5$\mu$m, where
the wavelength dependent emissivity of CO$_2$ ice produces a strong 
spectral feature.  In comparison,
the disk-averaged visible reflectivity of Mars did not show any 
observable changes in spectral
features, although clear increases in the disk-averaged brightness were 
seen at most visible
wavelengths.   Between the two Martian seasons, a consistent difference 
of $\sim20$\% was also seen in
the relative depths of the CO$_2$ absorption band between 14-16$\mu$m 
due primarily to the change in
overall atmospheric pressure.

This model, while reproducing Mars with a high degree of fidelity, can 
also be used to explore similar,
but different Mars-like planets.  Model runs for an increasingly frozen 
Mars showed a dramatic increase
in the relative prominence of the CO$_2$ ice feature between 
11-13\,$\mu$m, with the sub-observer polar
ice cap first visible at 30$^{\circ}$ extent.  Although in both the MIR 
and visible cases the overall
flux levels decreased or increased respectively, a similar detectability 
of an ice-related spectral
feature in the visible reflectivity was not observed until 60$^{\circ}$ 
extent.  The most notable
changes in the visible reflectivity occur between 0.5 and 1.0\,$\mu$m, 
and especially between 0.6 and
0.8\,$\mu$m, where the mostly flat spectrum of the CO$_2$ ice starts to 
smooth out the basalt
absorption features near 0.65 and 0.85$\mu$m in the disk average.    
Large changes in reflectivity and
spectral shape with the changing surface composition occur between 0.3 
and 0.5\,$\mu$m, and around 2.9
\,$\mu$m.   Instrument simulator models for representative TPF 
interferometer and coronograph designs
for Mars and a CO$_2$ ice-covered Mars-sized object indicate that 
atmospheric CO$_2$ is readily
detectable, and the 0.5-0.6 and 11-13$\mu$m CO$_2$ ice features might be 
detectable in the ice-covered
case, with sufficient integration even, at the relatively low spectral 
resolutions that are likely to
characterize TPF intrumentation.    These results provide the first 
thorough exploration of the likely range of observed brightnesses and 
spectral features as a function of season for the disk-averaged spectrum 
of a Mars-like planet, and should prove useful to planet detection and 
characterization missions such as NASA's TPF and ESA's Darwin missions.



\begin{figure}
\begin{center}
\includegraphics[width=10 cm]{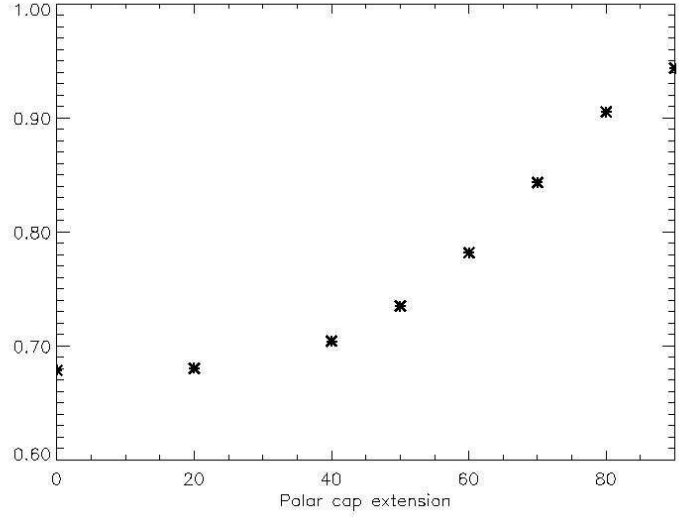}
\caption{ {\footnotesize \emph{Light-curves for an increasingly frozen Mars. 
 The quantities plotted are $ \frac{ \int_{\lambda_{1}}^{\lambda_{2}} \, \mathcal{I} \, \ud \lambda }{\int_{\lambda_{3}}^{\lambda_{4}}  \, \mathcal{I}  \,  \ud \lambda} $, where $\mathcal{I} (\lambda)$ are the disk-averaged radiation intensities shown in fig. \ref{fig:6ab} corresponding to different extensions of the polar-cap area. For this figure we have chosen  $\lambda_{1} = 0.5-\lambda_{2} = 0.7$
(a spectral range where the shape changes considerably), 
and $\lambda_{3} = 0.7-\lambda_{4} = 0.9$ (an interval where reflectivity increases but spectral shape doesn't change appreciably)    } } }  \label{fig:8acz}
\end{center}
\end{figure}


\section*{References}
\begin{description}
\item[[Beichman et al., 1999]]
C.A. Beichman, N. J. Woolf,  and C.A. Lindensmith, Eds. {\it The 
Terrestrial Planet 	Finder}, (JPL: Pasadena), JPL 99-3,1999. 
\item[[Beichman and Velusamy, 1999]]
 C.A. Beichman, and T. Velusamy,  \emph{Sensitivity of TPF Interferometer 
for Planet Detection}, {\it Optical and IR Interferometry 
from Ground and Space}. S.C.Unwin, and R.V. Stachnick, eds. ASP Conference 
Series, 
Vol.194 (San Francisco: ASP), 405, 1999.
\item[[Darwin website]]  http://www.esa.int/esaSC/120382\_index\_0\_m.html
\item[[Giorgini et al., 1997]]
       Giorgini JD, Yeomans DK, Chamberlin AB, Chodas PW, Jacobson RA, 
       Keesey MS, Lieske JH, Ostro SJ, Standish EM, Wimberly RN;
    BULLETIN OF THE AMERICAN ASTRONOMICAL SOCIETY (BAAS),v28,No.3,p.1158, 1996. \\
\emph{JPL's On-Line Solar System Data Service}, 
 http://ssd.jpl.nasa.gov/ 
\item[[G\'orski, et al., 1998]]
K. M. G\'orski, E. Hivon, B. D. Wandelt, 
\emph{Analysis issues for large CMB data sets },
Proceedings: Evolution of Large Scale Structure - Garching, August 1998; \\
 http://www.eso.org/science/healpix/ 
\item[[Goody, Yung, 1996]] R. M. Goody, Y. L. Yung, \emph{Atmospheric Radiation: Theoretical Basis}, Oxford University Press, 1996
\item[[Hanel et al., 1992]] R. A. Hanel, B. J. Conrath, D. E. Jennings, R. E. Samuelson,
\emph{Exploration of the Solar System by Infrared Remote Sensing},
Cambridge Planetary Science Series 7, 1992
\item[[James et al., 1992]] P. B. James, H. H. Kieffer, D. A. Paige,
\emph{ The Seasonal Cycle of Carbon Dioxide on Mars   },
chapter 27 of the book ``Mars'', H. H. Kieffer, B. M. Jakowsky, C. W. Snyder, M. S. Matthews,
The University of Arizona Press, 1992
\item[[JPL Planetquest website]]  http://planetquest.jpl.nasa.gov/
\item[[Kurucz, 1979]] Kurucz, RL, \emph{Model atmospheres for G-star, F-star, A-star, B-star and O-star}, Astrophys J Suppl S 40 (1): 1-340 1979 
\item[[Lellouch et al., 2000]] Lellouch, E., Encrenaz T., de Graauw T., Erard S., Morris
P., Crovisier J., Feuchtgruber H., Girard T., Burgdorf M.,
\emph{The 2.4-45 $\mu$m spectrum of Mars observed with the Infrared
Space Observatory}, Planetary and Space Science 48 (2000) 1393-1405
\item[[Meadows, Crisp, 1996]]  V. S. Meadows, D. Crisp, 
\emph{Ground-based near-infrared observations of the Venus nightside: The thermal structure and water abundance near the surface},
(Journal of Geophysical Research, vol. 101, 4595-4622, Feb. 25 1996) 
\item[[Rothman et al., 2003]]
L.S. Rothman; , A. Barbe, D. Chris Benner, L.R. Brown, C. Camy-Peyret, M.R. Carleer , K. Chance, C. Clerbaux, V. Dana, V.M. Devi, A. Fayt, J.-M. Flaud, R.R. Gamache, A. Goldman, D. Jacquemart, K.W. Jucks, W.J. Lafferty, J.-Y. Mandin, S.T. Massie, V. Nemtchinov, D.A. Newnham, A. Perrin, C.P. Rinsland, J. Schroeder, K.M. Smith, M.A.H. Smith, K. Tang, R.A. Toth, J. Vander Auwera , P. Varanasi, K. Yoshino, \\
\emph{The HITRAN molecular spectroscopic database: edition of 2000 including updates through 2001}, \\
\mbox{Journal of Quantitative Spectroscopy \& Radiative Transfer \S2 (2003) 5-44}
\item[[Tinetti et al., 2004a]]
Tinetti, G.; Meadows, V. S.; Crisp, D.; Fong, W.; Velusamy, T.; Fishbein, E.;
\emph{Detectability of surface and atmospheric properties of Earth-like and Mars-like planets from disk-averaged synthetic spectra}, 
International Journal of Astrobioloy, Vol. 3, Is. S1, January 2004, pp 34
\item[[Tinetti et al., 2004b]]
Tinetti, G.; Meadows, V. S.; Crisp, D.; Fong, W.; Velusamy, T.; Fishbein, E.,
\emph{Sensitivity to environmental properties in globally averaged synthetic spectra of Earth},
American Geophysical Union, Fall Meeting 2002, abstract P41B-0414
\item[[TPF website]] http://planetquest.jpl.nasa.gov/TPF/tpf\_index.html
\item[[Udry and Mayor, 2001]]
Udry, S., \& Mayor, M. 2001, in ESA proc., First European Workshop on Exo/Astrobiology
\item[[Wolszczan and Frail, 1992]]
 Wolszczan, A., \& Frail, D.; 1992, Nature,255,145
\item[[Yung, DeMore]] Y. L. Yung, W. B. DeMore, \emph{Photochemistry of Planetary Atmospheres}, Oxford University Press, 1999
\end{description}

\section*{Ackowledgements}
We would like to thank Eric Hivon for his support in the installation
and use of Healpix.

\end{document}